\documentclass[aps,twocolumn,showpacs,floatfix]{revtex4}

\usepackage[usenames]{color} 

\usepackage{amssymb,epsfig}

\usepackage{psfrag}

\newcommand{\bra}[1]{\left\langle #1 \right|} 
\newcommand{\ket}[1]{\left| #1 \right\rangle}

\newcommand{\zsl}{\!\!\not\!{z}} 
\newcommand{\beq}[1]{\begin{equation}\label{#1}} 
\newcommand{\eeq}{\end{equation}} 
\newcommand{\bea}[1]{\begin{eqnarray}\label{#1}} 
\newcommand{\eea}{\end{eqnarray}} 
\newcommand{\Dlr}{\stackrel{\leftrightarrow}{D}}

\newcommand{\lash}[1]{\not\! #1 \,}  
\newcommand{\nn}{\nonumber}

\newcommand{\dd}{{\rm d}}  
\newcommand{\Gl}[1]{Eq.~(\ref{#1})}

\newcommand{\vecl}[1]{\overleftarrow #1}  
\newcommand{\vecr}[1]{\overrightarrow #1}

\newcommand{\D}{{\cal D}}  
  
\newcommand{\V}{{\cal V}}  
\newcommand{\A}{{\cal A}}  
\newcommand{\T}{{\cal T}}

\newcommand{\al}{\alpha}  
\newcommand{\be}{\beta}  
\newcommand{\ep}{\varepsilon}  
\newcommand{\ga}{\gamma}  
\newcommand{\de}{\delta}  
  
\newcommand{\la}{\lambda}  
  
\newcommand{\si}{\sigma}  
\newcommand{\ro}{\varrho}

\psfrag{QF2F1}{{\footnotesize $\sqrt{Q^2}F_2^p/F_1^p$}}

\psfrag{Q2}{{\footnotesize $Q^2$}}

\psfrag{GMDELTA}{{\footnotesize $G^*_M/(3 G_D)$}}

\psfrag{REM}{{\footnotesize $R_{EM} = -G^*_E/G^*_M$}}

\psfrag{RSM}{{\footnotesize $R_{SM} \propto  -G^*_C/G^*_M$}}

\psfrag{F1p}{{\footnotesize $F_1^p/G_D$}}

\psfrag{F2p}{{\footnotesize $F_2^p/(\mu_p G_D)$}}

\begin{document}


\title{Nucleon Form Factors in QCD}

\author{V.M. Braun, A. Lenz and M. Wittmann\\}

\affiliation{Institut f{{\"u}}r Theoretische Physik, Universit{{\"a}}t
          Regensburg, D-93040 Regensburg, Germany }

\date{\today}

\begin{abstract}
We calculate the electromagnetic and the axial form factors of 
the nucleon within the framework of light cone sum rules (LCSR) to leading order in
QCD and including higher twist corrections. In particular
we motivate a certain choice for the interpolating nucleon field.
We find that a simple model of the nucleon distribution amplitudes which 
deviate from their asymptotic shape, but much less compared to 
the QCD sum rule estimates, allows one to describe the data remarkably well. 
\end{abstract}

\pacs{12.38.-t, 14.20.Dh; 13.40.Gp}




\maketitle

\section{Introduction}
\setcounter{equation}{0}  

{}Form factors play an extremely important role in the
studies of the internal structure of composite particles as the measure
of charge and current distributions.
In particular, the pioneering study  of the nucleon form factors by Hofstadter
and collaborators \cite{Mcallister:1956ng} demonstrated that the nucleons have a 
finite size of the order of a fermi.  
The behavior of the form factors at large momentum transfers is especially interesting.
Already in the pre-QCD times it was established that,
if one can treat the hadrons at high momentum transfer
as collinear beams of $N$ valence quarks located  at small transverse  
separations and exchanging  intermediate 
gluing particles with which  they interact via 
a dimensionless coupling constant,
then the spin-averaged form factor  behaves asymptotically as $1/(Q^2)^{N-1}$
\cite{Brodsky:1973kr}.  
This  hard-exchange picture and the resulting 
dimensional power  counting rules \cite{Brodsky:1973kr,Matveev:1973ra}
can be  formally  extended onto other hard exclusive processes.

After the advent of quantum chromodynamics, 
this hard-gluon-exchange picture was 
formalized with  the help of the
 QCD factorization approach  to  exclusive processes 
\cite{Chernyak:1977as,Radyushkin:1977gp,Lepage:1979zb}.
This approach introduces the concept of hadron distribution amplitudes (DAs). 
They are  fundamental nonperturbative  
functions describing the momentum  distributions  within rare parton configurations 
when the hadron is represented by   
a fixed  number of Fock constituents (quarks, antiquarks and gluons).  
It was shown that in the $Q^2 \to \infty$ limit,
form factors can be  written  in a factorized form, as a convolution 
of distribution amplitudes related to  hadrons in the initial and  final state  
times a ``short-distance'' coefficient function that 
is calculable in QCD perturbation theory.
The leading contribution corresponds to DAs with minimal possible number 
of constituents --- three for baryons and two for mesons.

\begin{figure*}[ht]
\centerline{\epsfxsize12.5cm\epsffile{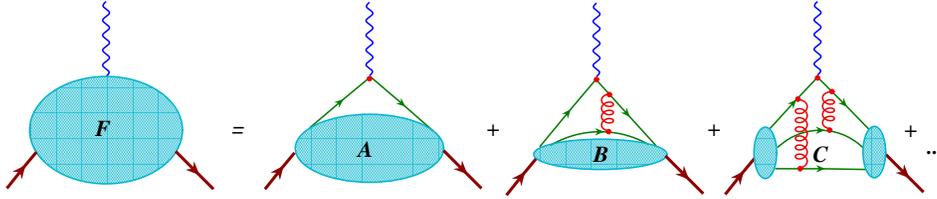}}
\caption{\label{figwan}\small
Structure of QCD factorization for baryon form factors.
}
\end{figure*}

The essential requirement for the 
applicability  of this approach is a high virtuality 
of the exchanged gluons and also of the quarks inside the 
short distance subprocess.
More  generally, in the case of the nucleon form factors
the hard perturbative QCD (pQCD) contribution  is only the third term of the factorization 
expansion. Schematically, one can envisage the expansion of, 
say, the Dirac electromagnetic nucleon form factor
$F_1(Q^2)$ of the form
\bea{schema}
F_1(Q^2) &~\sim~& A(Q^2)
+ \left ( \frac{\alpha_s(Q^2)}{\pi}\right )  \frac{B(Q^2)}{Q^2} 
\nonumber\\&&{}
+ \left ( \frac{\alpha_s(Q^2)}{\pi}\right ) ^2  \frac{C}{Q^4} + 
\ldots 
\eea
where $C$ is  a constant determined by the nucleon DAs, while $A(Q^2)$ and  $B(Q^2)$
are  form-factor-type functions generated by contributions
of low virtualities, see Fig.~\ref{figwan}.
The soft functions  $A(Q^2)$ and  $B(Q^2)$ are purely nonperturbative and cannot be 
further simplified e.g. factorized in terms of DAs. 
In the light-cone formalism, they are determined by 
overlap integrals of the soft parts of  hadronic wave functions corresponding to large 
transverse separations. 
%
%
Various estimates suggest that $A(Q^2)\lesssim 1/Q^6$, $B(Q^2)\lesssim 1/Q^4$ 
and at very large $Q^2$ they are further  suppressed by 
the Sudakov form factor. To be precise, in higher orders in $\alpha_s(Q)$ there exist double-logarithmic 
contributions $\sim 1/Q^4$ \cite{Duncan:1979hi} that are not 
factorized in the standard manner; however, also they are suppressed by the Sudakov mechanism
\cite{Duncan:1979ny,Milshtein:1981cy,Milshtein:1982js}.
Thus, the third term in (\ref{schema}) is formally 
the leading one at large $Q^2$ to power accuracy.

The main problem of the pQCD approach is a numerical suppression
of each hard gluon exchange by the $\alpha_s/\pi$ factor which is a standard perturbation theory 
penalty for each extra loop. 
If, say,  $\alpha_s/\pi \sim 0.1$, the pQCD 
contribution to baryon form factors is suppressed by a factor of 
100 compared to the purely soft term.  As the result, 
the onset of the perturbative regime is postponed to very large momentum transfers since 
the factorizable pQCD contribution ${O}(1/Q^4)$ has to win over nonperturbative effects 
that are suppressed by  extra powers of $1/Q^2$, but do not involve small coefficients.  
There is a growing consensus that ``soft'' contributions play the dominant role at present 
energies. Indeed, it is known for a long time that the use of 
QCD-motivated models for the  wave functions allows one to
obtain, without much effort,  soft contributions comparable in size 
to  experimentally observed values (see, e.g.~\cite{Isgur:1984jm,Isgur:1988iw,Kroll:1995pv}). 
A modern trend \cite{Radyushkin:1998rt,Diehl:1998kh} is to  use the concept of  
generalized parton distributions  (GPDs) to describe/parametrize soft contributions in various 
exclusive reactions, see  \cite{Goeke:2001tz,Diehl:2003ny,Belitsky:2005qn} for recent reviews,
and the models of GPDs usually are chosen such that the experimental data on form factors 
are described by the soft contributions alone, cf. Refs.~\cite{Belitsky:2003nz,Diehl:2004cx,Guidal:2004nd}. 
A subtle  point for these semi-phenomenological 
approaches is to avoid double counting of hard rescattering 
contributions  ``hidden'' in the model-dependent hadron wave functions
or GPD parametrizations. 

The dominant role  of the soft contribution for the  pion form factor at  
moderate momentum transfers, up to $Q^2\sim 2-3$~GeV$^2$, is supported by  its  
calculation \cite{Ioffe:1982qb,Nesterenko:1982gc}
within the QCD sum rule approach \cite{Shifman:1978bx}.
The application of the method at  higher $Q^2$ faces the 
problem  that the inclusion   of nonperturbative effects due to vacuum 
condensates through  the expansion over  inverse powers of the Borel parameter $M^2$ 
interferes with the large-$Q^2$ expansion of the form factors, 
producing an ill-behaved  series of the type $\sum_n c_n (Q^2/M^2)^n$.
{}For the nucleon form factors, the  QCD sum rule approach only works in the region 
 of small momentum transfers  $Q^2 <  1$~GeV$^2$ \cite{Belyaev:1992xf,Castillo:2003pt}. 
To extend the results  to higher $Q^2$, it was proposed \cite{Bakulev:1991ps} to resum 
the $(Q^2/M^2)^n$ contributions originating from  the Taylor expansion 
of simple models for nonlocal condensates.
 Another approach \cite{Nesterenko:1982gc,Nesterenko:1983ef}
is to use   the so-called local quark-hadron duality approximation,
in which the condensates are effectively neglected.
The parameter-free results for the pion and nucleon form factors obtained in this way are in 
a rather good agreement with the existing data.

We also have to mention the dispersion approach of Ref.~\cite{Meissner,Baldini:2005vn}
which allows to analyze form factors for all momenta (space- and time-like) in a largely
model-independent manner in terms of spectral functions on a hadronic level.
Also, in future, one expects that the rapid development of lattice QCD will allow one to calculate
baryon form factors to sufficient precision from first principles, see e.g. 
\cite{Gockeler:2003ay,Alexandrou:2004xn,Gockeler:2004vx,Gockeler:2004mn}. 
Such studies are necessary and interesting in its own right, but do not add to our understanding of 
how QCD actually ``works'' to transfer the large momentum along the nucleon constituents, the quarks and gluons.
The main motivation to study ``hard''processes has always been to understand hadron properties in terms of 
quark and gluon degrees of freedom; for example, the rationale for the continuing measurements
of the total inclusive cross section in deep inelastic scattering is to extract quark and gluon parton distributions. 
Similar, experimental 
measurements of the form factors at large momentum transfers should eventually allow one to determine baryon distribution 
amplitudes and this task is obscured by the presence of large ``soft'' contributions which have to be subtracted.

In Ref.~\cite{Braun:2001tj} we have suggested to calculate baryon form factors
for moderately large $Q^2$ using light-cone  sum rules (LCSR) \cite{Balitsky:1989ry,Chernyak:1990ag}.  
This technique is attractive because in LCSRs  ``soft'' contributions to the form factors are calculated in 
terms of the same DAs that
enter the pQCD calculation and there is no double counting. Thus, the LCSRs provide one with the most 
direct relation of the hadron form factors
and distribution amplitudes that is available at present, with no other nonperturbative parameters.  
 
\begin{figure}[ht]
\centerline{\epsfxsize5cm\epsffile{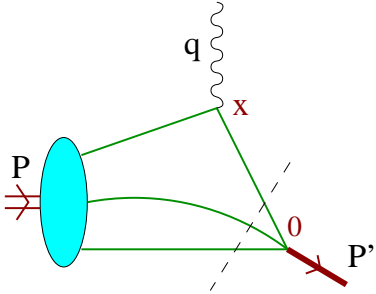}}
\caption{\label{figsum}\small
Schematic structure of the light-cone sum rule for baryon form factors.
}
\end{figure}

The basic object of the LCSR approach is the 
correlation function $$\int\! dx\, e^{-iqx}\langle 0| T \{ \eta (0) j(x) \} | N(P) \rangle $$ in which
$j$  represents the electromagnetic (or weak) probe and $\eta$
is a suitable operator with nucleon quantum numbers.    
 The other (in this example, initial state) nucleon  is explicitly represented by its state vector 
 $| N(P)\rangle $, see a schematic representation in Fig.~\ref{figsum}.
 When both  the momentum transfer  $Q^2$ and 
 the momentum $(P')^2 = (P-q)^2$ flowing in the $\eta$ vertex are large and negative,
 the asymptotics of the correlation function is governed by the light-cone kinematics $x^2\to 0$ and
 can be studied using the operator product expansion (OPE)   
$T \{ \eta(0) j(x) \} \sim \sum C_i(x) {\cal O}_i(0)$ on the 
light-cone $x^2=0$.   The  $x^2$-singularity  of a particular perturbatively calculable
short-distance factor  $C_i(x)$  is determined by the twist of the relevant
composite operator ${\cal O}_i$, whose matrix element $\langle 0|  {\cal O}_i(0)| N(P) \rangle $
is given by an appropriate moment of the nucleon DA.
Next, one can represent the answer in form of the dispersion integral in $(P')^2$ and define the nucleon contribution
by the cutoff in the quark-antiquark invariant mass, the so-called interval of duality $s_0$ (or continuum threshold).
The main role of the interval of duality is that it does not allow large momenta $|k^2| > s_0$ to flow through the 
 $\eta$-vertex; to the lowest order $O(\alpha_s^0)$ one obtains a purely soft    
contribution to the form factor as a sum of terms ordered by twist of the relevant operators and
hence including both the leading- and the higher-twist nucleon DAs. Note that, in difference to the hard mechanism, the 
contribution of higher-twist DAs is only suppressed by powers of 
$|(P')^2|\sim 1-2$~GeV$^2$ (which is translated to the suppression
by powers of the Borel parameter after applying the usual QCD sum rule machinery), but not by powers of $Q^2$. This feature is
in agreement with the common wisdom that soft contributions are not constrained to small transverse separations.  

The LCSR expansion also  contains terms  
generating the asymptotic pQCD contributions. They   appear 
at proper order in $\alpha_s$, i.e., in  the $O(\alpha_s)$ term for the
pion form factor, at the $\alpha_s^2$ order for the nucleon form factors, etc. 
In the pion case, it was explicitly demonstrated 
\cite{Braun:1999uj,Bijnens:2002mg} that the contribution of hard  
rescattering is correctly reproduced in the LCSR   
approach as a part of the $O(\alpha_s)$ correction.
It should be noted that  the  diagrams of LCSR that 
contain the ``hard'' pQCD  contributions also possess ``soft'' parts,
i.e., one should perform  a separation  of ``hard'' and ``soft''
terms inside each diagram.  As a result, 
the distinction between ``hard'' and ``soft'' contributions appears to 
be scale- and scheme-dependent \cite{Braun:1999uj}. 
During the  last years there have been numerous applications of LCSRs  
to mesons, see \cite{Braun:1997kw,Colangelo:2000dp} for a review.
{}Following the work \cite{Braun:2001tj} nucleon form factors 
were further considered in this framework in Refs.~\cite{Lenz:2003tq,Wang:2006uv,Wang:2006su} 
and the weak decay $\Lambda_b\to p\ell\nu_\ell$ in
\cite{Huang:2004vf}.  The generalization to the $N\gamma\Delta$ transition form factor
was worked out in \cite{Braun:2005be}.

In this paper, we go beyond the original work  \cite{Braun:2001tj} in several important aspects.
First, we present a detailed study using different interpolating currents  for the nucleon and  
choose one which appears to be the optimal. Second, we calculate both the electromagnetic and 
weak decay form factors. Third, we make an update of the parameters of higher-twist DAs which 
feature prominently in this approach and are important numerically. We then formulate a simple model 
for the DAs that provides a good description of the available experimental data.
Finally, we include a 
complete summary of higher-twist  DAs and work out the light-cone expansion of three-quark currents
for all Lorentz structures, which extends the results given in \cite{Braun:2001tj}.
The presentation is organized as follows. Section 2 is introductory and contains the form factor 
definitions and some general discussion. Section 3 is devoted to the construction of the sum rules.
We present here our numerical results and the conclusions. The final section 4 
is reserved for a short summary and an outlook. The paper contains five appendices devoted to 
a summary of correlation functions to tree-level accuracy, summary of three-quark
nucleon distribution amplitudes, the OPE of generic three-quark
amplitudes to twist-5 accuracy, and the QCD sum rule estimates for the parameters of higher-twist DAs.

\section{Preliminaries}

\subsection{Electromagnetic Form Factors of the Nucleon}\label{emff.sse}
The matrix element of the electromagnetic current 
\beq{em}
 j_{\mu}^{\rm em}(x) = e_u \bar{u}(x) \ga_{\mu} u(x) + e_d \bar{d}(x) \ga_{\mu} d(x)
\eeq
taken between nucleon states is conventionally written in terms of the 
{Dirac} and {Pauli form factors} $F_1(Q^2)$ and $F_2(Q^2)$:
\bea{F1F2}
\lefteqn{\bra{N(P')}j_{\mu}^{\rm em}(0)\ket{N(P)}=}
\nonumber\\&=&
\bar{N}(P')\left[\ga_{\mu}F_1(Q^2)-i\frac{\sigma_{\mu\nu}q^{\nu}}{2m_N}F_2(Q^2)\right]N(P),
\eea
where $P_\mu$ is the initial nucleon momentum, $P^2 =m_N^2$, $P'=P-q$, $Q^2 :=-q^2$, 
$\sigma_{\mu\nu}=\frac{i}{2}[\ga_{\mu},\ga_{\nu}]$  and $N(P)$ is the nucleon spinor. 

Experimental data on the  scattering of electrons off nucleons, e.g. $e^- + p \to e^- + p$,
is usually presented in terms of the {electric} $G_E(Q^2)$  and {magnetic} $G_M(Q^2)$
Sachs form factors which are related to $F_{1,2}(Q^2)$ as
\bea{GMGE}
G_M(Q^2) & = & F_1(Q^2)+F_2(Q^2),
\\
G_E(Q^2) & = & F_1(Q^2)-\frac{Q^2}{4m_N^2}F_2(Q^2).
\eea
In the Breit frame the form factors
$G_E(Q^2)$ and $G_M(Q^2)$ can be thought of, loosely speaking, as the Fourier transforms of the 
charge distribution and magnetization density in the nucleon.

The normalization of the form factors at $Q^2=0$ is given by the nucleon charges
and magnetic moments
\bea{normal}
 G^{p}_{E}(0)=1,&~&G_M^p(0)=\mu_p=2.792847337(29)\,, 
\nonumber\\
G^{n}_{E}(0)=0,&&G_M^n(0)=\mu_n=-1.91304272(45) \; 
\eea
for the proton and the neutron, respectively \cite{PDG}.

Experimentally one finds 
\cite{Walker94,Andivahis94,Litt70,Berger71,Janssens66,Arrington:2003df,Christy:2004rc,Qattan:2004ht,Lung93,Kubon02,Anklin98}
that the magnetic form factors of both the proton
and the neutron can be described by the famous dipole formula:
\begin{eqnarray}
G_M^p(Q^2) &=&  \frac{\mu_p}{\left(1 + \frac{Q^2}{\mu_0^2}\right)^2}\,,
\\
G_M^n(Q^2) &=&  \frac{\mu_n}{\left(1 + \frac{Q^2}{\mu_0^2}\right)^2}\,,
\end{eqnarray}
with $\mu_0^2 = 0.71$~GeV$^2$.
For the electric form factor of the neutron the measured values are close to zero \cite{Zhou01,Rohe99}.
In the case of the electric form factor of the proton the situation was
controversial for some time, with the experimental measurements using the classical method
of Rosenbluth separation producing very different results compared to the ones obtained using
the method of polarization transfer.  Recently it was 
argued
\cite{Maximon:2000hm,Blunden:2003sp,Arrington:2003ck,Chen:2004tw,Arrington:2004ae,Tomasi-Gustafsson:2004ms,Afanasev:2005mp,Kondratyuk:2005kk,Blunden:2005ew}
 that the former approach is not applicable for sufficiently large 
momentum transfers, as the contribution of the electric form factor to the spin-averaged 
cross section is strongly contaminated by contributions of the two-photon exchange. 
The existing estimates 
\cite{Maximon:2000hm,Blunden:2003sp,Arrington:2003ck,Chen:2004tw,Arrington:2004ae,Tomasi-Gustafsson:2004ms,Afanasev:2005mp,Kondratyuk:2005kk,Blunden:2005ew} 
indicate that the two-photon exchange corrections 
have the right sign and order of magnitude to bring the values obtained via the 
Rosenbluth separation to the ones measured using the polarization transfer, 
although the situation is not finally settled. In this work we rely on the 
polarization transfer data \cite{JLab1, JLab2, JLab3}.

\subsection{Charged Weak Form Factors}\label{ccweakff.sse}

In order to describe charged current (CC) neutrino reactions like
\begin{eqnarray}
\nu_\mu + n & \to & \mu^- + p,
\\
\bar \nu_\mu + p & \to & \mu^+ + n,
\end{eqnarray}
one has to deal with matrix elements between nucleon states of the 
vector $V_{\mu}^{CC}$ and the axial-vector current $A_{\mu}^{CC}$:
\bea{current3}
V_{\mu}^{CC} (x) & = & \bar{\Psi} \ga^{\mu} \tau^+ \Psi (x)\, ,
\\
A_{\mu}^{CC} (x) & = & \bar{\Psi} \ga^{\mu} \ga_5 \tau^+ \Psi (x)\, , 
\eea
where $\Psi$ is an (iso)spinor consisting of an up- and a down-quark and 
$\tau^+ = 1/2 (\tau_1 + i \tau_2)$ is a linear combination of the familiar 
Pauli matrices.

One defines the corresponding vector and the axial-vector form factors as
\bea{formfactor3}  
\lefteqn{\bra{N(P')} V_\mu^{CC}(0) \ket{N(P)} =}
\nonumber\\&=&  
\bar{N}(P') \left[ 
\ga_\mu  F_1^{CC}(Q^2) -  i \frac{\si_{\mu\nu}  q^\nu}{2 m_N}  F_2^{CC}(Q^2) 
\right]N(P),  
\nonumber\\
\lefteqn{\bra{N(P')} A_\mu^{CC}(0) \ket{N(P)} =
\bar{N}(P') \Big[ \ga_\mu G_A^{CC}(Q^2) }
\nn \\ && {}
- \frac{q_\mu}{2 m_N}  G_P^{CC}(Q^2)
-  i \frac{\si_{\mu\nu}  q^\nu}{2 m_N}  G_T^{CC}(Q^2) 
\Big] \ga_5 N(P)\, .  
\nonumber
\\
\eea  
The vector form factors can be related with
the electromagnetic ones with the help of isospin symmetry, to wit
\bea{iso2}
\langle p| \bar u\gamma^\mu d |n\rangle &=&
\langle p|j^\mu_{em}|p\rangle - \langle n|j^\mu_{em}|n\rangle,
\nonumber\\
\langle p| \bar u\gamma^\mu\gamma_5 d |n\rangle
&=&  \langle p|\big(\bar{u} \gamma^{\mu} \gamma_5 u - \bar{d} \gamma^{\mu} \gamma_5 d\big) |p\rangle.
\eea
The first relation gives, e.g.
\bea{vecff}
  F^{CC}_{1}(Q^2) &=& F_1^p(Q^2)-F_1^n(Q^2)\,, 
\nonumber\\
  F^{CC}_{2}(Q^2) &=&  F_2^p(Q^2)-F_2^n(Q^2)\,. 
\eea
The axial form factor $G_A^{CC}(Q^2)$ can be determined either from quasi-elastic 
neutrino scattering or from pion electroproduction (with the help of 
current algebra).
The neutrino data are available for $Q^2$ values up to 3 GeV$^2$ 
\cite{GACC1, GACC2,GACC3,GACC4}. They were reanalyzed recently in \cite{Budd}.
The pion electroproduction data exist for $Q^2 < 1$ GeV$^2$,  e.g. \cite{Mainz}. 
After the inclusion of a finite pion mass correction in the analysis \cite{Bernard}, 
the extracted form factor agrees well with the determinations in neutrino scattering. 
All the existing experimental data for $G_A^{CC}(Q^2)$ at $Q^2< 1$ GeV$^2$
are very well described by the dipole formula:  
\bea{dipole3}  
G_A^{CC}(Q^2)  & = & \frac{g_A}{\left(1+\frac{Q^2}{M_A^2} \right)^2}, 
\eea  
with $g_A=1.267\pm 0.004$. The mass parameter is fitted to be $M_A  =  1.001 \pm 0.020$~{GeV} and 
$M_A  =  1.013 \pm 0.015$~{GeV} from neutrino scattering and  pion electroproduction,
respectively.

The pseudoscalar form factor $G_P^{CC}(Q^2)$ can be extracted separately from muon capture of the proton 
$\mu^- + p \to \nu_\mu + n$ or from pion electroproduction.
In this case only the  data for $Q^2$-values below 0.2 GeV$^2$ exist \cite{GPdata}, 
which is to low for the application of our method.
Using PCAC and the pion pole dominance model one can express the
pseudoscalar form factor $G_P^{CC}$ in terms of the axial form factor
$G_A^{CC}$:
\begin{equation}
G_P^{CC}(Q^2) = \frac{4 m_N^2 G_A^{CC}(Q^2)}{Q^2 + m_\pi^2}\,.
\label{PCAC}
\end{equation} 
This form is consistent with the conservation of the flavor nonsinglet axial current 
in the chiral limit $(m_\pi^2 \to 0)$.

Finally, the tensor form factor  $G_T^{CC}(Q^2)$ must vanish by virtue of the 
isospin symmetry and T-invariance,
so it is normally not included. The reason why we leave it in Eq.~(\ref{formfactor3}) is
that in our approach the initial and the final state nucleons are treated differently, 
so that T-invariance is not manifest.

\subsection{Neutral Weak Form Factors}\label{ncweakff.sse}

The cross section for elastic neutrino-proton and neutrino-neutron  scattering can be 
expressed in terms of matrix elements of a vector $V_{\mu}^{NC}$ and 
an axial-vector $A_{\mu}^{NC}$ neutral currents:
\bea{current2}
V_{\mu}^{NC} (x)& = & \frac{1}{2} 
\Big[
\left( 1- \frac{8}{3} \sin^2 \theta_W \right) \bar{u} \ga^{\mu} u (x)
\nonumber\\&&{}-
\left( 1- \frac{4}{3} \sin^2 \theta_W \right) \bar{d} \ga^{\mu} d(x)
\Big],
\\
A_{\mu}^{NC} (x)& = & \frac{1}{2} 
\left[\bar{u} \ga^{\mu} \ga_5 u (x)- \bar{d} \ga^{\mu} \ga_5 d (x)\right],
\eea
where  $\theta_W$ is the Weinberg angle. 
The matrix elements of neutral  currents over the nucleon states are conventionally  
written as
\bea{formfactor2}  
\lefteqn{\bra{N(P')} V_\mu^{NC}(0) \ket{N(P)} =}
\nonumber\\&=&  
\bar{N}(P') \left[ 
\ga_\mu  F_1^{NC}(Q^2) -  i \frac{\si_{\mu\nu}  q^\nu}{2 m_N}  F_2^{NC}(Q^2) 
\right]N(P)  
\nonumber\\
\lefteqn{\bra{N(P')} A_\mu^{NC}(0) \ket{N(P)} =
\bar{N}(P') \Big[ \ga_\mu G_A^{NC}(Q^2) }
\nonumber \\ && {}
- \frac{q_\mu}{2 m_N}  G_P^{NC}(Q^2)
-  i \frac{\si_{\mu\nu}  q^\nu}{2 m_N}  G_T^{NC}(Q^2) 
\Big] \ga_5 N(P)\, .  
\nonumber 
\\ 
\eea  
The vector form factors $F_1^{NC}$ and $ F_2^{NC}$ are, again,  just linear
combinations of the electromagnetic form factors of the nucleon. 
For the axial neutral weak form factors ($G_A^{NC}$ and $G_P^{NC}$) there is little
data, and only for $Q^2 < 1$~GeV$^2$ \cite{Ahrens} which is below 
the region we try to describe theoretically. The tensor form factor $G_T^{NC}(Q^2)$ must vanish 
by virtue of T-invariance; the reason we include it will become clear later.

\subsection{Light-Cone Kinematics}\label{lckin.sse}

Having in mind the practical construction of light-cone sum rules that involve nucleon DAs,
we define a light-like vector $z_\mu$ by the condition
\beq{z}
       q\cdot z =0\,,\qquad z^2 =0
\eeq
and introduce the second light-like vector 
vector 
\bea{vectors}
p_\mu &=& P_\mu  - \frac{1}{2} \, z_\mu \frac{m_N^2}{P\cdot z}\,,~~~~~ p^2=0\,, 
\eea
so that $P \to p$ if the nucleon mass can be neglected, $m_N \to 0$.
The photon momentum can be written as 
\begin{eqnarray}
q_{\mu}=q_{\bot \mu}+ z_{\mu}\frac{P\cdot q}{P\cdot z}\, .
\end{eqnarray}
We also need the projector onto the directions orthogonal to $p$ and $z$,
\begin{equation}
       g^\perp_{\mu\nu} = g_{\mu\nu} -\frac{1}{pz}(p_\mu z_\nu+ p_\nu z_\mu),
\end{equation}
and use the notation
\begin{equation}
    a_z\equiv a_\mu z^\mu, \qquad  a_p\equiv a_\mu p^\mu\,,
\end{equation}
for arbitrary Lorentz vectors $a_\mu$.
In turn, $a_\perp$ denotes the generic component of $a_\mu$ orthogonal to
$z$ and $p$, in particular
\beq{qperp}
   q_{\perp\mu} = q_\mu -\frac{pq}{pz} z_\mu\,.
\eeq

We use the standard Bjorken--Drell
convention \cite{BD65} for the metric and the Dirac matrices; in particular,
$\gamma_{5} = i \gamma^{0} \gamma^{1} \gamma^{2} \gamma^{3}$,
and the Levi-Civita tensor $\epsilon_{\mu \nu \lambda \sigma}$
is defined as the totally antisymmetric tensor with $\epsilon_{0123} = 1$.

Assume for a moment that the nucleon moves in the positive 
${\bf e_z}$ direction, then $p^+$ and $z^-$ are the only nonvanishing 
components of $p$ and $z$, respectively. 
The infinite momentum frame can be visualized
as the limit $p^+ \sim Q \to \infty$ with fixed $P\cdot z = p \cdot z \sim 1$ 
where $Q$ is the large scale in the process.
Expanding the matrix element in powers of $1/p^+$ introduces
the power counting in $Q$. In this language,  twist counts the
suppression in powers of $p^+$. Similarly, 
the nucleon spinor $N_\ga(P,\la)$ has to be decomposed 
in ``large'' and ``small'' components as
\bea{spinor}
N_\ga(P,\la) &=& \frac{1}{2 p\cdot z} \left(\!\not\!{p}\! \!\not\!{z} +
\!\not\!{z}\!\!\not\!{p} \right) N_\ga(P,\la)
\nonumber\\&=& N^+_\ga(P,\la) + N^-_\ga(P,\la)\,,
\eea
where we have introduced two projection operators
\beq{project}
\Lambda^+ = \frac{\!\not\!{p}\! \!\not\!{z}}{2 p\cdot z} \quad ,\quad
\Lambda^- = \frac{\!\not\!{z}\! \!\not\!{p}}{2 p\cdot z}
\eeq
that project onto the ``plus'' and ``minus'' components of the spinor.
Note  the useful relations
\beq{bwgl}
\lash{p} N(P) = m_N N^+(P)\,,\quad \lash{z} N(P) = \frac{2 p \cdot z}{m_N} N^-(P)
\eeq
that are a consequence of the Dirac equation \hfill\break 
$\not\hspace*{-0.10cm}P N(P) = m_N N(P)$.
Using the explicit expressions for $N(P)$ it is easy to see 
that $\Lambda^+N = N^+ \sim \sqrt{p^+}$ while $\Lambda^-N = N^- \sim 1/\sqrt{p^+}$.

Note that all expressions are invariant under the reparametrization $z_\mu \to \alpha z_\mu$
where $\alpha$ is a real  number; we will use this freedom to set $z_\mu$ equal to the 
``minus'' component of the distance between the currents in the operator product.

%
%
%
\section{Light-cone sum rules for baryon form factors}

%
%
\subsection{Choice of the current}

As already mentioned, in the LCSR approach one of the participating nucleons
is replaced by a suitable interpolating current for which there are several
choices. Altogether, there exist three local operators with isospin $I=1/2$  
numbers that do not involve derivatives \cite{Ioffe:1982ce}. They can be chosen as
\bea{currents}
\eta_1(x) &=& \ep^{ijk} \left[u^i(x) C\ga_\mu u^j(x)\right]\,\ga_5 \ga^\mu d^k(x)\,,
\\
\eta_2(x) &=& \ep^{ijk} \left[u^i(x) C\si_{\mu\nu} u^j(x)\right]\,\ga_5 \si^{\mu\nu} d^k(x)\,,
\\
\eta_3(x) &=&\frac{2}{3} \epsilon^{ijk} \Big( \left[u^i(x) C \zsl u^j(x)\right] 
\ga_5 \zsl d^k(x) 
\nn\\ &&{}
-  \left[u^i(x) C \zsl d^j(x)\right]\ga_5 \zsl u^k(x)\Big)\,, 
\label{CZcurrent}
\eea 
where $u(x)$ and $d(x)$ are the $u$-quark and the $d$-quark field operators, respectively, 
$i,j,k$ are color indices, $C$ is the charge conjugation matrix \cite{BD65} and  $z$ is a light-like vector, $z^2=0$. 
The corresponding couplings  
\bea{coupling} 
\bra{0} \eta_1(0)\ket{N(P)}  &=& \lambda_1 m_N N(P) \,,  
\nn \\
\bra{0} \eta_2(0)\ket{N(P)}  &=& \lambda_2 m_N N(P) \,,  
\nn \\
\bra{0} \eta_3(0) \ket{N(P)} &=&   f_{\rm N} (Pz) \not\!{z} N(P)\,  
\label{eta3}
\eea  
are well known, albeit with limited precision, from the vast QCD sum rule 
literature, see Eq.~(\ref{numerics1}) in Appendix B for the current estimates. 

The operator $\eta_1$ is known as the Ioffe-current \cite{Ioffe:1981kw}. There is overwhelming
evidence (see, however \cite{Dosch:1988vv}) 
that this current gives rise to more accurate and reliable sum rules compared to  
$\eta_2$ \cite{Chung:1981cc}. By this reason we do not use $\eta_2$ in the construction of the 
LCSRs below. The both constants $\lambda_1$ and $\lambda_2$ do appear in the sum rules, however,
as they determine the normalization of the higher-twist-4 nucleon DAs.  

In turn, the operator $\eta_3$ is twist-3 and the corresponding coupling $f_N$ determines 
the normalization of the leading twist nucleon DA \cite{Che84}.
Using the currents of lower twist is in general advantageous, as the corresponding correlation functions have 
lower dimension and are less affected by the model-dependent continuum subtraction. 
LCSRs obtained with this current are most close in spirit to pQCD factorization.   
The price to pay is that this current couples both to the spin $J=1/2$ and spin $J=3/2$ baryons
and it is unclear whether the unwanted $I=3/2$ contributions to the correlation function
are sufficiently suppressed. This current has been used more rarely in the practice of QCD sum rule
calculations, so that there is less experience.

In Ref.~\cite{Braun:2001tj} a modification of the current (\ref{CZcurrent}) was used
\bea{eta4}
&&\eta_4(x) = \epsilon^{ijk} \left[u^i(x) C \zsl u^j(x)\right] 
\ga_5 \zsl d^k(x)\,,
\nn\\
&& \bra{0} \eta_4(0) \ket{N(P)} =   f_{\rm N} (Pz) \not\!{z} N(P)\,,   
\eea 
which, in difference to $\eta_3$, also couples to the isospin $I=3/2$ states
(e.g. the $\Delta$-isobar). 

A priory, it is not obvious whether using the pure isospin-1/2 or, similarly, pure spin-1/2 currents 
improves the accuracy of the sum rules. It is conceivable that the summation over quantum numbers 
makes the duality approximation for taking into account contributions of heavy resonances and the continuum more
accurate and also suppresses poorly known contributions of higher-dimension(twist) operators.
In a different context, there have been various proposals to add together correlation functions of opposite parity, 
\cite{Zhitnitsky:1985dd}, use chirally projected quark fields, e.g. \cite{Huang:1998gp},  etc.

We believe that there is no general recipe; one has to consider each case separately and the conclusions
can vary. For the problem at hand, it was noticed in \cite{Lenz:2003tq} that using the currents $\eta_3$ and $\eta_4$ 
one obtains numerical results for the nucleon form factors that differ significantly 
from one another. In this work we demonstrate that
this difference is due to the contamination of the sum rules \cite{Braun:2001tj} 
by the contributions of isospin $I=3/2$ states and, therefore, 
the use of the pure-isospin current $\eta_3$ is strongly preferred compared to $\eta_4$.
On the other hand, the sum rules obtained using $\eta_1$ and $\eta_3$
are complementary to a large extent and both of them are useful.  
Nevertheless, using the Ioffe current $\eta_1$ produces the sum rules 
that are more stable and seem to be superior in all respects, 
which makes this current to be our final choice.

\subsection{Correlation functions}

We consider the set of correlation functions
\bea{correlator} 
T_{\nu}^{i,\mathrm{em}}(P,q) &=& i\! \int\! \dd^4 x \, e^{i q x}  
\bra{0} T\left[\eta_i(0) j_{\nu}^{\mathrm{em}}(x)\right] \ket{N(P)},
\nn\\
T_{\nu}^{i,\mathrm{a}}(P,q) &=& i\! \int\! \dd^4 x \, e^{i q x}  
\bra{0} T\left[\eta_i(0) j_{\nu}^{\mathrm{a}}(x)\right] \ket{N(P)},
\nn\\
T_{\nu}^{i,\mathrm{v}}(P,q) &=& i\! \int\! \dd^4 x \, e^{i q x}  
\bra{0} T\left[\eta_i(0) j_{\nu}^{\mathrm{v}}(x)\right] \ket{N(P)},
\nn\\[-2mm]
\eea
where $T$ denotes time-ordering, 
$\ket{N(P)}$ is the proton state with four-momentum $P_\mu$, 
$P^2 = m_N^2$,
$\eta_i$ with $i=1,3,4$ are the nucleon currents defined in
(\ref{currents}). Further,  $j^{\mathrm{em}}_{\nu}$ is the electromagnetic 
current defined (\ref{em}) whereas $j^{\mathrm{a}}_{\nu}$ and $j^{\mathrm{v}}_{\nu}$ are the 
isospin-one axial and vector currents, respectively: 
\bea{axial}  
j_{\nu}^{\mathrm{a,nc}}(x) &=& 
\frac{1}{2}[\bar{u}(x) \ga_{\nu} \ga_5 u(x) - \bar{d}(x) \ga_{\nu} \ga_5 d(x)]\,,  
\nn \\
j_{\nu}^{\mathrm{a,cc}}(x) &=&  \bar{u}(x) \ga_{\nu}\gamma_5 d(x)\,,
\nn \\
j_{\nu}^{\mathrm{v,nc}}(x) &=& 
\frac{1}{2}[\bar{u}(x) \ga_{\nu} u(x) - \bar{d}(x) \ga_{\nu} d(x)],  
\nn \\
j_{\nu}^{\mathrm{v,cc}}(x) &=&  \bar{u}(x) \ga_{\nu} d(x)\,.
\eea
The correlation functions in (\ref{correlator}) involve several invariant functions
that can be separated by the appropriate projections. Lorentz structures that are 
most useful for writing the LCSRs are usually those containing the maximum 
power of the large momentum $p^+ \sim pz$. We define, for the 
Ioffe current
\bea{project4}
  \Lambda_+ T^{1,\mathrm{em}}_z &=& (pz)\left\{ m_N \mathcal{A}^{\mathrm{em}}_1 + 
    \!\not\!q_\perp \mathcal{B}^{\mathrm{em}}_1\right\}N^+(P)\,,
\nn\\
  \Lambda_+ T^{1,\mathrm{a}}_z &=& (pz)\left\{ m_N \mathcal{A}^{\mathrm{a}}_1 + 
    \!\not\!q_\perp \mathcal{B}^{\mathrm{a}}_1\right\}\gamma_5N^+(P)\,,
\nn\\
  \Lambda_+ T^{1,\mathrm{v}}_z &=& (pz)\left\{ m_N \mathcal{A}^{\mathrm{v}}_1 + 
  \!\not\!q_\perp \mathcal{B}^{\mathrm{v}}_1\right\}N^+(P)\,,
\eea
where $\mathcal{A}$ and  $\mathcal{B}$ 
depend on the Lorentz-invariants $Q^2=-q^2$ and $P'^2 = (P-q)^2$. For the 
leading-twist current $\eta_3$ we use instead
\bea{project3}
  T^{3,\mathrm{em}}_z &=& \frac{2(pz)^3}{m_N^2} \left\{m_N\mathcal{A}^{\mathrm{em}}_3 + 
      \!\not\!q_\perp \mathcal{B}^{\mathrm{em}}_3\right\}N^-(P)\,,
\nn\\
  T^{3,\mathrm{a}}_z &=& \frac{2(pz)^3}{m_N^2} \left\{m_N\mathcal{A}^{\mathrm{a}}_3 + 
     \!\not\!q_\perp \mathcal{B}^{\mathrm{a}}_3\right\}\gamma_5 N^-(P)\,,
\nn\\
  T^{3,\mathrm{v}}_z &=& \frac{2(pz)^3}{m_N^2} \left\{m_N\mathcal{A}^{\mathrm{v}}_3 + 
  \!\not\!q_\perp \mathcal{B}^{\mathrm{v}}_3\right\}N^-(P)\,,
\eea
and similarly for $\eta_4$:
\bea{project5}
  T^{4,\mathrm{em}}_z &=& \frac{2(pz)^3}{m_N^2} \left\{m_N\mathcal{A}^{\mathrm{em}}_4 + 
      \!\not\!q_\perp \mathcal{B}^{\mathrm{em}}_4\right\}N^-(P)\,,
\nn\\
  T^{4,\mathrm{a}}_z &=& \frac{2(pz)^3}{m_N^2} \left\{m_N\mathcal{A}^{\mathrm{a}}_4 + 
      \!\not\!q_\perp \mathcal{B}^{\mathrm{a}}_4\right\}\gamma_5 N^-(P)\,,
\nn\\
  T^{4,\mathrm{v}}_z &=& \frac{2(pz)^3}{m_N^2} \left\{m_N\mathcal{A}^{\mathrm{v}}_4 + 
  \!\not\!q_\perp \mathcal{B}^{\mathrm{v}}_4\right\}N^-(P)\,,
\eea
The elastic nucleon form factor contribution of interest corresponds to a pole 
term in the variable $P'^2$. For the relevant projections we get
\begin{widetext}
\bea{projections-lhs}
&&{\cal A}_1^{\rm em} = \frac{2 \lambda_1 F_1^{\rm em}}{m_N^2- P'^2},
\hspace{0.5cm} 
{\cal B}_1^{\rm em} = \frac{\lambda_1  F_2^{\rm em}}{m_N^2- P'^2},
\hspace{0.25cm} \hspace{0.25cm}
{\cal A}_1^{\rm a,nc} = \frac{2 \lambda_1  G_A^{NC}}{m_N^2- P'^2},
\hspace{0.5cm} 
{\cal B}_1^{\rm a,nc} = \frac{\lambda_1 G_T^{NC}}{m_N^2- P'^2},
\nn
\\
&&{\cal A}_1^{\rm v,cc} = \frac{2 \lambda_1  F_1^{\rm v,cc}}{m_N^2- P'^2},
\hspace{0.5cm} 
{\cal B}_1^{\rm v,cc} = \frac{\lambda_1 F_2^{\rm v,cc}}{m_N^2- P'^2},
\hspace{0.25cm} \hspace{0.25cm}
{\cal A}_1^{\rm a,cc} = \frac{2 \lambda_1 G_A^{CC}}{m_N^2- P'^2},
\hspace{0.5cm} 
{\cal B}_1^{\rm a,cc} = \frac{\lambda_1 G_T^{CC}}{m_N^2- P'^2}.
\nn\\
&&{\cal A}_{3,4}^{\rm em} = \frac{2 f_{\rm N} F_1^{\rm em} }{m_N^2- P'^2},
\hspace{0.5cm} 
{\cal B}_{3,4}^{\rm em} = \frac{-f_{\rm N} F_2^{\rm em}}{m_N^2- P'^2},
\hspace{0.25cm}  \hspace{0.25cm}
{\cal A}_{3,4}^{\rm a,nc} = \frac{-2 f_{\rm N} G_A^{NC}}{m_N^2- P'^2},
\hspace{0.5cm} 
{\cal B}_{3,4}^{\rm a,nc} = \frac{f_{\rm N}G_T^{NC}}{m_N^2- P'^2},
\nn
\\
&&{\cal A}_{3,4}^{\rm v,cc} = \frac{2 f_{\rm N} F_1^{\rm v,cc} }{m_N^2- P'^2},
\hspace{0.5cm} 
{\cal B}_{3,4}^{\rm v,cc} = \frac{-f_{\rm N}F_2^{\rm v,cc}}{m_N^2- P'^2},
\hspace{0.25cm} \hspace{0.25cm}
{\cal A}_{3,4}^{\rm a,cc} = \frac{-2 f_{\rm N} G_A^{CC}}{m_N^2- P'^2},
\hspace{0.5cm} 
{\cal B}_{3,4}^{\rm a,cc} = \frac{f_{\rm N}G_T^{CC}}{m_N^2- P'^2},
\eea
\end{widetext}
Note that the pseudoscalar form factor $G_P$ does not contribute to the $z_\nu T_\nu$ 
projection because of the condition $qz=0$. It can be extracted from the LCSR for other structures, 
or through the relation to the axial form factor in Eq.~(\ref{PCAC}) which is exact  
to our accuracy.

On the other hand, the correlation functions can be calculated in QCD for 
sufficiently large negative $P'^2$ and $q^2 = -Q^2$ in terms of nucleon DAs using the OPE.
The corresponding expressions (to tree-level accuracy) are collected in Appendix~A.
Matching between the two representations one obtains the light-cone sum rule that 
relates the nucleon form factors with nucleon DAs. The precise procedure was described
many times in the literature (see e.g. \cite{Braun:2001tj}) so we omit the technical 
steps. The resulting sum rules depend on two parameters: the continuum threshold
$s_0 \simeq (1.5$~GeV$)^2$ and Borel parameter $M^2$ which defines the scale at which 
the matching between the two representations is done. The dependence on the Borel 
parameter is rather weak. For definiteness, in the plots shown below we take $M^2=2$~GeV$^2$.

\subsection{Results: Ioffe current}
\label{results-ioffe}

In this section we present LCSR predictions for the nucleon form factors that 
are obtained using the Ioffe interpolating current $\eta_1$ for the proton.
The form factors are plotted in the range of the momentum transfers $1\le Q^2 \le 10$~GeV$^2$;
for smaller $Q^2$ our approach is not applicable, for larger $Q^2$ we expect that radiative 
corrections to the sum rules (that include in particular the  usual pQCD contribution) will become 
dominant. The calculations are done using two representative sets of nucleon distribution amplitudes:
asymptotic DAs (solid curves) and including the corrections estimated using QCD sum rules
(dashed curves), see Appendices B,C,D for the definitions. At this stage we do not attempt
to fit the form factors by tuning the parameters of DAs, the difference between the solid and the 
dashed curves gives more or less the range of form factor values that can be obtained with 
the DAs of ``reasonable'' shape. 

\begin{figure}[t]
  \includegraphics[width=0.45\textwidth,angle=0]{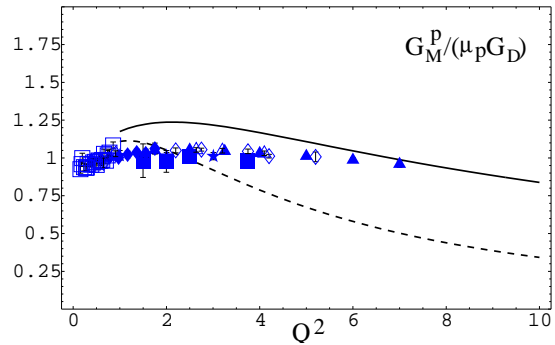}
\caption{LCSR prediction (Ioffe current) for the magnetic form factor of the proton normalized 
to the dipole form factor $G_M^p/(\mu_p G_D)$. 
The data points:
$\bigstar$: SLAC 1994 \cite{Walker94};
$\blacktriangle$: SLAC 1994 \cite{Andivahis94};
$\blacksquare$: SLAC 1970 \cite{Litt70}*;
$\blacklozenge$: Bonn 1971 \cite{Berger71}*;
$\Box$        : Stanford 1966 \cite{Janssens66}*;
$\Diamond$: JLab 2004 \cite{Christy:2004rc};
$\triangle$: JLab 2005 \cite{Qattan:2004ht}.
($*$: Data actually taken from \cite{Arrington:2003df}).}
\label{GMpIoffe}
\end{figure}

The prediction for the proton magnetic form factor normalized 
to the dipole form factor, $G_M^p/(\mu_p G_D)$, where 
$$G_D = (1+Q^2/0.71\, \mbox{GeV}^2)^2,$$
is shown in Fig.~\ref{GMpIoffe}.
Both the $Q^2$ dependence and the magnitude of the form factor is reproduced 
rather well, especially if using asymptotic DAs. 

\begin{figure}[ht]
  \includegraphics[width=0.45\textwidth,angle=0]{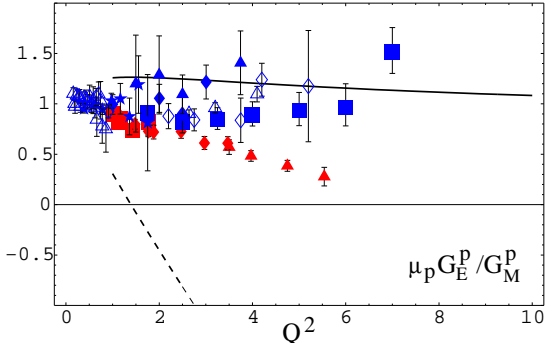}
\caption{LCSR prediction (Ioffe current) for the electric form factor of the proton normalized 
to the magnetic form factor: $\mu_p G_E^p/G_M^p$. 
The data points shown in red are obtained via Polarization transfer:  
{ $\blacktriangle$: Jefferson LAB 2002 \cite{JLab3}};
 { $\blacksquare$: Jefferson LAB 2001 \cite{JLab2}};
 { $\blacklozenge$: Jefferson LAB 2000 \cite{JLab1}};
The data points shown in blue are obtained via Rosenbluth separation:  
 { $\blacksquare$: SLAC 1994 \cite{Andivahis94}};
 { $\blacklozenge$: SLAC  1994 \cite{Walker94}}; 
 { $\blacktriangle$: SLAC 1970 \cite{Litt70}  *};
 { $\bigstar$: Bonn 1971 \cite{Berger71}*}; 
 { $\Box$: Stanford 1966 \cite{Janssens66}*};
   $\Diamond$: JLab 2004 \cite{Christy:2004rc};
   $\triangle$: JLab 2005 \cite{Qattan:2004ht}.
($*$: Data actually taken from \cite{Arrington:2003df}).
(Color identification refers to the online version)}
\label{GEpIoffe}
\end{figure}

The result for the ratio of the proton electric form factor to the magnetic for factor, $\mu_p G_E^p/G_M^p$,
is plotted in Fig.~\ref{GEpIoffe}.
 For completeness, we include on this plot the data 
obtained both via Rosenbluth separation and the polarization transfer techniques, although the former one
is most likely flawed.  
Most interestingly, this ratio appears to be very sensitive 
to the shape of nucleon DAs. Whereas the experimental data obtained via Rosenbluth separation could nicely 
be described by asymptotic DAs alone, the polarization transfer data require considerable corrections. 

\begin{figure}[ht]
  \includegraphics[width=0.45\textwidth,angle=0]{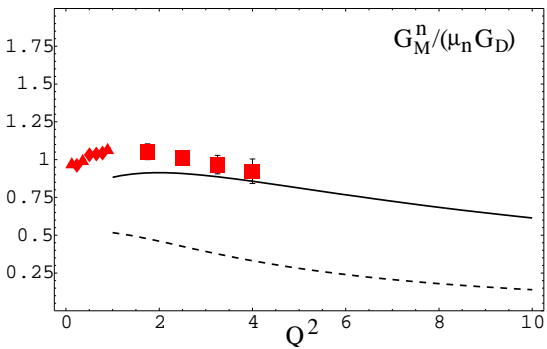}
\caption{LCSR prediction (Ioffe current) for the magnetic form factor of the neutron normalized 
to the dipole form factor $G_M^n/(\mu_n G_D)$.
The data points:
 { $\blacksquare$:  SLAC 1993 \cite{Lung93}};
 { $\blacktriangle$:  Mainz 2002 \cite{Kubon02}}; 
 { $\blacklozenge$:  Mainz 1998 \cite{Anklin98}}.}
\label{GMnIoffe}
\end{figure}

\begin{figure}[ht]
  \includegraphics[width=0.45\textwidth,angle=0]{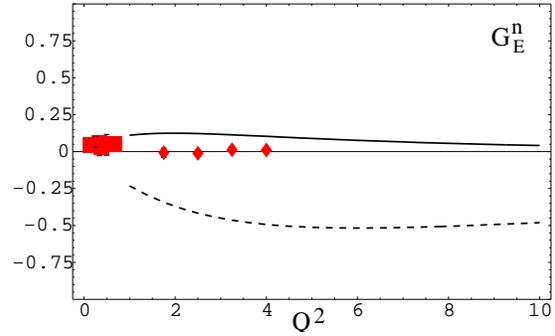}
\caption{LCSR prediction (Ioffe current) for the electric form factor of the neutron $G_E^n(Q^2)$.
The data points:
{\it Red symbols}: experimental values:
{ $\blacklozenge$:  SLAC 1993 \cite{Lung93}};
{ $\blacktriangle$: Jefferson Lab 2001 \cite{Zhou01}};
{ $\blacksquare$: Mainz 1999 \cite{Rohe99}}.}
\label{GEnIoffe}
\end{figure}
\begin{figure}[ht]
  \includegraphics[width=0.45\textwidth,angle=0]{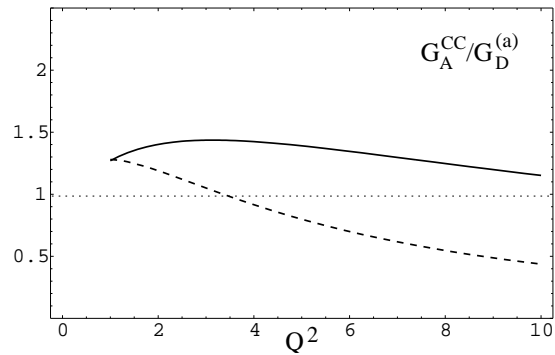}
\caption{LCSR prediction (Ioffe current) for the axial form factor $G_A^{CC}$ normalized to $G_D^{(a)}=g_A/(1+Q^2)^2$.
}
\label{GApIoffe}
\end{figure} 
\begin{figure}[ht]
  \includegraphics[width=0.45\textwidth,angle=0]{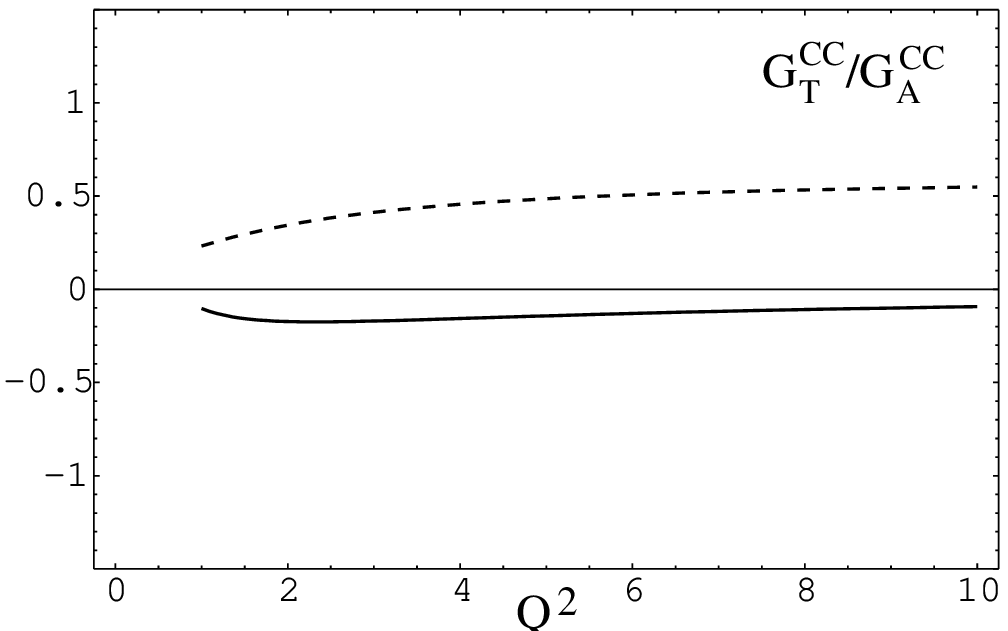}
\caption{LCSR prediction (Ioffe current) for the tensor form factor $G_T^{CC}$ normalized to $G_A^{CC}$.
}
\label{GTpIoffe}
\end{figure} 

The LCSR predictions for the neutron are shown in Fig.~\ref{GMnIoffe} and Fig.~\ref{GEnIoffe} for the magnetic 
and the electric form factors, respectively. In this case, again, the magnetic form factor is described reasonably well
by asymptotic DAs, while the magnitude and even the sign of the electric form factor depends on their shape. 
Further, the LCSR prediction for the axial form factor of the proton $G_A^{CC}$ normalized to $G_D^{(a)}=1.267/(1+Q^2)^2$
is shown in Fig.~\ref{GApIoffe}. Experimentally, this ratio is close to one.
A more steep $Q^2$-dependence of the axial form factor compared to the electromagnetic ones 
seems to be correctly reproduced, and also the normalization
agrees within 50\% accuracy.
 
Last but not least, in Fig.~\ref{GTpIoffe} we show the LCSR result for the form factor 
$G_T^{CC}$ normalized to $G_A^{CC}$. As mentioned above, this form factor is forbidden by T-invariance so that
this ratio has to be zero. In the LCSR approach the final and the initial state nucleons are treated differently 
and the T-invariance is not manifest. Smallness of $G_T$ is therefore an indication of how good the nucleon state 
is separated from the continuum by the simple duality assumption. We observe that $G_T^{CC}/G_A^{CC}$ strongly
depends on the shape of the nucleon DAs. It is small and negative for asymptotic DAs 
but becomes  positive if the DAs acquire large corrections. 

\subsection{Results: Leading-twist currents}
\label{results-eta34}

\subsubsection{Checking isospin relations}

One of the main motivations for our study is to find out the optimal 
nucleon current for the calculation of nucleon
form factors within the framework of LCSR.

\begin{figure*}[ht]
  \includegraphics[width=0.34\textwidth,angle=0]{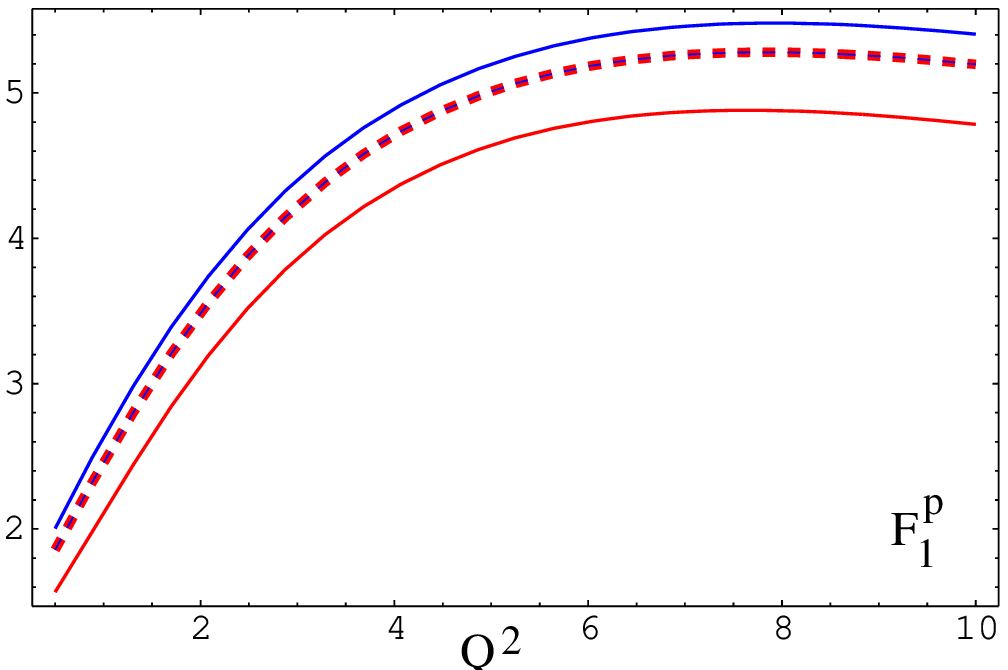}
  \includegraphics[width=0.35\textwidth,angle=0]{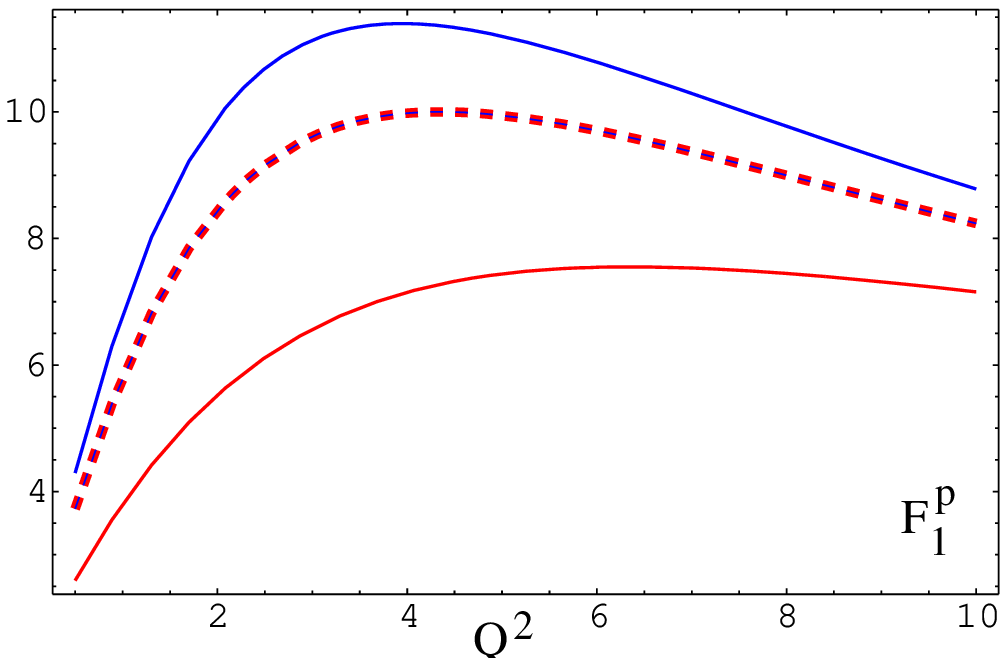}\\[1mm]
   \hspace*{-2mm}\includegraphics[width=0.35\textwidth,angle=0]{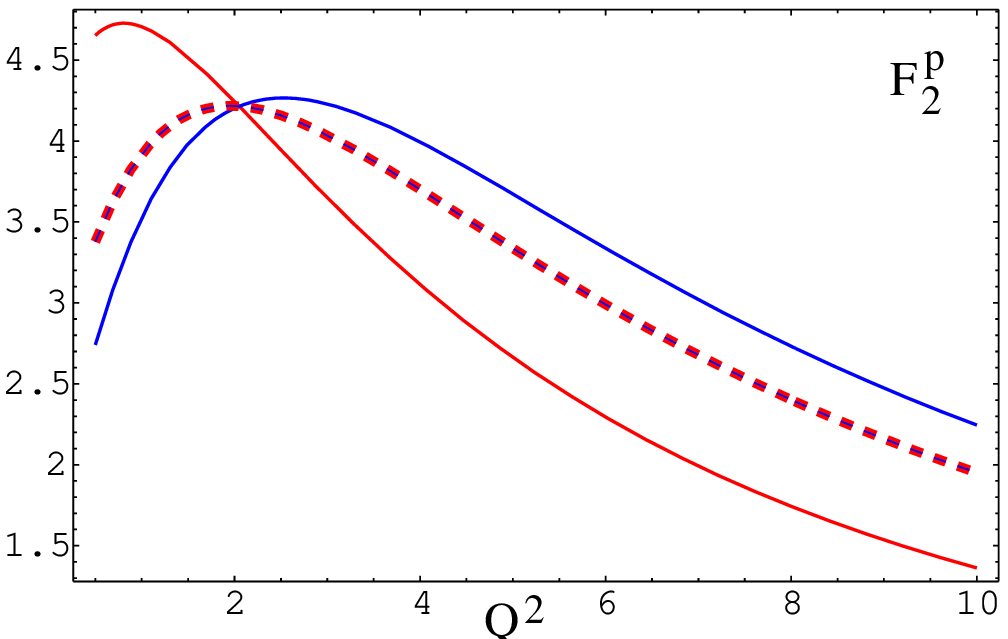}~
  \includegraphics[width=0.34\textwidth,angle=0]{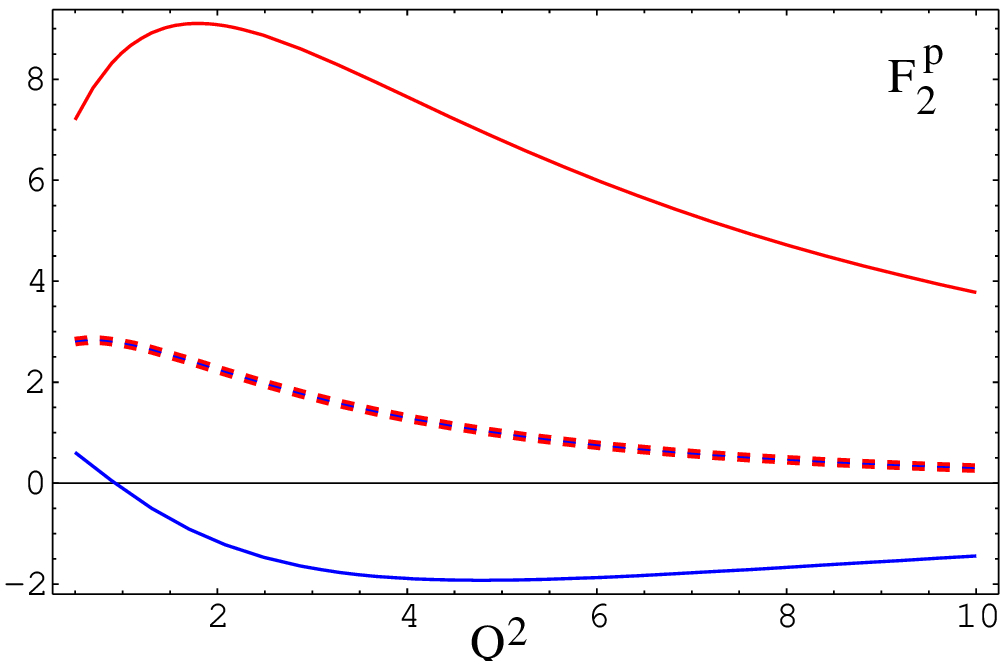}
\caption{\label{plotisoF11}
Test of the isospin relations for the form factor $F_1$ (upper panels) 
and $F_2$ (lower panels) using the asymptotic form of the nucleon DAs
(left panels) and including corrections estimated from QCD sum rules 
(right panels). 
The red (lower) and the blue (upper) solid curves
show the r.h.s and the l.h.s. of the relation in (\ref{isospin10}), obtained
from the LCSRs using the current $\eta_4$ and normalized to the dipole form factor $G_D$.
The blue-red thick dashed curves shows the LCSR result obtained with the $\eta_3$ 
current, in which case the isospin relation is satisfied identically.
(Color identification refers to the online version.)}
\label{fig:isospin1}
\end{figure*}

In Ref.~\cite{Braun:2001tj} the  current $\eta_4$, Eq.~(\ref{eta4}),  was used which  in difference to 
$\eta_3$, Eq.~(\ref{CZcurrent}),
couples both to isospin $I=1/2$ and $I=3/2$ states.  A priory, it is not obvious which current is better 
since the summation over quantum numbers may improve the accuracy of the duality approximation for the continuum and
also it usually suppresses poorly known contributions of higher-dimension(twist) operators. The new observation of this work 
is that this argumentation can be tested by checking the isospin relations e.g. for the vector current:
\bea{isospin10}
    F_1^{CC}(Q^2) &=& F_1^p(Q^2) - F_1^n(Q^2)\,,
\nn\\
    F_2^{CC}(Q^2) &=& F_2^p(Q^2) - F_2^n(Q^2)\,,
\eea 
cf. (\ref{iso2}). These relations  are fulfilled identically if the $\eta_3$ (or $\eta_1$) current is used, 
because in this case the correlation functions (\ref{correlator}) satisfy isospin relations by themselves. 
However, when using the $\eta_4$ current, the extracted form factors satisfy the relations in (\ref{isospin10}) only approximately,
within the sum rule accuracy. In particular, their violation provides one with a direct measure of the contamination of the nucleon 
contribution by isospin $I=3/2$ states.  

The results are shown in Fig.~\ref{fig:isospin1} for $F_1$ and $F_2$ on the two upper and two lower panels, respectively.
We see that (unphysical) isospin breaking is  relatively moderate in case that 
asymptotic nucleon DAs are chosen, but it explodes if the DAs acquire 
significant corrections. The situation with axial form factors proves to be similar.
Since the ultimate goal of our study is to determine nucleon DAs 
from the comparison to the data, such a behavior presents a crucial disadvantage.
We conclude that the LCSRs with the current $\eta_4$ do not pass the test; they are strongly 
contaminated with the isospin $I=3/2$ contributions and do not allow for any quantitative
form factor determinations. Hence, hereafter we drop the $\eta_4$ current and  only consider $\eta_3$
(and $\eta_1$).

\subsubsection{Nucleon form factors for $\eta_3$}
\begin{figure*}
  \includegraphics[width=0.45\textwidth,angle=0]{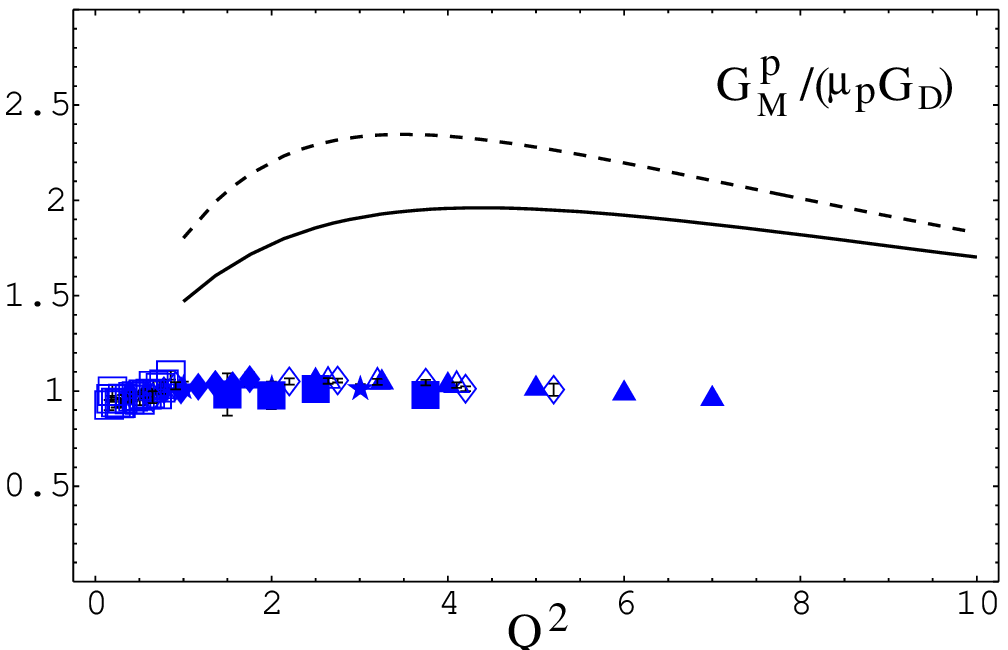}
  \includegraphics[width=0.45\textwidth,angle=0]{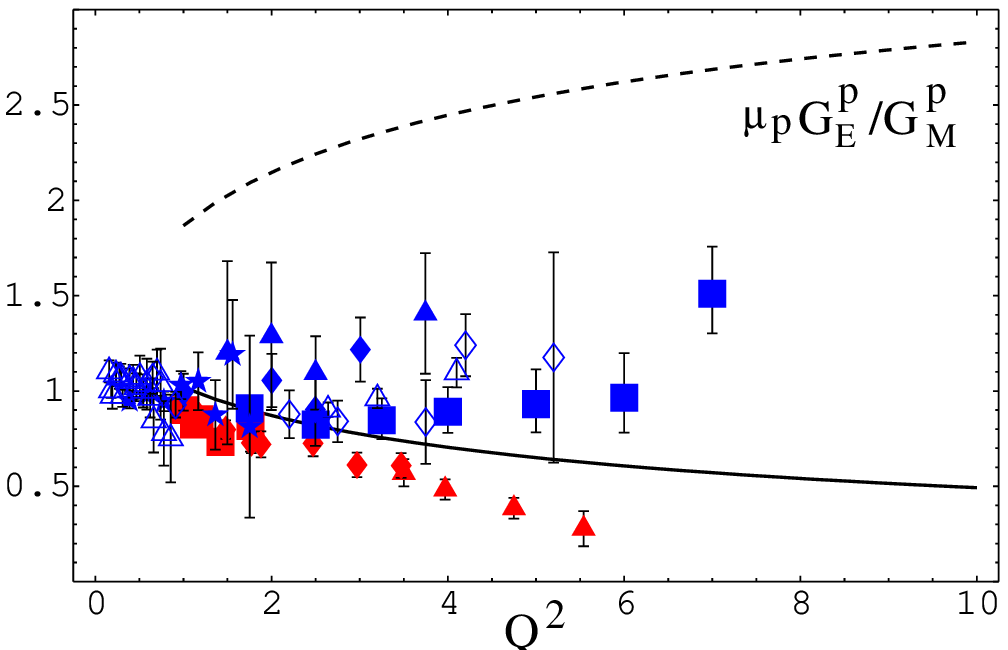}\\[1mm]
  \includegraphics[width=0.45\textwidth,angle=0]{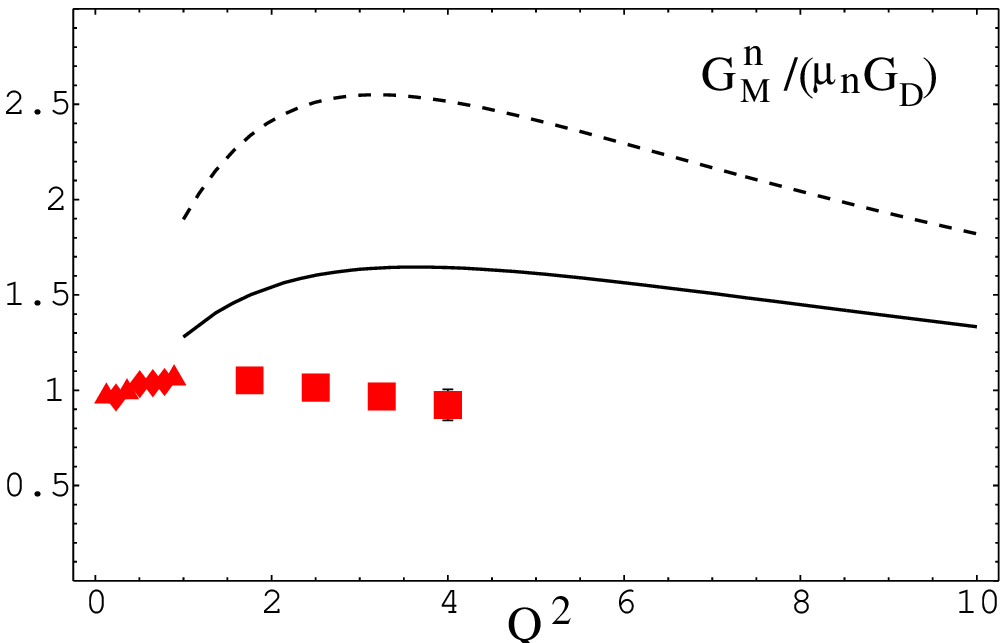}
  \includegraphics[width=0.45\textwidth,angle=0]{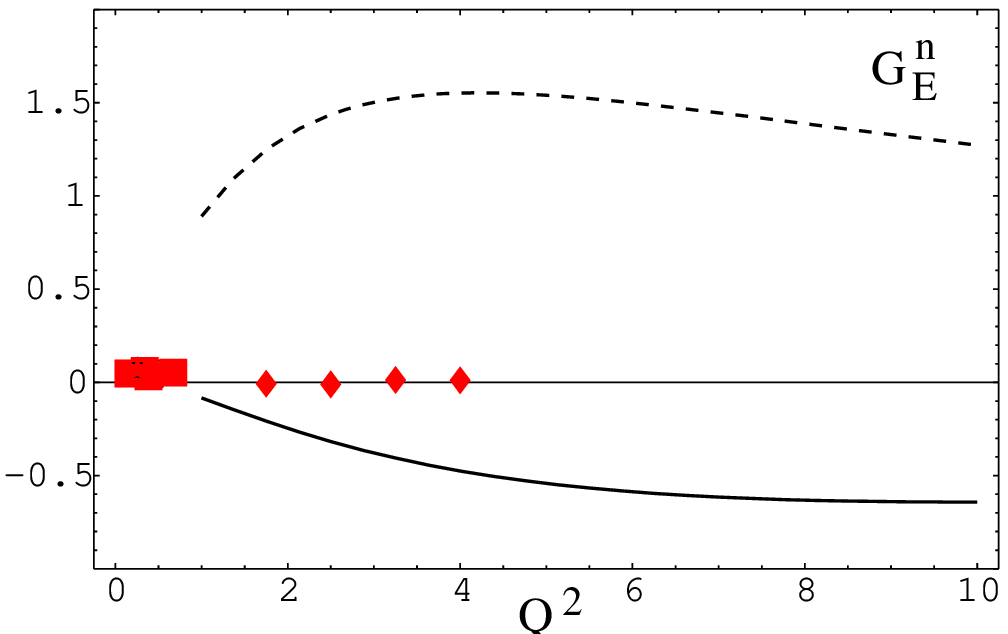}
\caption{LCSR results  for the electromagnetic form factors of the 
 nucleon, obtained using the leading-twist-3 interpolating current $\eta_3$.
Identification of the curves and the data points is the same as in 
Figs.~\ref{GMpIoffe}--\ref{GEnIoffe}.
}
\label{fig:Femeta3}
\end{figure*}

\begin{figure*}
  \includegraphics[width=0.44\textwidth,angle=0]{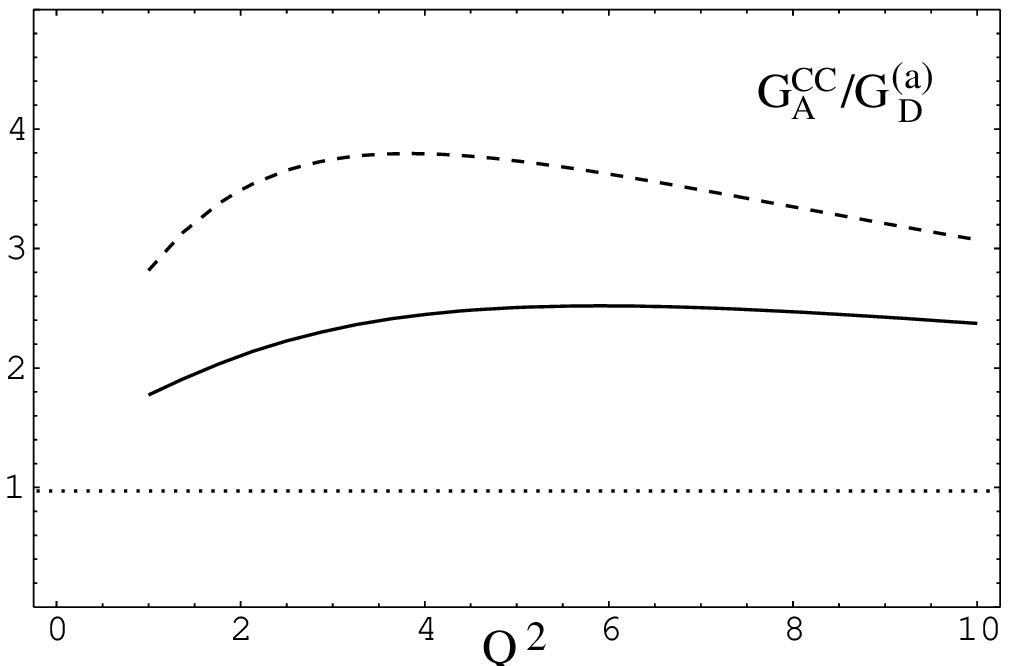}
  \includegraphics[width=0.45\textwidth,angle=0]{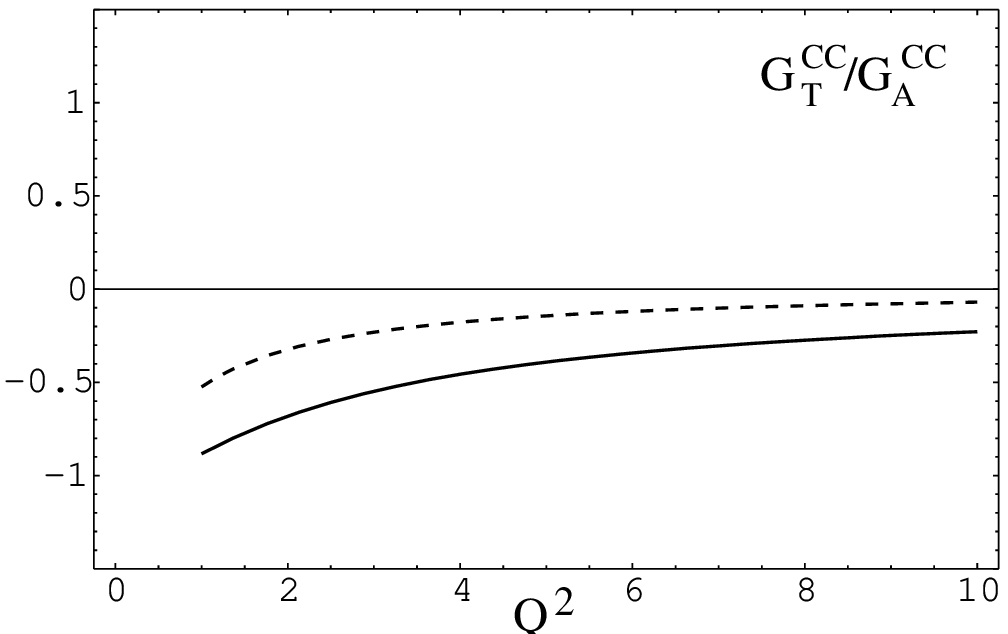}
\caption{LCSR results for the axial form factor  of the proton
$G_A^{CC}$ normalized to $G_D^{(a)}=g_A/(1+Q^2)^2$ (left panel) and 
tensor form factor $G_T^{CC}$ normalized $G_A^{CC}$ (right panel),
obtained using the leading-twist-3 interpolating current $\eta_3$.
}
\label{fig:FATeta3}
\end{figure*}

Here we present LCSR results for the nucleon form factors, obtained using the 
interpolating current current $\eta_3$. Electromagnetic form factors are 
shown in Fig.~\ref{fig:Femeta3}, and weak form factors in Fig.~\ref{fig:FATeta3},
respectively. As above, the calculations are done using 
asymptotic DAs (solid curves) and including the corrections estimated using QCD sum rules
(dashed curves), see Appendices B,C,D for details.
The agreement with the data is in general somewhat worse compared to the calculations using 
the Ioffe current and, most interestingly, the corrections to asymptotic DAs  ``work'' in opposite
direction. We repeat that nonzero values obtained for the tensor form factor $G_T$ are artifact 
of our approach and can be used to quantify the error estimates. 


As a matter of principle, sum rules using all interpolating currents have to produce the same results.
In practice it has never been the case and the optimal choice of the interpolating current is a very 
important part of the QCD sum rule method. In our case  it is possible that the difference between 
predictions based on $\eta_1$ and $\eta_3$ currents will 
decrease when radiative corrections to the sum
rules are included. Still, on the basis of information that we have now and the experience of QCD sum 
rule calculations with baryons in general, we believe that the Ioffe current $\eta_1$ provides the 
best option for the construction of the LCSRs. The LCSRs based on the leading twist current $\eta_3$ are 
valid and useful for making a consistency check since their structure and the relative 
weight of DAs of different twist is very different. 

\begin{figure*}
  \includegraphics[width=0.35\textwidth,angle=0]{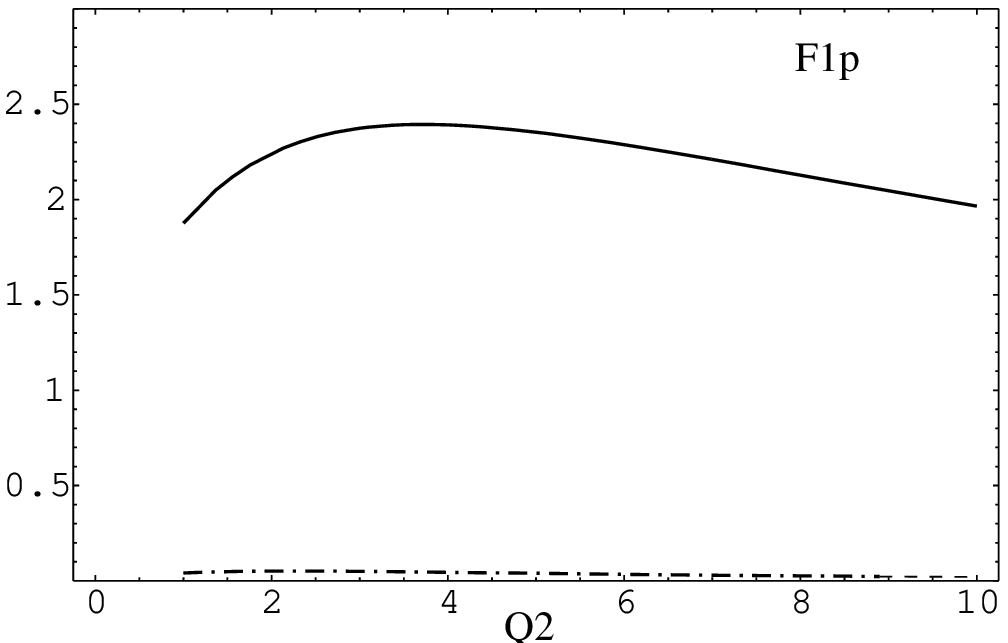}
  \includegraphics[width=0.35\textwidth,angle=0]{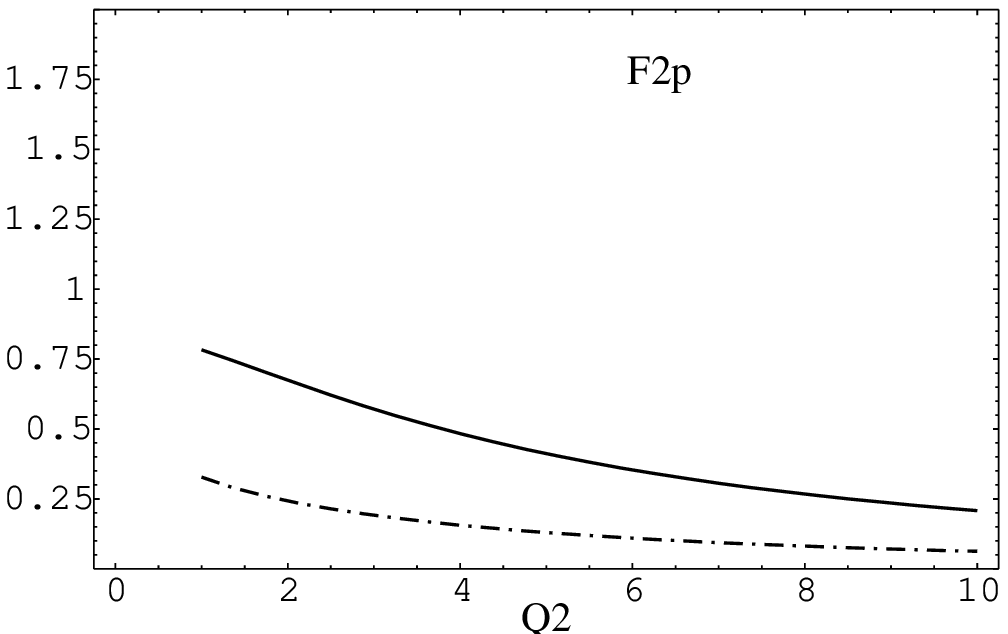}\\[1mm]
  \includegraphics[width=0.35\textwidth,angle=0]{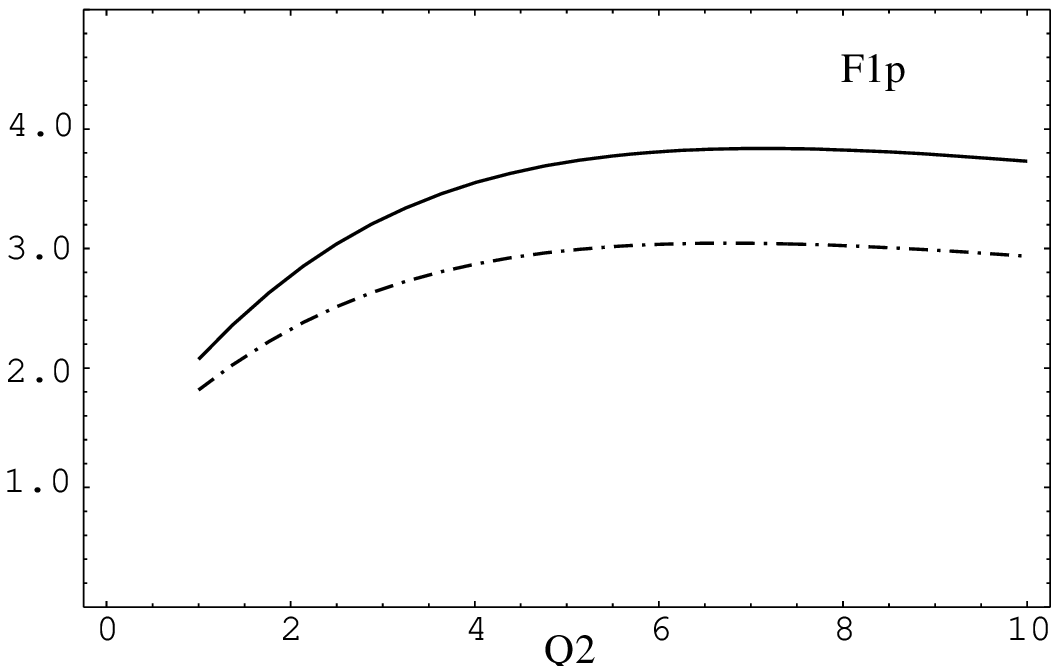}
  \includegraphics[width=0.35\textwidth,angle=0]{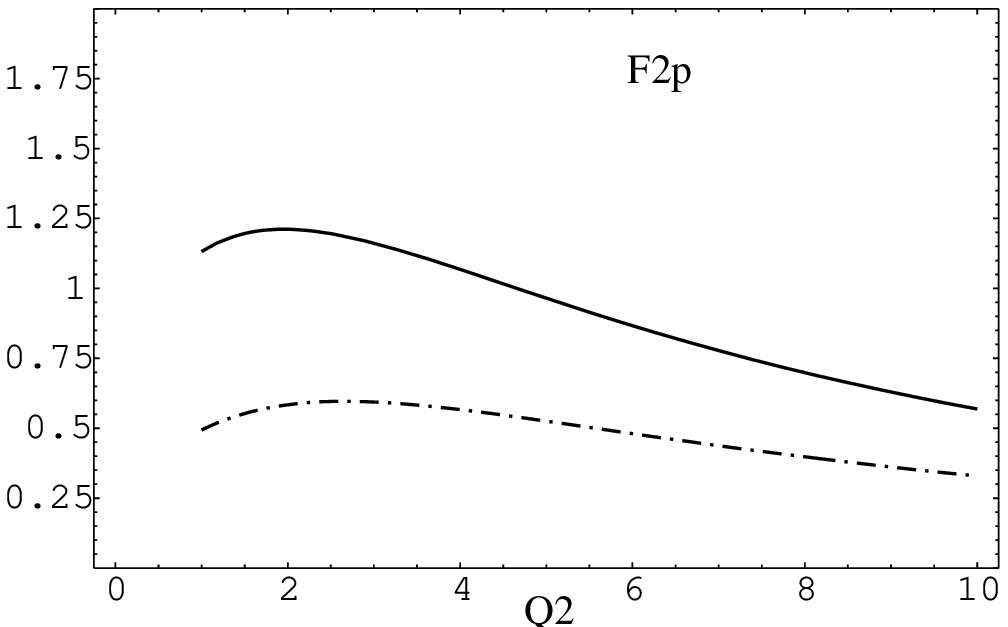}
\caption{LCSR results (solid curves) for the electromagnetic form factors of the 
nucleon, $F_1^p$ (left) and $F_2^p$ (right) obtained using the interpolating current $\eta_1$
(upper two panels) and $\eta_3$ (lower two panels) and asymptotic nucleon DAs.
On each plot, the dashed-dotted curves show the contribution of leading-twist 
DAs only, including the corresponding nucleon mass corrections.  
}
\label{fig:twist3}
\end{figure*}

To illustrate this issue we show in Fig.~\ref{fig:twist3} the 
LCSR results for $F_1^p$ and $F_2^p$ 
obtained using the interpolating current $\eta_1$ (upper two panels) and $\eta_3$ (lower two panels) 
and asymptotic nucleon DAs.
On each plot, solid curves correspond to the sum of contributions of all twists and 
the dashed-dotted curves show the contribution of leading-twist 
DAs only, including the corresponding nucleon mass corrections.
Notice that the $Q^2$ dependence of the leading-twist and the higher-twist contributions is
almost the same. This is to be expected, since higher-twist corrections to the sum rules are only 
suppressed by a power of the Borel parameter, not a power of $Q^2$. On the other hand, the relative weight of 
the leading-twist and the higher-twist terms  depends strongly on the current:    
E.g. for Ioffe current  $F_1$ is almost entirely higher-twist, whereas for the $\eta_3$  current the 
leading-twist contribution is dominant. Also for $F_2$ the sum rules based on the $\eta_3$ current 
are more sensitive to the leading-twist DAs and may be more useful to obtain restrictions 
on the corresponding parameters. We expect that such differences will be moderated upon inclusion 
of the radiative corrections to the sum rules.
However, the Ioffe current-based sum rules will most likely
still provide higher accuracy for the form factors.  

\subsection{A model for the nucleon distribution amplitudes}

\begin{figure*}
  \includegraphics[width=0.45\textwidth,angle=0]{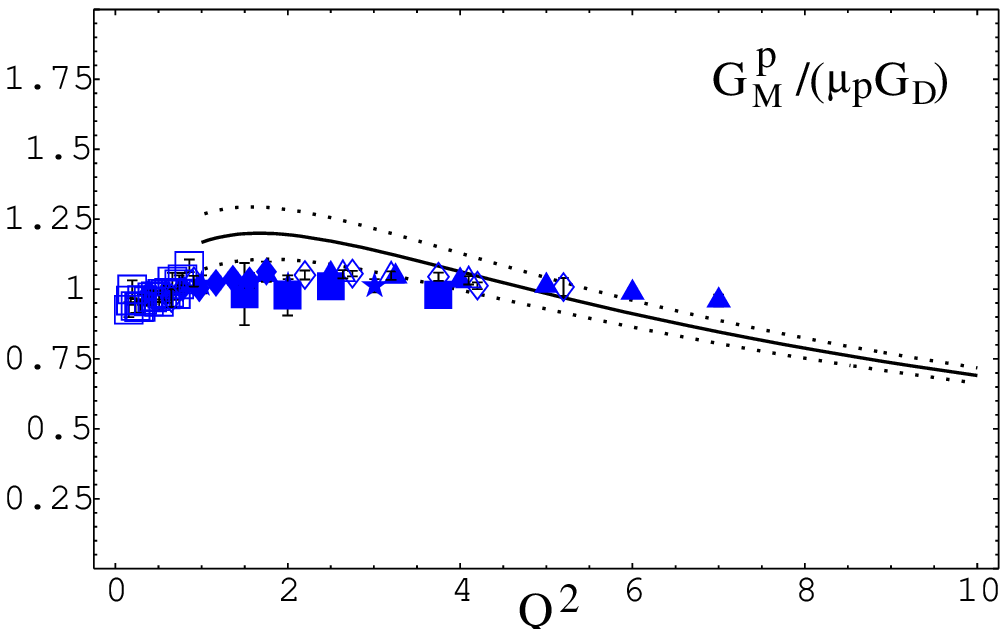}
  \includegraphics[width=0.45\textwidth,angle=0]{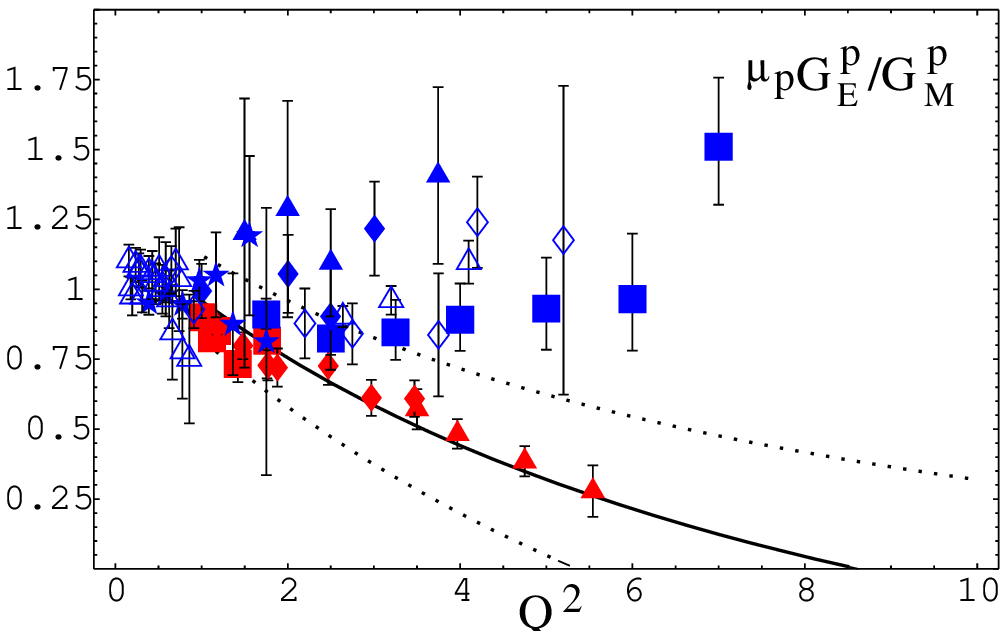}\\[1mm]
  \includegraphics[width=0.45\textwidth,angle=0]{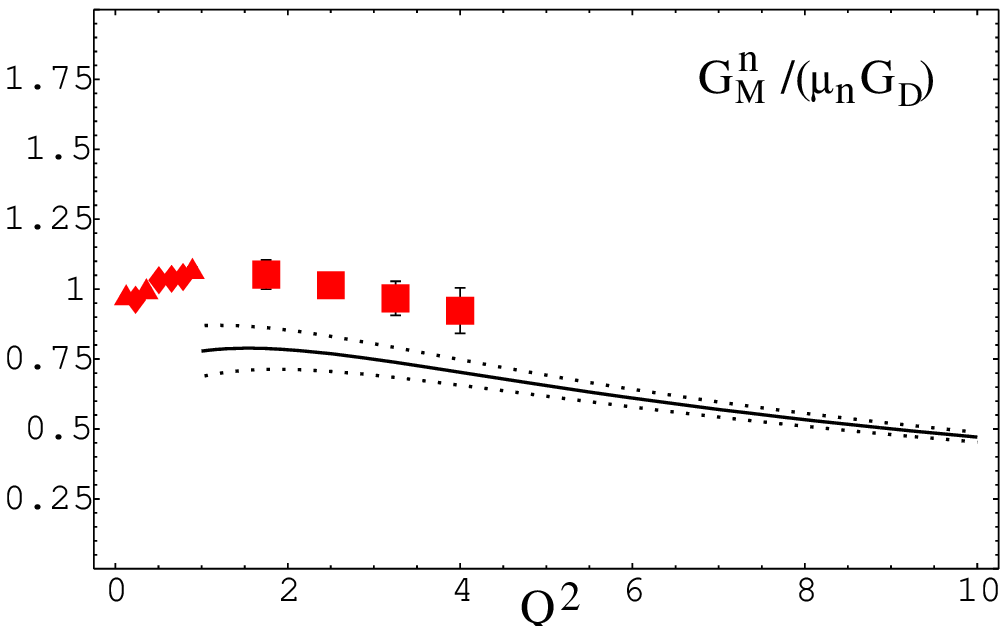}
  \includegraphics[width=0.45\textwidth,angle=0]{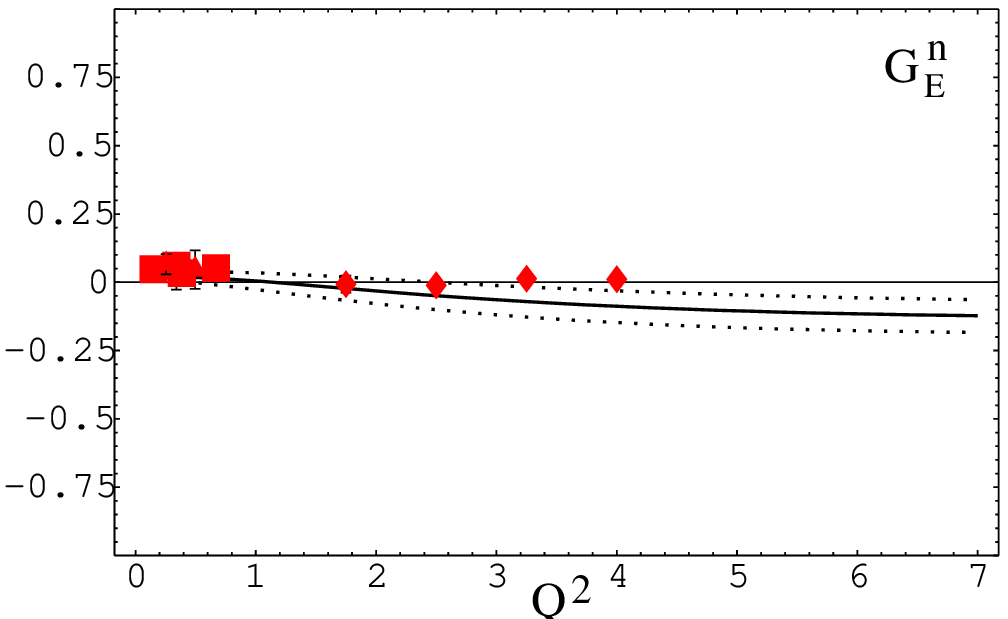}
\caption{LCSR results (solid curves) for the electromagnetic form factors of the 
nucleon, obtained using the model of the nucleon DAs (\ref{superset}) and the Ioffe current $\eta_1$.
The dotted curves show the effect of the variation of the ratio $f_N/\lambda_1$ by 30\%.
Identification of the data points is the same as in 
Figs.~\ref{GMpIoffe}--\ref{GEnIoffe}.
}
\label{fig:Femtune}
\end{figure*}
\begin{figure*}
  \includegraphics[width=0.44\textwidth,angle=0]{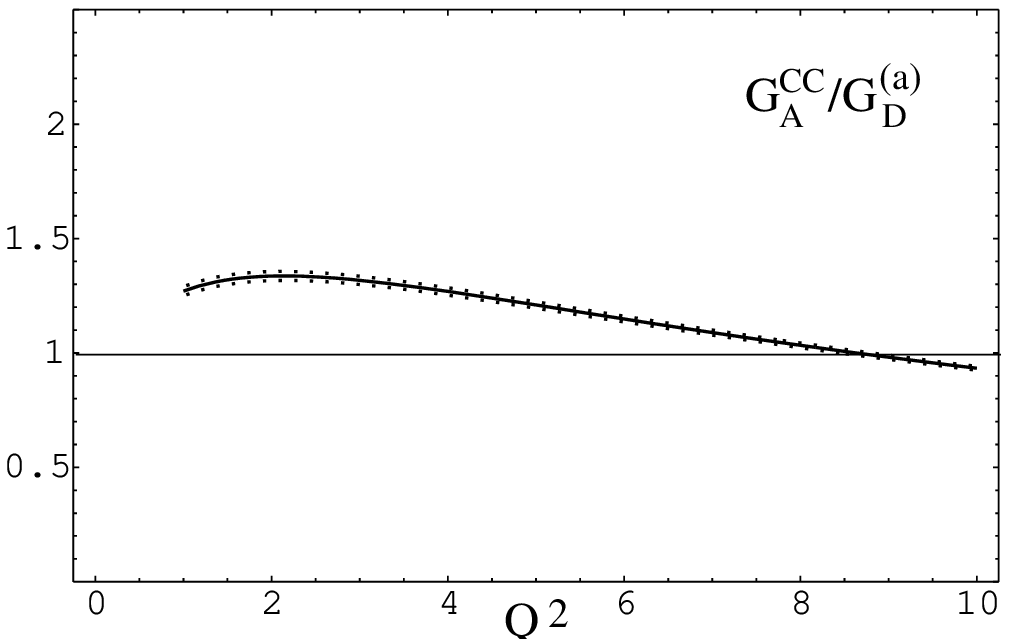}
  \includegraphics[width=0.45\textwidth,angle=0]{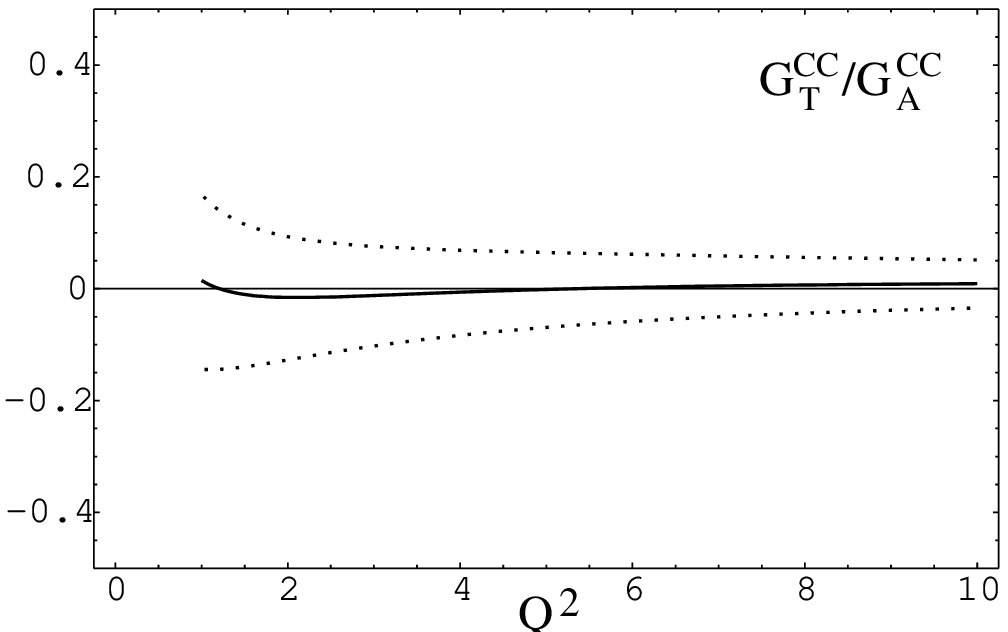}
\caption{LCSR results (solid curves) for the axial form factor  of the proton
$G_A^{CC}$ normalized to $G_D^{(a)}=g_A/(1+Q^2)^2$ (left panel) and 
tensor form factor $G_T^{CC}$ normalized to $G_A^{CC}$ (right panel),
obtained using obtained using the model of the nucleon DAs (\ref{superset}) and the 
Ioffe current $\eta_1$.
The dotted curves show the effect of the variation of the ratio $f_N/\lambda_1$ by 30\%.
}
\label{fig:FATtune}
\end{figure*} 

The nucleon DAs provide the principal nonperturbative input to the LCSRs. 
As we have seen, in many cases experimental data are in between the LCSR calculations that
asymptotic and QCD sum rule-based DAs. This suggests that a good description of the 
data is possible by tuning the parameters of the DAs. 
As a demonstration, we present here the
results obtained using a simple model in which the deviation from the asymptotic DAs
is taken to be one third of that suggested by the QCD sum rule estimates. 

The corresponding parameters are:
\bea{superset}
&&A_1^u  =    0.13 \, ,\quad 
V_1^d  =    0.30 \, ,
\nn \\
&&f_1^d  =    0.33 \, ,\quad
f_1^u  =    0.09 \, ,\quad
f_2^d =     0.25\,.
\eea
where the first two refer to the leading twist-3 and the rest correspond to twist-4.
These values  are not unreasonable, since QCD sum rules are known to overestimate the 
matrix elements of  higher conformal spin operators, and we just made the simplest assumption that 
all sum rule results have to be rescaled by the same factor. Our leading-twist parameter $V_1$
is very close to the phenomenological Bolz-Kroll model \cite{Bolz:1996sw}; $A_1$ is
somewhat bigger but the dependence of the leading-order sum rules on this parameter
is weak. To this accuracy, the sum rules also do not depend on the parameters $\lambda_2$ and $f_2^d$;
this dependence is present, however, in the transition form factors like $\gamma^* N \to \Delta$.    

\begin{figure}[ht]
  \includegraphics[width=0.45\textwidth,angle=0]{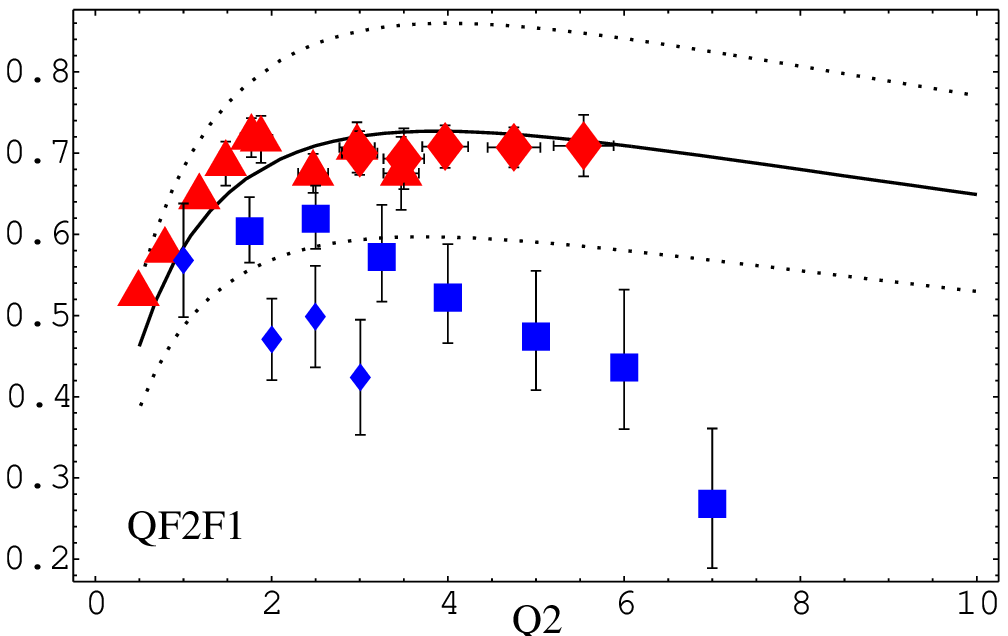}
\caption{LCSR results (solid curves) for the ratio  $\sqrt{Q^2} F_2^p/F_1^p$ 
obtained using the model of the nucleon DAs (\ref{superset}) and the Ioffe current $\eta_1$.
Data points:
{\it Red symbols}: experimental values obtained via Polarization transfer:  
{ $\blacktriangle,\blacklozenge$:  M. Jones (private communication)};
{\it Blue symbols}: experimental values obtained via Rosenbluth
separation:  
{ $\blacksquare$: SLAC 1994 \cite{Andivahis94}}; 
{ $\blacklozenge$: SLAC 1994 \cite{Walker94}}.
(Color identification refers to the online version)}
\label{fig:F2/F1}
\end{figure}

The calculations using this model are shown by solid curves  for the electromagnetic form factors 
in Fig.~\ref{fig:Femtune} and weak form factors in Fig.~\ref{fig:FATtune},
respectively. In addition, in Fig.~\ref{fig:F2/F1} we plot the corresponding $\sqrt{Q^2} F_2^p/F_1^p$ ratio.  
On the same plots we show by the dotted curves the effect of the variation of the couplings ratio 
$f_N/\lambda_1$  (see Eqs.~(\ref{eta3}) within a conservative 30\% error range.
This ratio determines the overall normalization of the leading-twist DAs compared to
higher twist so that the sensitivity to $f_N/\lambda_1$ is a good indication 
of the relative size of the leading-twist contributions to the LCSRs.

One sees that the experimental data on the electromagnetic form factors are reproduced very well, 
and, most welcome, the unphysical tensor form factor $G_T$ becomes consistent with zero. Also for the axial form factor there is
a good agreement, both in shape and normalization.

\begin{figure*}[ht]
  \includegraphics[width=0.44\textwidth,angle=0]{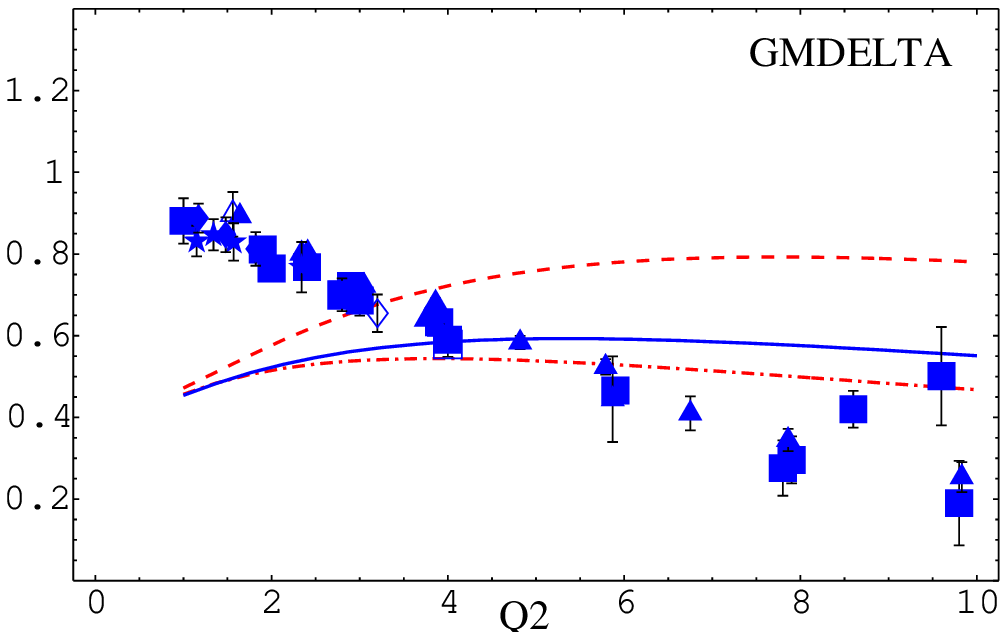}
  \includegraphics[width=0.45\textwidth,angle=0]{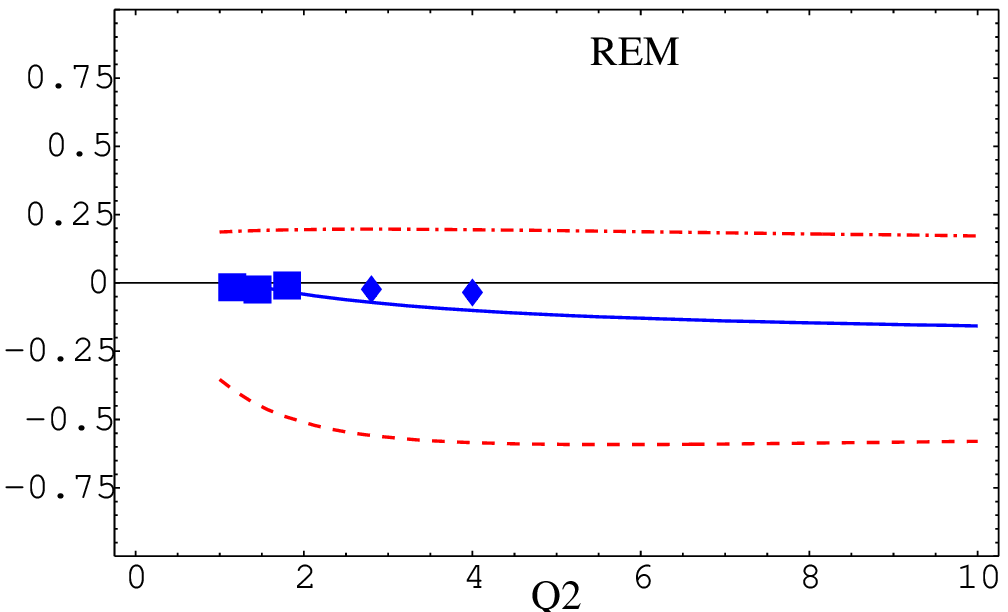}
  \includegraphics[width=0.45\textwidth,angle=0]{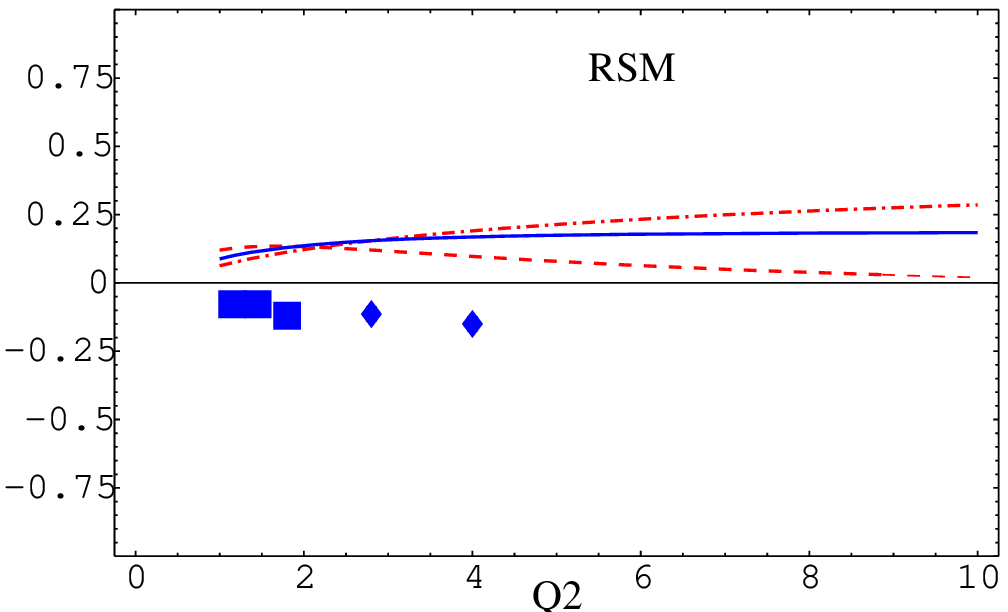}
\caption{$\gamma* N \to \Delta$ transition form factors in  the LCSR approach~\cite{Braun:2005be}.
The blue solid curve corresponds to the calculation using the model in Eq.~(\ref{superset}).
The red dash-dotted  and dashed curves are obtained using the asymptotic and the QCD sum rule motivated 
DAs, respectively. For the references to the experimental data see ~\cite{Braun:2005be}.
(Color identification refers to the online version)}
\label{fig:Delta}
\end{figure*}

Last but not least, we can use the same set of DAs to calculate the 
$\gamma^* N \to \Delta$ transition form factors within the LCSR approach, following Ref.~\cite{Braun:2005be}.
The results are shown in Fig.~\ref{fig:Delta}. In this case we also get a much better agreement with the experimental data 
on the electric form factor compared to the calculations that use asymptotic or sum rule-based DAs.   

We should warn that the model in Eq.~(\ref{superset}) is not based on any systematic attempt to fit
the data and in fact we believe that such any fitting would be premature before the radiative 
corrections to the LCSR are calculated. In addition, one has to take into account the 
scale dependence of the parameters of the DAs and study in more detail the dependence of the sum rules 
on the Borel parameter.  Still, the very
possibility to describe many different form factors using the same set of DAs is nontrivial
and indicates the selfconsistency of our approach. The true parameters of the DAs are probably not far 
from the numbers quoted in Eq.~(\ref{superset}), although at this stage we cannot give any error estimates.

%
\section{Conclusions}
%

{}Following the proposal in  Ref.~\cite{Braun:2001tj} we have made a comprehensive  study 
of leading order light-cone sum rules for the electromagnetic and weak nucleon form factors.
We presented detailed results that are obtained using different interpolating currents for the nucleon and  
argue that the Ioffe current appears to be the optimal. 
We make an update of the QCD sum rule estimates of the shape parameters of higher-twist DAs and also 
present a simple model that seems to reproduce main features of the experimental data remarkably well.
In addition we included a complete summary of higher-twist  DAs 
and some new expressions for the OPE of three-quark operators which extend the results given in \cite{Braun:2001tj}.

Our main conclusion in this work is that the LCSR approach to baryon form factors seems to be 
sufficiently accurate to allow one to get a quantitative description of hard exclusive reactions with baryons.   
{}From the theory point of view, this technique is attractive because in LCSRs  ``soft'' contributions to the form factors 
are calculated in terms of the same DAs that
enter the pQCD calculation and there is no double counting. The asymptotic pQCD limit, in fact, formally 
corresponds to a part of the two-loop $\alpha_s^2$ radiative correction to the sum rules.
Thus, the LCSRs provide one with the most direct relation of the hadron form factors
and distribution amplitudes that is available at present, with no other nonperturbative parameters.  

We remind that the sum rules considered in this work are tree-level.
{}Further progress in this direction requires the calculation of the radiative one-loop correction
to contributions of twist-3 and twist-4 operators.  
It would also be very valuable to have lattice evaluations for at least some of the parameters that enter 
the DAs, most importantly the $f_N/\lambda_1$ ratio.

\section*{Acknowledgement}
We would like to thank N.~Mahnke for collaboration in early stages 
of this work and Daowei Wang for pointing out misprints in Appendix B of \cite{Braun:2001tj} 
and for comparing the results for $\mathcal{A}_1^M$ and $\mathcal{T}_1^M$.


\appendix  
\renewcommand{\theequation}{\Alph{section}.\arabic{equation}} 

\section*{Appendices}  
\section{Summary of Correlation Functions}
\setcounter{equation}{0}
In this Appendix we present the tree-level results for the correlation functions 
defined in Eq.~(\ref{project4}), (\ref{project3}). The correlation functions are expressed
in terms of nucleon DAs that are summarized in Appendix B below. We use the following
notations:
\bea{notat1}
\widetilde{F}(x_3)
 &=& \int_1^{x_3}\!\!dx'_3\int_0^{1- x^{'}_{3}}\!\!\!dx_1\, F(x_1,1-x_1-x'_3,x'_3)\,,
\nn\\
\lefteqn{\hspace*{-0.9cm}\widetilde{\!\widetilde{F}}(x_3)= \int_1^{x_3}\!\!dx'_3 \int_1^{x'_3}\!\!dx^{''}_3  }
\nn\\[-2mm]&&\hspace*{-0.2cm}\times 
\int_0^{1- x^{''}_{3}}\!\!\!\!dx_1\, F(x_1,1-x_1-x^{''}_3,x^{''}_3)
\eea
and
\bea{notat2}
\widehat{F}(x_2) 
&=& \int_1^{x_2}\!\!dx'_2\int_0^{1-x^{'}_{2}}\!\!\!dx_1 F(x_1,x'_2,1-x_1-x'_2)\,,
\nn\\
\lefteqn{\hspace*{-0.9cm}\widehat{\!\widehat{F}}(x_2)= \int_1^{x_2}\!\!dx'_2 \int_1^{x'_2}\!\!dx^{''}_2}
\nn\\[-2mm] &&{}\hspace*{-0.2cm} \times
\int_0^{1-x^{''}_{2}}\!\!\!dx_1 F(x_1,x^{''}_2,1-x_1-x^{''})\,,
\eea
where $F=A,V,T$ is a generic nucleon DA that depends on the three valence quark momentum fractions,
and also shorthand notations for the combinations of the DAs:
\bea{notat31}
&&\hspace*{-3mm}V_{43} = V_4-V_3\,,
\nn\\
&&\hspace*{-3mm}V_{123} = V_1-V_2-V_3\,,
\nn\\
&&\hspace*{-3mm}V_{1345} = -2V_1+V_3+V_4+2 V_5\,,
\nn\\
&&\hspace*{-3mm} V_{12345} = 2 V_1-V_2-V_3-V_4-V_5\,,
\nn\\
&&\hspace*{-3mm}  V_{123456} = -V_1+V_2+V_3+V_4+V_5-V_6\,,
\eea
\bea{notat32}
&&\hspace*{-3mm}A_{34} = A_3-A_4\,,
\nn\\
&&\hspace*{-3mm}A_{123} = -A_1+A_2-A_3\,,
\nn\\
&&\hspace*{-3mm} A_{1345} = -2A_1-A_3-A_4+2 A_5\,,
\nn\\
&&\hspace*{-3mm} A_{12345} = 2 A_1-A_2+A_3+A_4-A_5\,,
\nn\\
&&\hspace*{-3mm}  A_{123456} = A_1-A_2+A_3+A_4-A_5+A_6
\eea
and also 
\bea{notat33}
&&\hspace*{-3mm}T_{137}   = T_1 - T_3 - T_7\,,
\nn\\
&&\hspace*{-3mm}T_{13478}  = 2T_1 -T_3 - T_4 - T_7 - T_8\,,
\nn\\
&&\hspace*{-3mm}T_{134678} = T_1 -T_3 - T_4 + T_6 - T_7 - T_8 \,.
\eea
In addition, in this Appendix we use
$$q_3 \equiv q-x_3 P\,,\qquad q_2 \equiv q-x_2 P\,.$$

In this notation we obtain, for the correlation functions involving the Ioffe current $\eta_1$:
\begin{widetext}
%
%
\bea{Ioffe-em}
  \mathcal{A}^{\mathrm{em}}_1 &= & \phantom{-}2e_d\int_0^1 dx_3\left\{
     \frac{Q^2+q_3^2}{q_3^4} \widetilde{V}_{123}+\frac{x_3}{q_3^2}\int_0^{\bar x_3} dx_1 V_3(x_i)+
      \frac{x_3^2m_N^2}{q_3^4}\widetilde{V}_{43}\right\}
\nn\\
&&{} + 2 e_u \int_0^1 dx_2 \left\{
      \frac{x_2}{q_2^2} \int_0^{\bar x_2}dx_1 \left[ -2 V_1 + 3V_3+A_3 \right](x_i)
       - \frac{2x_2 m_N^2}{q_2^4}\mathcal{V}_1^{M(u)} +\frac{Q^2-q_2^2}{q_2^4}\widehat{V}_{123}
\right.
\nn\\
&&{}\left. \hspace{0.0cm}
       + \frac{Q^2+q_2^2}{q_2^4}\widehat{A}_{123}
       -\frac{x_2^2 m_N^2}{q_2^4}\left[\widehat{V}_{1345} - 2\widehat{V}_{43}+ \widehat{A}_{34}\right] 
      -\frac{2x_2 m_N^2}{q_2^4}\widehat{\widehat{V}}_{123456}
\right\},
\nn\\
 \mathcal{B}^{\mathrm{em}}_1 &=& -2e_d\int_0^1 dx_3\left\{
    \frac{1}{q_3^2}\int_0^{\bar x_3} dx_1 V_1(x_i) +\frac{m_N^2}{q_3^4}\mathcal{V}_1^{M(d)}
     - \frac{x_3 m_N^2}{q_3^4} \left[\widetilde{V}_{123} -  \widetilde{V}_{43} \right]
     \right\}
\nn\\ &&{}
 + 2 e_u \int_0^1 dx_2 \left\{
    \frac{1}{q_2^2}\int_0^{\bar x_2}dx_1 \left[V_1+A_1\right](x_i)
      +\frac{m_N^2}{q_2^4}\left[\mathcal{V}_1^{M(u)}+\mathcal{A}_1^{M(u)}\right]
        \right.
\nn\\ &&{}\left. 
\hspace{0.0cm}
      +\frac{x_2m_N^2}{q_2^4} \left[ 
\widehat{V}_{1345} +  \widehat{V}_{123}+\widehat{A}_{123} - 2\widehat{V}_{43}+\widehat{A}_{34}\right]
\right\},
\eea
\bea{Ioffe-a}
  \mathcal{A}^{\mathrm{a,nc}}_1 &=& \int_0^1 dx_3\left\{
     \frac{Q^2+q_3^2}{q_3^4} \widetilde{V}_{123}+\frac{x_3}{q_3^2}\int_0^{\bar x_3} dx_1 V_3(x_i)+
      \frac{x_3^2m_N^2}{q_3^4}\widetilde{V}_{43}\right\}
\nn\\
&&{}+   \int_0^1 dx_2 \left\{
       \phantom{-}
      \frac{x_2}{q_2^2} \int_0^{\bar x_2}dx_1 \left[2 A_1 + 3A_3+V_3\right](x_i) 
      + \frac{2 x_2 m_N^2}{q_2^4}\mathcal{A}_1^{M(u)}
      + \frac{Q^2-q_2^2}{q_2^4}\widehat{A}_{123}
\right.
\nn\\
&&{}\left. \hspace{0.0cm}
       +\frac{Q^2+q_2^2}{q_2^4}\widehat{V}_{123} 
      +\frac{x_2^2 m_N^2}{q_2^4} \left[ \widehat{A}_{1345}- 2\widehat{A}_{34}+ \widehat{V}_{43}\right] 
-\frac{2x_2 m_N^2}{q_2^4}\,\widehat{\!\widehat{A}}_{123456}
\right\},
\nn\\
 \mathcal{B}^{\mathrm{a,nc}}_1 &=& \int_0^1 dx_3\left\{
    \frac{1}{q_3^2}\int_0^{\bar x_3} dx_1 V_1(x_i) +\frac{m_N^2}{q_3^4}\mathcal{V}_1^{M(d)}
     -\frac{x_3 m_N^2}{q_3^4}\left[ \widetilde{V}_{123} -\widetilde{V}_{43} \right]
      \right\}
\nn\\ &&{}
 +  \int_0^1 dx_2 \left\{
    \frac{1}{q_2^2}\int_0^{\bar x_2}dx_1 \left[V_1+A_1\right](x_i)
      +\frac{m_N^2}{q_2^4}\left[\mathcal{V}_1^{M(u)}+\mathcal{A}_1^{M(u)}\right]
\right.
\nn\\ &&{}\left. \hspace{0.0cm}
      +\frac{x_2m_N^2}{q_2^4}\left[
\widehat{A}_{1345} - \widehat{V}_{123}-\widehat{A}_{123} - 2\widehat{A}_{34}+\widehat{V}_{43}\right]
\right\}.
\eea
{}For the correlation functions involving the leading-twist current $\eta_3$ we get
%
%
\bea{CZ-em}
{\cal A}_3^{\rm em} & = &
-\frac{4}{3} e_u\int_0^1 \dd x_2 
\left\{\frac{x_2}{q_2^2} 
\int_0^{\bar x_2}\!\!dx_1 \left[V_1+2 T_1 \right](x_i) 
+ \frac{x_2m_N^2  }{q_2^4} \left( {\cal V}_1^{M(u)}+2{\cal T}_1^{M(u)} \right)
-\frac{x_2^2m_N^2}{q_2^4}\left[\widehat{V}_{12345} +2 \widehat{T}_{13478}
 \right] \right.
\nn\\&& \left.\hspace*{0.0cm}{}
-\frac{2x_2^3m_N^4}{q_2^6}\left[\widehat{\widehat{V}}_{123456} -2\, \widehat{\!\widehat{T}}_{134678} \right]
\right\} 
- \frac{2}{3} e_d \int_0^1 \dd x_3 \left\{ (x_2 \to x_3, \widehat{F} \to \widetilde{F} ) \right\},
\nn\\
{\cal B}_3^{\rm em} & = &
\phantom{-} \frac{4}{3} e_u\int_0^1\!\! \dd x_2 
\left\{ 
\frac{x_2m_N^2}{q_2^4}\left[ \widehat{V}_{123} +2 \widehat{T}_{137}\right] 
+ \frac{2x_2^2 m_N^4}{q_2^6} 
\left[ \widehat{\widehat{V}}_{123456} -2\, \widehat{\!\widehat{T}}_{134678}
        \right]
\right\}
\nn \\ &&{}  
+ \frac{2}{3} e_d  \int_0^1 \dd x_3 \left\{ (x_2 \to x_3), \widehat{F} \to \widetilde{F} \right\},
\eea
\bea{CZ-a}
{\cal A}_3^{\mathrm{a,nc}} &=& 
\frac{2}{3} \int_0^1\!\!\!\! \dd x_2  \left\{ 
\frac{x_2}{q_2^2}\int_0^{\bar x_2}\!\!dx_1 \left[A_1 +2 T_1\right](x_i)
+ \frac{x_2m_N^2}{q_2^4}
\left[ {\cal A}_1^{M(u)}+2{\cal T}_1^{M(u)} \right]
- \frac{x_2^2m_N^2 }{q_2^4} \left[\widehat{A}_{12345}  +2 \widehat{T}_{13478} \right] 
\right.
\nn\\
&&
\left.{}\hspace*{0.0cm}
+ \frac{ 2 x_2^3m_N^4}{q_2^6}\left[ \,\widehat{\!\widehat{A}}_{123456} +2\, \widehat{\!\widehat{T}}_{134678} \right]
\right\} 
-\frac{1}{3} \int_0^1\!\! \dd x_3 \left\{ \frac{x_3}{q_3^2}
\int_0^{\bar x_3}\!\!dx_1 \left[V_1 (x_i) -2 T_1 (x_i) \right]\right.  
\nn\\
&&{}
\left.
+\frac{x_3m_N^2  }{q_3^4}
 \left[{\cal V}_1^{M(d)}-2{\cal T}_1^{M(d)} \right]
-\frac{x_3^2m_N^2}{q_3^4} \left[ \widetilde{V}_{12345}  -2 \widetilde{T}_{13478} \right] 
- \frac{ 2 x_3^3m_N^4}{q_3^6} \left[\widetilde{\widetilde{V}}_{123456} +2\,\widetilde{\!\widetilde{T}}_{134678} \right]
\right\},
\nn \\
{\cal B}_3^{\mathrm{a,nc}} &=& 
-\frac{2}{3} \int_0^1 \dd x_2 \left\{ \frac{x_2m_N^2}{q_2^4} 
  \left[\widehat{A}_{123} -2 \widehat{T}_{137}\right] + \frac{2 x^2_2 m_N^4}{q_2^6}
\left[\,\widehat{\!\widehat{A}}_{123456} + 2\, \widehat{\!\widehat{T}}_{134678} \right]
\right\}
\nn\\
&&{}-
\frac{1}{3} \int_0^1 \dd x_3 \left\{ \frac{x_3m_N^2}{q_3^4}
\left[\widetilde{V}_{123}  - 2 \widetilde{T}_{137}\right] 
+\frac{2 x^2_3 m_N^4}{q_3^6}\left[\widetilde{\widetilde{V}}_{123456} +2\, \widetilde{\!\widetilde{T}}_{134678} \right]
  \right\},
\eea  
and, finally, for the correlation functions involving $\eta_4$:
%
%
%
%
%
%
\bea{CZold-em}
{\cal A}_4^{\rm em} & = &
- 4 e_u\int_0^1 \dd x_2 \left\{\frac{x_2}{q_2^2} 
\int_0^{\bar x_2}\!\!dx_1 V_1(x_i) + \frac{x_2m_N^2}{q_2^4} {\cal V}_1^{M(u)}
-\frac{x_2^2m_N^2}{q_2^4} \widehat{V}_{12345} 
-\frac{2x_2^3m_N^4}{q_2^6} \widehat{\widehat{V}}_{123456} 
\right\} 
\nn\\&&{}
- 2 e_d  \int_0^1 \dd x_3 \left\{ (x_2 \to x_3, \widehat{F} \to \widetilde{F} ) \right\},
\nn\\
{\cal B}_4^{\rm em} & = &
\phantom{-} 4 e_u \int_0^1\!\! \dd x_2 
\left\{ \frac{x_2m_N^2}{q_2^4} \widehat{V}_{123} 
+ \frac{2x_2^2 m_N^4}{q_2^6} \widehat{\widehat{V}}_{123456} 
\right\}  
+ 2 e_d  \int_0^1 \dd x_3 \left\{ (x_2 \to x_3), \widehat{F} \to \widetilde{F} \right\},
\eea
\bea{CZold-a}
{\cal A}_4^{\mathrm{a,nc}} &=&\phantom{-} 
2 \int_0^1\!\!\!\! \dd x_2  \left\{ 
\frac{x_2}{q_2^2}\int_0^{\bar x_2}\!\!dx_1 A_1 (x_i)
+ \frac{x_2m_N^2}{q_2^4}{\cal A}_1^{M(u)}
- \frac{x_2^2m_N^2 }{q_2^4} \widehat{A}_{12345}  
+ \frac{ 2 x_2^3m_N^4}{q_2^6} \widehat{\!\widehat{A}}_{123456} 
\right\} 
\nn\\ &&{}-
 \int_0^1\!\! \dd x_3 \left\{ \frac{x_3}{q_3^2}
\int_0^{\bar x_3}\!\!dx_1 V_1 (x_i) 
+\frac{x_3m_N^2  }{q_3^4} {\cal V}_1^{M(d)}
-\frac{x_3^2m_N^2}{q_3^4} \widetilde{V}_{12345}  
- \frac{ 2 x_3^3m_N^4}{q_3^6} \widetilde{\widetilde{V}}_{123456} 
\right\},
\nn \\
{\cal B}_4^{\mathrm{a,nc}} &=& 
-2 \int_0^1 \dd x_2 \left\{ \frac{x_2m_N^2}{q_2^4} \widehat{A}_{123}  + \frac{2 x^2_2 m_N^4}{q_2^6}
\widehat{\!\widehat{A}}_{123456} 
\right\}
-  \int_0^1 \dd x_3 \left\{ \frac{x_3m_N^2}{q_3^4} \widetilde{V}_{123} 
+\frac{2 x^2_3 m_N^4}{q_3^6} \widetilde{\widetilde{V}}_{123456} 
  \right\},
\eea  
\bea{CZold-V}
{\cal A}_4^{\mathrm{v,cc}} &=& 
-2  \int_0^1\!\!\!\! \dd x_2  \left\{ 
\frac{x_2}{q_2^2}\int_0^{\bar x_2}\!\!dx_1 \left[V_1 -  A_1 - 2 T_1\right](x_i)
+ \frac{x_2m_N^2}{q_2^4} \left[  {\cal V}_1^{M(u)} - {\cal A}_1^{M(u)} - 2{\cal T}_1^{M(u)} \right]
\right.
\nn\\
&&
\left.{}\hspace*{0.0cm}
+ \frac{x_2^2m_N^2 }{q_2^4} \left[- \widehat{V}_{12345} + \widehat{A}_{12345}  +2 \widehat{T}_{13478} \right] 
- \frac{ 2 x_2^3m_N^4}{q_2^6}\left[ \widehat{\widehat{V}}_{123456}
                          +  \,\widehat{\!\widehat{A}}_{123456} 
                          +2\, \widehat{\!\widehat{T}}_{134678} \right]
\right\}, 
\nn\\ 
{\cal B}_4^{\mathrm{v,cc}} &=& 2 
\int_0^1 \dd x_2 \left\{ \frac{x_2m_N^2}{q_2^4} 
  \left[\widehat{V}_{123} + \widehat{A}_{123} + 2 \widehat{T}_{137}\right] 
+ \frac{2 x^2_2 m_N^4}{q_2^6}
\left[\widehat{\widehat{V}}_{123456} + \,\widehat{\!\widehat{A}}_{123456} 
+ 2\, \widehat{\!\widehat{T}}_{134678} \right]
\right\},
\eea  
\bea{CZold-Acc}
{\cal A}_4^{\mathrm{a,cc}} &=& 
2  \int_0^1\!\!\!\! \dd x_2  \left\{ 
\frac{x_2}{q_2^2}\int_0^{\bar x_2}\!\!dx_1 \left[V_1 -  A_1 + 2 T_1\right](x_i)
+ \frac{x_2m_N^2}{q_2^4} \left[  {\cal V}_1^{M(u)} - {\cal A}_1^{M(u)} + 2{\cal T}_1^{M(u)} \right]
\right.
\nn\\
&&
\left.{}\hspace*{0.0cm}
+ \frac{x_2^2m_N^2 }{q_2^4} \left[- \widehat{V}_{12345} + \widehat{A}_{12345}  -2 \widehat{T}_{13478} \right] 
- \frac{ 2 x_2^3m_N^4}{q_2^6}\left[ \widehat{\widehat{V}}_{123456}
                          +  \,\widehat{\!\widehat{A}}_{123456} 
                          -2\, \widehat{\!\widehat{T}}_{134678} \right]
\right\}, 
\nn\\ 
{\cal B}_4^{\mathrm{a,cc}} &=& 2 
\int_0^1 \dd x_2 \left\{ \frac{x_2m_N^2}{q_2^4} 
  \left[ \widehat{V}_{123} + \widehat{A}_{123} - 2 \widehat{T}_{137}\right] 
+ \frac{2 x^2_2 m_N^4}{q_2^6}
\left[\widehat{\widehat{V}}_{123456} + \,\widehat{\!\widehat{A}}_{123456} 
- 2\, \widehat{\!\widehat{T}}_{134678} \right]
\right\}.
\eea  
In all expressions the functions with a ``tilde'' and a ``hat'' have $x_3$ and $x_2$ as
an argument, respectively, cf. (\ref{notat1}), (\ref{notat2}).
Also, in the terms involving two integrations over the momentum fractions, the remaining
momentum fraction is replaced by using $x_1+x_2+x_3=1$. The results in (\ref{CZold-em}) agree with 
the corresponding expressions in \cite{Braun:2001tj} up to two misprints: a factor two in the $\mathcal{V}^{M(u)}$-term
and the sign of the $V_{123456}$ contribution; 
the other expressions are new. 

The answers for the neutral vector current $j_\nu^{\mathrm{v,nc}}$
are easily obtained from the corresponding expressions for the electromagnetic current by
a substitution $e_u \to 1/2$ and $e_d \to -1/2$. Also, since the currents $\eta_1$ and $\eta_3$ 
are pure isospin $I=1/2$, the correlation functions involving the flavor-changing charged
currents $j_\nu^{\mathrm{v,cc}}$ and $j_\nu^{\mathrm{a,cc}}$ are given in terms of the corresponding 
correlation functions involving the flavor-conserving  currents by exact isospin relations
\bea{isospin3}
 && \mathcal{A}_i^{\mathrm{v,cc}} = 2 \mathcal{A}_i^{\mathrm{v,nc}}\,,\qquad  
    \mathcal{B}_i^{\mathrm{v,cc}} = 2 \mathcal{B}_i^{\mathrm{v,nc}}\,,\qquad
     \mathcal{A}_i^{\mathrm{a,cc}} = 2 \mathcal{A}_i^{\mathrm{a,nc}}\,,\qquad  
    \mathcal{B}_i^{\mathrm{a,cc}} = 2 \mathcal{B}_i^{\mathrm{a,nc}}\,\qquad i=1,3\,.
\eea   
These relations are not manifest because they involve isospin relations between different 
nucleon DAs and provide a nontrivial check of the calculation. 

The Borel transformation and the continuum subtraction  
are performed by using the following substitution rules:  
\bea{borel0}  
\int \dd x \frac{\varrho(x) }{(q-x  P)^2} &=&  
- \int_0^1 \frac{\dd x }{x} \frac{\varrho(x)}{(s - {P'}^2)}  
  \,\to\,  
-   \int_{x_0}^1 \frac{\dd x}{x} \varrho(x)  
\exp{ \left( - \frac{\bar x Q^2}{x M^2}  - \frac{\bar x m_N^2}{M^2}\right)}\, ,    
\nonumber\\
\int \dd x \frac{\varrho(x) }{(q-x  P)^4} &=&  
\int_0^1 \frac{\dd x }{x^2} \frac{\varrho(x)}{(s - {P'}^2)^2}  
\,\to\,  
\frac{1}{M^2} \int_{x_0}^1 \frac{\dd x}{x^2} \varrho(x)  
\exp{  
\left( - \frac{\bar x Q^2}{x M^2}  - \frac{\bar x m_N^2}{M^2}\right)}    
+  
\frac{\varrho(x_0)\,e^{-s_0 /M^2} }{Q^2 + x_0^2 m_N^2} \, ,    
\nonumber\\
\int \dd x \frac{\varrho(x) }{(q-x  P)^6} &=&  
- \int_0^1 \frac{\dd x }{x^3} \frac{\varrho(x)}{(s - {P'}^2 )^3}  
 \,\to\,  -  
\frac{1}{2 M^4} \int_{x_0}^1 \frac{\dd x}{x^3} \varrho(x)  
\exp{  
\left( - \frac{\bar x Q^2}{x M^2} - \frac{\bar x m_N^2}{M^2}  
\right)}  
\nonumber \\  &&  
- \frac12  
\frac{\varrho(x_0)\,e^{-s_0 /M^2} }{x_0\left(Q^2 + x_0^2 m_N^2\right) M^2}  
+ \frac12  \frac{x_0^2}{Q^2 + x_0^2 m_N^2} \left[\frac{d}{dx_0}  
\frac{\varrho(x_0)}{x_0\left(Q^2 + x_0^2 m_N^2\right)} \right]  
\,e^{-s_0 /M^2} \, \nonumber \\  
\eea  
where $M$ is the Borel parameter, 
$s = \frac{1-x}{x} Q^2  + (1-x) m_N^2$ and 
$x_0$ is the solution of the corresponding quadratic equation for $s = s_0$:
\bea{x0}
   x_0 &=&\bigg[ \sqrt{(Q^2+s_0-m_N^2)^2+ 4 m_N^2 Q^2}-(Q^2+s_0-m_N^2)\bigg]
   /(2m_N^2)\,.
\eea  
The contributions $\sim e^{-s_0 /M^2}$ in \Gl{borel0}
correspond to the ``surface terms'' arising from  successive  
partial integrations to reduce the power in the denominators  
$(q - x P)^{2N} = (s  - {P'}^2 )^{2N} (-x)^{2N}$  
with $N > 1$ to the usual dispersion representation with the denominator  
$\sim (s - {P'}^2 )$. Without continuum subtraction, i.e. in the 
limit  $s_0 \to \infty$ these terms vanish.  

In addition, in the hadronic representation for the same correlation 
functions one has to make the substitution
\beq{borel5}
 \frac{1}{m_N^2-P'^2} \to e^{-m_N^2/M^2}.
\eeq
{}As an example, we present here the final sum rules for $F_1^p$, $F_2^p$ and $G_A^{\mathrm{NC}}$
  obtained using the Ioffe current:
\bea{LCSRIoffe}
F_1^{\rm em}(Q^2)  & = & \frac{1}{2 \lambda_1}
\left[ \int_{x_0}^1 \dd x \left( - \frac{\varrho_2^a(x)}{x}  + 
\frac{\varrho_4^a(x)}{x^2M^2} \right)
        \exp{ \left( - \frac{\bar x Q^2}{x M^2}  + \frac{x 
m_N^2}{M^2}\right)}\,
+
\frac{\varrho_4^a(x_0)\,e^{-(s_0-m_N^2) /M^2} }{Q^2 + x_0^2 m_N^2} \right],
\nn
\\
F_2^{\rm em}(Q^2) & = & \frac{1}{\lambda_1}
\left[ \int_{x_0}^1 \dd x \left( - \frac{\varrho_2^b(x)}{x}  + 
\frac{\varrho_4^b(x)}{x^2M^2} \right)
        \exp{ \left( - \frac{\bar x Q^2}{x M^2}  + \frac{x 
m_N^2}{M^2}\right)}\,
+
\frac{\varrho_4^b(x_0)\,e^{-(s_0-m_N^2) /M^2} }{Q^2 + x_0^2 m_N^2} \right],
\nn
\\
G_A^{\rm NC}(Q^2) & = & \frac{1}{2 \lambda_1}
\left[ \int_{x_0}^1 \dd x \left( - \frac{\varrho_2^c(x)}{x}  + 
\frac{\varrho_4^c(x)}{x^2M^2} \right)
        \exp{ \left( - \frac{\bar x Q^2}{x M^2}  + \frac{x 
m_N^2}{M^2}\right)}\,
+
\frac{\varrho_4^c(x_0)\,e^{-(s_0-m_N^2) /M^2} }{Q^2 + x_0^2 m_N^2} \right],
\nn
\\
\eea
with
\bea{abbrev}
\varrho_2^a(x) & = &\phantom{-}
       2e_d \bigg\{ \widetilde{V}_{123} + x \int_0^{\bar x} dx_1 V_3(x_i)+ 
\bigg\}
  + 2 e_u \bigg\{ x \int_0^{\bar x}dx_1 \left[ -2 V_1 + 3V_3+A_3 
\right](x_i)
        - \widehat{V}_{123} + \widehat{A}_{123} \bigg\},
\nn \\
\varrho_4^a(x) & = &\phantom{-}
       2e_d \bigg\{Q^2 \widetilde{V}_{123}+ x^2 m_N^2 
\widetilde{V}_{43}\bigg\}
+ 2 e_u  \bigg\{ Q^2 \left( \widehat{V}_{123} + \widehat{A}_{123} \right)
- x^2 m_N^2 \left[\widehat{V}_{1345} - 2\widehat{V}_{43}+ \widehat{A}_{34}\right]
 \nn \\ && {}
- 2x m_N^2  \left( \mathcal{V}_1^{M(u)} + \widehat{\widehat{V}}_{123456} \right) \bigg\},
\nn \\
\varrho_2^b(x) & = &
- 2 e_d \left\{ \int_0^{\bar x} dx_1 V_1(x_i)    \right\}
+ 2 e_u \left\{ \int_0^{\bar x}dx_1 \left[V_1+A_1\right](x_i) \right\},
\nn \\
\varrho_4^b(x) & = &  -2e_d m_N^2
\bigg\{ \mathcal{V}_1^{M(d)} - x \left[\widetilde{V}_{123} - 
\widetilde{V}_{43} \right] \bigg\}
  + 2 e_u m_N^2 \bigg\{ \left[\mathcal{V}_1^{M(u)}+\mathcal{A}_1^{M(u)}\right]
\nn \\ &&{}
  + x \left[\widehat{V}_{1345} + 
\widehat{V}_{123}+\widehat{A}_{123}-2\widehat{V}_{43}+\widehat{A}_{34}\right]
\bigg\},
\nn \\
\varrho_2^c(x) & = &
\left\{\widetilde{V}_{123}+ x \int_0^{\bar x} dx_1 V_3(x_i) \right\}
+\left\{ x  \int_0^{\bar x}dx_1 \left[2 A_1 + 
3A_3+V_3\right](x_i)
       - \widehat{A}_{123} +\widehat{V}_{123}  \right\},
\nn \\
\varrho_4^c(x) & = &
  \bigg\{ Q^2 \widetilde{V}_{123}+ x^2m_N^2 \widetilde{V}_{43}\bigg\}
+\bigg\{
     Q^2 \left( \widehat{A}_{123} +\widehat{V}_{123} \right)
    + x^2 m_N^2 \left[ \widehat{A}_{1345}- 2\widehat{A}_{34}+ 
\widehat{V}_{43}\right]
\nn \\ &&{}
    + 2x  m_N^2 \left( \mathcal{A}_1^{M(u)} - 
\widehat{\!\widehat{A}}_{123456} \right)
\bigg\}.
\eea
Form factors of the neutron are obtained by the substitution $e_u\leftrightarrow e_d$.

\section{Nucleon Distribution Amplitudes}
\setcounter{equation}{0}

In the following we give a summary of the three-quark distribution amplitudes 
from twist-3 to twist-6 as obtained in \cite{BFMS}. The DAs are defined 
by the matrix element of the renormalized three-quark operator at light-like separations
\bea{defdisamp}
\lefteqn{ 4 \bra{0} \ep^{ijk} u_\al^i(a_1 z) u_\be^j(a_2 z) d_\ga^k(a_3 z) 
\ket{P} =}
\nn \\
&=&  
S_1 m_N C_{\al \be} \left(\ga_5 N^+\right)_\ga + 
S_2 m_N C_{\al \be} \left(\ga_5 N^-\right)_\ga + 
P_1 m_N \left(\ga_5 C\right)_{\al \be} N^+_\ga + 
P_2 m_N \left(\ga_5 C \right)_{\al \be} N^-_\ga  
\nn \\
&& + 
V_1  \left(\!\not\!{p}C \right)_{\al \be} \left(\ga_5 N^+\right)_\ga + 
V_2  \left(\!\not\!{p}C \right)_{\al \be} \left(\ga_5 N^-\right)_\ga + 
\frac{V_3}{2} m_N  \left(\ga_\perp C \right)_{\al \be}\left( \ga^{\perp} 
\ga_5 N^+\right)_\ga 
\nn \\ 
&& +
\frac{V_4}{2} m_N  \left(\ga_\perp C \right)_{\al \be}\left( \ga^{\perp} 
\ga_5 N^-\right)_\ga + 
V_5  \frac{m_N^2}{2 p z} 
\left(\!\not\!{z}C \right)_{\al \be} \left(\ga_5 N^+\right)_\ga + 
\frac{m_N^2}{2 pz} V_6  \left(\!\not\!{z}C \right)_{\al \be} \left(\ga_5 N^-\right)_\ga 
\nn \\
&& + 
A_1  \left(\!\not\!{p}\ga_5 C \right)_{\al \be} N^+_\ga + 
A_2  \left(\!\not\!{p}\ga_5 C \right)_{\al \be} N^-_\ga + 
\frac{A_3}{2} m_N  \left(\ga_\perp \ga_5 C \right)_{\al \be}\left( \ga^{\perp} 
N^+\right)_\ga 
\nn \\ 
&& +
\frac{A_4}{2} m_N  \left(\ga_\perp \ga_5 C \right)_{\al \be}\left( \ga^{\perp} 
N^-\right)_\ga + 
A_5  \frac{m_N^2}{2 p z} 
\left(\!\not\!{z}\ga_5 C \right)_{\al \be} N^+_\ga + 
\frac{m_N^2}{2 pz}  A_6  \left(\!\not\!{z}\ga_5 C \right)_{\al \be} N^-_\ga 
\nn \\
&& +
T_1 \left(i \si_{\perp p} C\right)_{\al \be} 
\left(\ga^\perp\ga_5 N^+\right)_\ga + 
T_2 \left(i \si_{\perp\, p} C\right)_{\al \be} 
\left(\ga^\perp\ga_5 N^-\right)_\ga + 
T_3 \frac{m_N}{p z}
\left(i \si_{p\, z} C\right)_{\al \be} 
\left(\ga_5 N^+\right)_\ga  
\nn \\
&& 
+ T_4 \frac{m_N}{p z}\left(i \si_{z\, p} C\right)_{\al \be} 
\left(\ga_5 N^-\right)_\ga + 
T_5 \frac{m_N^2}{2 p z}  \left(i \si_{\perp\, z} C\right)_{\al \be} 
\left(\ga^\perp\ga_5 N^+\right)_\ga + 
\frac{m_N^2}{2 pz}  T_6 \left(i \si_{\perp\, z} C\right)_{\al \be} 
\left(\ga^\perp\ga_5 N^-\right)_\ga  
\nn \\ 
&& + m_N \frac{T_7}{2} \left(\si_{\perp\, \perp'} C\right)_{\al \be} 
\left(\si^{\perp\, \perp'} \ga_5 N^+\right)_\ga +
m_N \frac{T_8}{2} \left(\si_{\perp\, \perp'} C\right)_{\al \be} 
\left(\si^{\perp\, \perp'} \ga_5 N^-\right)_\ga\,, 
\eea
\end{widetext}
where for brevity we do not show the Wilson lines that make this operator gauge-invariant;
$\alpha,\beta,\gamma$ are Dirac indices and $C$ is the charge-conjugation matrix \cite{BD65}.
Each DA $F = V_i,A_i,T_i,S_i,P_i$ can be 
represented as 
\bea{fourier}
F(a_j, Px) = \int \! {\cal D} x\, e^{-i Px \sum_i x_i a_i} F(x_i)\,,
\eea
where the functions $F(x_i)$ depend on the dimensionless
variables $x_i,\, 0 < x_i < 1, \sum_i x_i = 1$ which 
correspond to the longitudinal momentum fractions 
carried by the quarks inside the nucleon.  
The integration measure is defined as 
\bea{integration}
{}\hspace*{-3mm}\int\! {\cal D} x  = \!\int_0^1\!\!\! \dd x_1 \dd x_2 \dd x_3\, 
\de (x_1 + x_2 + x_3 - 1)\,.\hspace*{5mm}\phantom{-}
\eea 

Distribution amplitudes can be expanded in contributions of operators with a given 
conformal spin \cite{BFMS}. This is convenient since operators with different spin do not 
mix under renormalization in one loop. More importantly, only the operators with the same spin can be 
related by equations of motion so that the truncation of the conformal spin expansion at a certain 
order produces a selfconsistent approximation. In Ref.~\cite{BFMS} the contributions of the leading 
and the next-to-leading order in conformal expansion have been taken into account. 

To this accuracy one obtains twist-3 DAs:
\bea{twist-3} 
\label{All-twist-3}
V_1(x_i,\mu) &=& 120 x_1 x_2 x_3 \left[\phi_3^0(\mu) + 
\phi_3^+(\mu) (1- 3 x_3)\right]\,,
\nn \\
A_1(x_i,\mu) &=& 120 x_1 x_2 x_3 (x_2 - x_1) \phi_3^-(\mu) \,,
\nn \\
T_1(x_i,\mu) &=& 120 x_1 x_2 x_3 
\Big[\phi_3^0(\mu) 
\nonumber\\&&{}- \frac12\left(\phi_3^+ - \phi_3^-\right)(\mu) 
(1- 3 x_3)\Big]\,.
\eea
{Twist-4 distribution amplitudes:}
\bea{twist-4} 
\label{All-twist-4}
V_2(x_i,\mu)  &=& 24 x_1 x_2 
\left[\phi_4^0(\mu)  + \phi_4^+(\mu)  (1- 5 x_3)\right] \,,
\nn\\
A_2(x_i,\mu)  &=& 24 x_1 x_2 (x_2 - x_1) \phi_4^-(\mu) \,,
\nn \\
T_2(x_i,\mu) 
&=& 24 x_1 x_2 \left[
\xi_4^0(\mu) + \xi_4^+(\mu)( 1- 5 x_3)\right]\,,
\nn \\
V_3(x_i,\mu)  &=& 
12 x_3 \left[
\psi_4^0(\mu)(1-x_3)  \right.\nn\\
&&{}\hspace*{-1.2cm}
+ \psi_4^+(\mu)( 1-x_3 - 10 x_1 x_2) \nn\\
&& \left.\hspace*{-1.2cm}+ \psi_4^-(\mu) (x_1^2 + x_2^2 - x_3 (1-x_3) ) \right]\,,
\nn \\
A_3(x_i,\mu) &=& 12 x_3 (x_2-x_1) 
\left[\left(\psi_4^0 + 
\psi_4^+ \right)(\mu)
\right. \nn\\&&\left.{}\hspace*{-1.2cm}
+ \psi_4^-(\mu) (1-2 x_3) \right] \,,
\nn \\
T_3(x_i,\mu)  &=& 
6 x_3 \left[
(\phi_4^0 + \psi_4^0 + \xi_4^0)(\mu)(1-x_3) \right.\nn\\ 
&&\hspace*{-1.2cm}\left.+
(\phi_4^+ + \psi_4^+ + \xi_4^+)(\mu) ( 1-x_3 - 10 x_1 x_2)
\right. \nn \\ 
&&\left. \hspace*{-1.2cm}+ 
(\phi_4^- - \psi_4^- + \xi_4^-)(\mu) (x_1^2 + x_2^2 - x_3 (1-x_3) )
\right],
\nn \\
T_7(x_i,\mu)  &=& 
6 x_3 \left[
(\phi_4^0 + \psi_4^0 - \xi_4^0)(\mu)(1-x_3) \right.\nn\\ 
&&\left.\hspace*{-1.2cm}+
(\phi_4^+ + \psi_4^+ - \xi_4^+)(\mu) ( 1-x_3 - 10 x_1 x_2)
\right. \nn \\ 
&&\left.\hspace*{-1.2cm} + 
(\phi_4^- - \psi_4^- - \xi_4^-)(\mu) (x_1^2 + x_2^2 - x_3 (1-x_3) )
\right],
\nn \\
S_1(x_i,\mu) &=& 
6 x_3 (x_2-x_1) \left[
(\phi_4^0 + \psi_4^0 + \xi_4^0 + \phi_4^+ + \psi_4^+ 
\right. \nn\\
&& \left.\hspace*{-1.2cm} + \xi_4^+)(\mu)+ (\phi_4^- - \psi_4^- + \xi_4^-)(\mu)(1-2 x_3) \right],
\nn \\
P_1(x_i,\mu) &=& 
6 x_3 (x_1-x_2) \left[
(\phi_4^0 + \psi_4^0 -\xi_4^0 + \phi_4^+ + \psi_4^+ \right. \nn\\
&& \left.\hspace*{-1.2cm}- \xi_4^+)(\mu) + (\phi_4^- - \psi_4^- - \xi_4^-)(\mu)(1-2 x_3) \right].
\nn\\{}
\eea
Note that $T_3$ and $T_7$ differ only in the sign of the $\xi$-contributions, while
$P_1$ and $S_1$ differ only in the sign of the $\phi$- and $\psi$-contributions.

{Twist-5 distribution amplitudes:}
\bea{twist-5} 
V_4(x_i,\mu) &=& 3 \left[
\psi_5^0(\mu)(1-x_3) 
\right.\nn\\&&\left.\hspace*{-1.2cm}{}
 + \psi_5^+(\mu)(1-x_3 - 2 (x_1^2 +  x_2^2))
\right.\nn\\&&\left.\hspace*{-1.2cm}{}
+ \psi_5^-(\mu)\left(2 x_1x_2 - x_3(1-x_3)\right) \right],
\nn\\
A_4(x_i,\mu) &=& 3 (x_2 -x_1)
\left[- \psi_5^0(\mu) + \psi_5^+(\mu)(1- 2 x_3) 
\right.\nn\\&&\left.\hspace*{-1.2cm}{}
+ \psi_5^-(\mu) x_3  \right],
\nn \\
T_4(x_i,\mu) &=& \frac32 \left[
(\phi_5^0 +  \psi_5^0 + \xi_5^0) (\mu)(1-x_3)
\right.\nn\\&&\left.\hspace*{-1.2cm}{}
+ \left(\phi_5^+ + \psi_5^+ + \xi_5^+ \right)(\mu)(1-x_3 - 2 (x_1^2 +  x_2^2))
\right.\nn\\&&\left.\hspace*{-1.2cm}{}
+\left(\phi_5^- - \psi_5^- + \xi_5^- \right) (\mu)\left(2 x_1x_2 - x_3(1-x_3)\right)\right],
\nn \\
T_8(x_i,\mu) &=& \frac32 \left[
(\phi_5^0 +  \psi_5^0 - \xi_5^0) (\mu)(1-x_3)
\right.\nn\\&&\left.\hspace*{-1.2cm}{}
+ \left(\phi_5^+ + \psi_5^+ - \xi_5^+ \right)(\mu)(1-x_3 - 2 (x_1^2 +  x_2^2))
\right.\nn\\&&\left.\hspace*{-1.2cm}{}
+\left(\phi_5^- - \psi_5^- - \xi_5^- \right) (\mu)\left(2 x_1x_2 - x_3(1-x_3)\right)
\right],
\nn \\
V_5(x_i,\mu) &=& 6 x_3 
\left[\phi_5^0(\mu)  + \phi_5^+(\mu)(1- 2 x_3)\right], 
\nn\\
A_5(x_i,\mu) &=& 6 x_3 (x_2-x_1) \phi_5^-(\mu) \,, 
\nn\\
T_5(x_i,\mu) &=& 6 x_3 \left[ 
\xi_5^0(\mu) + \xi_5^+(\mu)( 1- 2 x_3)\right],
\nn \\
S_2(x_i,\mu) &=& \frac32 (x_2 -x_1) 
\left[- \left(\phi_5^0 + \psi_5^0 + \xi_5^0\right)(\mu)
\right.\nn\\&&\left.\hspace*{-1.2cm}{} 
+ \left(\phi_5^+ + \psi_5^+ + \xi_5^+\right)(\mu) (1- 2 x_3)
\right.\nn\\&&\left.\hspace*{-1.2cm}{}
+ \left(\phi_5^- - \psi_5^- + \xi_5^- \right)(\mu) x_3 
\right],
\nn \\
P_2(x_i,\mu) &=& \frac32 (x_2 -x_1)
\left[- \left(-\phi_5^0 - \psi_5^0 + \xi_5^0\right)(\mu)
\right.\nn\\&&\left.\hspace*{-1.2cm}{}  
+ \left(-\phi_5^+ - \psi_5^+ + \xi_5^+\right)(\mu) (1- 2 x_3)
\right.\nn\\&&\left.\hspace*{-1.2cm}{} 
+ \left(-\phi_5^- + \psi_5^- + \xi_5^- \right)(\mu) x_3 
\right]\,,
\eea
Note that $T_4$ and $T_8$ differ only in the sign of the $\xi$-contributions, while
$P_2$ and $S_2$ differ only in the sign of the $\phi$- and $\psi$-contributions.
Note that the results for  $S_2$ and $P_2$ quoted in \cite{BFMS} contain misprints.

{Finally, twist-6 distribution amplitudes:}
\bea{twist-6} 
\label{All-twist-6}
V_6(x_i,\mu) &=& 2 \left[\phi_6^0(\mu) +  \phi_6^+(\mu) (1- 3 x_3)\right],
\nn \\
A_6(x_i,\mu) &=& 2 (x_2 - x_1) \phi_6^- \,, \\
T_6(x_i,\mu) &=& 2 \Big[\phi_6^0(\mu) - 
\frac12\left(\phi_6^+-\phi_6^-\right) (1\!-\! 3 x_3)\Big].
\nn
\eea
In all cases $\mu$ is the renormalization scale.

The coefficients in the above expansions can be expressed in terms of eight
non-perturbative parameters $f_N, \lambda_1, \lambda_2, f_1^u, f_1^d, f_2^d, A_1^u, V_1^d$ 
which are defined in Appendix D, see also \cite{BFMS}, section 3.2.
The corresponding relations read, for the leading conformal spin:
\bea{local-coefficients}
\phi_3^0 = \phi_6^0 = f_N \,,\hspace{0.15cm} 
&~&  
\phi_4^0 = \phi_5^0 = 
\frac{1}{2} \left(f_N + \la_1 \right) \,, 
\nn \\
\xi_4^0 = \xi_5^0 = 
\frac{1}{6} \la_2\,,
&~&  
\psi_4^0  = \psi_5^0 =          
\frac12\left(f_N - \la_1 \right). 
\eea
{}For the next-to-leading spin, 
for twist-3:
\bea{twist3-parameter}
\phi_3^- = \frac{21}{2} f_N \, A_1^u, &~&
\phi_3^+ = \frac{7}{2}  f_N \, (1 - 3 V_1^d),
\eea
for twist-4:
\bea{twist4-parameter}
\phi_4^+ &=& \frac{1}{4} \left[ 
 f_N( 3 - 10  V_1^d ) + \la_1(3- 10 f_1^d) 
\right],
\nn \\
\phi_4^- &=& - \frac{5}{4} \left[
f_N( 1 - 2 A_1^u ) - \la_1(1- 2 f_1^d -4 f_1^u)
\right],
\nn \\
\psi_4^+ &=& - \frac{1}{4} \left[
f_N( 2\! +\! 5 A_1^u \!-\! 5 V_1^d) - \la_1 (2 \!-\! 5 f_1^d \!-\! 5 f_1^u) 
\right],
\nn \\
\psi_4^- &=&  \frac{5}{4} \left[
f_N(2 - A_1^u - 3 V_1^d ) - \la_1(2- 7 f_1^d + f_1^u)
\right],
\nn \\
\xi_4^+ &=& \frac{1}{16} \la_2 (4\!-\! 15 f_2^d)\,, ~
\xi_4^- = \frac{5}{16} \la_2(4\!-\! 15 f_2^d),
\eea
for twist-5: 
\bea{twist5-parameter}
\phi_5^+ &=& 
- \frac{5}{6} \left[
f_N( 3 + 4   V_1^d) - \la_1 (1 - 4 f_1^d )
\right],
\nn \\
\phi_5^- &=& 
- \frac{5}{3} \left[
f_N( 1 - 2 A_1^u ) - \la_1(f_1^d - f_1^u) 
\right],
\nn \\
\psi_5^+ &=& 
-\frac{5}{6} \left[
f_N(5\! + \!2 A_1^u \!-\!2 V_1^d) - \lambda_1 (1 \!-\! 2 f_1^d \!-\! 2 f_1^u)
\right],
\nn \\
\psi_5^- &=& \phantom{-}
\frac{5}{3} \left[
f_N( 2 - A_1^u - 3 V_1^d) + \la_1 (f_1^d - f_1^u) 
\right],
\nn \\
\xi_5^+ &=& \phantom{-}\frac{5}{36} \lambda_2 (2- 9 f_2^d) \,,
\; \; \; \;
\xi_5^- = - \frac{5}{4} \la_2 f_2^d\,,
\eea
and for twist-6: 
\bea{twist6-parameter}
\phi_6^+ &=& \frac{1}{2}\left[
f_N ( 1 - 4 V_1^d ) - \la_1  (1 - 2 f_1^d) 
\right],
 \\
\phi_6^- &=& \phantom{-}\frac{1}{2} \left[
f_N (1 +  4 A_1^u ) + \la_1 (1- 4 f_1^d - 2 f_1^u)
\right] \,.
\nn
\eea
The normalization of all DAs  is determined by three dimensionful parameters  
$f_N, \lambda_1, \lambda_2$ that are well known from the QCD sum rule 
literature and correspond to nucleon couplings to the existing three different 
three-quark local operators with the correct spin and isospin, see \cite{Ioffe:1982ce}. 
The numerical values (at the scale $\mu = 1$ GeV) are \cite{BFMS,Che84}:
\bea{numerics1}
 f_N  &=&         (5.0 \pm 0.5) \cdot 10^{-3} \mbox{GeV}^2 \; 
\nonumber \\
\lambda_1  &=&  - (2.7 \pm 0.9) \cdot 10^{-2} \mbox{GeV}^2 \; 
\nonumber \\
\lambda_2  &=&    (5.4 \pm 1.9) \cdot 10^{-2} \mbox{GeV}^2 \; 
\eea
see also Appendix D. The remaining five parameters determine the shape of the 
DAs (deviation from the asymptotic form) and their values are much more
controversial. Asymptotic DAs correspond to the choice
\bea{asymptotic}
 &&V_1^d = \frac{1}{3}\,,~A_1^u =0\,,\nonumber\\&& f_1^d = \frac{3}{10}\,, ~f_1^u = \frac{1}{10}\,, ~f_2^d = \frac{4}{15}\,. 
\eea  
The leading-twist-3 parameters $V_1^d$ and $A_1^u$ were calculated  
using QCD sum rules in Refs~\cite{Che84,KS87,COZ88} with the result 
\cite{COZ88}
\bea{numericsCZ}
A_1^u  =    0.38 \pm 0.15\,,  
\nn \\
V_1^d  =    0.23 \pm 0.03\,,  
\eea
while the remaining twist-4 shape parameters $f_1^d, f_1^u, f_2^d$ were estimated 
by the same method in \cite{BFMS}. In this work we reconsider the corresponding sum rules
(see Appendix D) and obtain a new estimate  
\bea{numericsCZ2}
f_1^d  =    0.40 \pm 0.05\,,  
\nn \\
f_2^d  =    0.22 \pm 0.05\,,  
\nn \\ 
f_1^u  =    0.07 \pm 0.05\,. 
\eea
The set of nucleon DAs obtained using the parameters in (\ref{numericsCZ}) and 
(\ref{numericsCZ2}) is sometimes referred to as the Chernyak-Zhitnitsky-like model of the DAs.  

Alternatively, there exists a phenomenological model for the leading-twist DA \cite{Bolz:1996sw} which was obtained
by modelling the soft contribution by a convolution of light-cone wave functions. This model is much closer 
to the asymptotic DA compared to the CZ-model and corresponds to the choice 
\cite{Bolz:1996sw}
\bea{numericsBK}
A_1^u  =    1/14\,, 
\nn \\
V_1^d  =    13/42\,, 
\eea
Estimates of the higher-twist DAs in the same technique are not available. 

In the main text we suggest one more model that allows one to obtain good agreement with the data
on the nucleon form factors within the LCSR approach. The corresponding parameters are given in 
Eq.~(\ref{superset}).


\begin{widetext}

%
\section{Operator Product expansion of three-quark currents}
%
\setcounter{equation}{0}

In this Appendix we present the tree-level expansion of the nucleon matrix element 
of the three-quark operator in terms of nucleon DAs to the twist-5 accuracy.
The general Lorentz decomposition reads \cite{BFMS}
\bea{zerl}
\lefteqn{ 4 \bra{0} \ep^{ijk} u_\al^i(a_1 x) u_\be^j(a_2 x) d_\ga^k(a_3 x) \ket{P} }
\nn \\
&=&  
{\cal S}_1 m_N C_{\al \be} \left(\ga_5 N\right)_\ga + 
{\cal S}_2 m_N^2 C_{\al \be} \left(\!\not\!{x} \ga_5 N\right)_\ga + 
{\cal P}_1 m_N \left(\ga_5 C\right)_{\al \be} N_\ga + 
{\cal P}_2 m_N^2 \left(\ga_5 C \right)_{\al \be} \left(\!\not\!{x} N\right)_\ga 
\nn \\
&& + 
\left(\mathcal{V}_1+\frac{x^2m_N^2}{4}\mathcal{V}_1^M \right)
\left(\!\not\!{P}C \right)_{\al \be} \left(\ga_5 N\right)_\ga + 
\V_2 m_N \left(\!\not\!{P} C \right)_{\al \be} \left(\!\not\!{x} \ga_5 N\right)_\ga  + 
\V_3 m_N  \left(\ga_\mu C \right)_{\al \be}\left(\ga^{\mu} \ga_5 N\right)_\ga 
\nn \\ 
&& +
\V_4 m_N^2 \left(\!\not\!{x}C \right)_{\al \be} \left(\ga_5 N\right)_\ga +
\V_5 m_N^2 \left(\ga_\mu C \right)_{\al \be} \left(i \si^{\mu\nu} x_\nu \ga_5 
N\right)_\ga 
+ \V_6 m_N^3 \left(\!\not\!{x} C \right)_{\al \be} \left(\!\not\!{x} \ga_5 N\right)_\ga  
\nn \\ 
&& + 
\left(\mathcal{A}_1+\frac{x^2m_N^2}{4}\mathcal{A}_1^M\right)
\left(\!\not\!{P}\ga_5 C \right)_{\al \be} N_\ga + 
\A_2 m_N \left(\!\not\!{P}\ga_5 C \right)_{\al \be} \left(\!\not\!{x} N\right)_\ga  + 
\A_3 m_N  \left(\ga_\mu \ga_5 C \right)_{\al \be}\left( \ga^{\mu} N\right)_\ga 
\nn \\ 
&& +
\A_4 m_N^2 \left(\!\not\!{x} \ga_5 C \right)_{\al \be} N_\ga +
\A_5 m_N^2 \left(\ga_\mu \ga_5 C \right)_{\al \be} \left(i \si^{\mu\nu} x_\nu  
N\right)_\ga 
+ \A_6 m_N^3 \left(\!\not\!{x} \ga_5 C \right)_{\al \be} \left(\!\not\!{x} N\right)_\ga  
\nn \\
&& +
\left(\mathcal{T}_1+\frac{x^2m_N^2}{4}\mathcal{T}_1^M\right)
\left(P^\nu i \si_{\mu\nu} C\right)_{\al \be} \left(\ga^\mu\ga_5 N\right)_\ga 
+ 
\T_2 m_N \left(x^\mu P^\nu i \si_{\mu\nu} C\right)_{\al \be} \left(\ga_5 N\right)_\ga 
\nn \\
&& 
+ \T_3 m_N \left(\si_{\mu\nu} C\right)_{\al \be} \left(\si^{\mu\nu}\ga_5 N\right)_\ga 
+ \T_4 m_N \left(P^\nu \si_{\mu\nu} C\right)_{\al \be} \left(\si^{\mu\ro} x_\ro \ga_5 N\right)_\ga 
+ \T_5 m_N^2 \left(x^\nu i \si_{\mu\nu} C\right)_{\al \be} \left(\ga^\mu\ga_5 N\right)_\ga 
\nn \\
&& 
+\T_6 m_N^2 \left(x^\mu P^\nu i \si_{\mu\nu} C\right)_{\al \be} \left(\!\not\!{x} \ga_5 N\right)_\ga  
+\T_{7} m_N^2 \left(\si_{\mu\nu} C\right)_{\al \be} \left(\si^{\mu\nu} \!\not\!{x} \ga_5 N\right)_\ga
+ \T_{8} m_N^3 \left(x^\nu \si_{\mu\nu} C\right)_{\al \be} \left(\si^{\mu\ro} x_\ro \ga_5 N\right)_\ga \,.
\nn \\
\eea
where it is assumed that the ``calligrafic'' functions depend on $x^2$ at most logarithmically.
Leaving aside the terms in $x^2$,  $\mathcal{V}_1^M,\mathcal{A}_1^M$ and $\mathcal{T}_1^M$, the rest  
of the functions can be expressed, to the tree-level accuracy, in terms of the nucleon DAs at the renormalization scale 
$\mu^2 \sim |1/x^2|$:
 \bea{deftw}
 &&{\cal S}_1 = S_1 \,, \quad  
2 Px \, {\cal S}_2 = S_1-S_2 \,, \quad 
 {\cal P}_1 = P_1 \,, \quad 
2 Px \, {\cal P}_2 = P_2-P_1 \,,\quad 
 \V_1 = V_1 \,, \quad  
2 Px \, \V_2 = V_1 - V_2 - V_3  \,, 
\nonumber \\&& 
2 \V_3 = V_3 \,, \quad  
4 Px \, \V_4  = - 2 V_1 + V_3 + V_4  + 2 V_5\,,\quad 
4 Px \V_5 = V_4 - V_3\,, \quad 
\nonumber \\&&
4 \left(Px\right)^2  \V_6 = - V_1 + V_2 +  V_3 +  V_4 + V_5 - V_6\,,\quad
 \A_1 = A_1\,, \quad
 2 Px \A_2 = - A_1 + A_2 -  A_3\,, \quad
\nonumber \\&&
2 \A_3 = A_3\,, \quad 
 4 Px \A_4 = - 2 A_1 - A_3 - A_4  + 2 A_5\,, \quad
 4 Px \A_5 = A_3 - A_4\,, 
\nonumber \\&&
 4 \left(Px\right)^2  \A_6 =  A_1 - A_2 +  A_3 +  A_4 - A_5 + A_6\,,\quad
\T_1 = T_1\,,  \quad
2 Px \T_2 = T_1 + T_2 - 2 T_3\,, \quad
2 \T_3 = T_7\,, 
\nonumber \\&&
 2 Px \T_4 = T_1 - T_2 - 2  T_7\,,\quad 
2 Px \T_5 = - T_1 + T_5 + 2  T_8\,, \quad
4 \left(Px\right)^2 \T_6 = 2 T_2 - 2 T_3 - 2 T_4 + 2 T_5 + 2 T_7 + 2 T_8\,,
\nn \\&&
4 Px \T_7 = T_7 - T_8\,, \quad
4 \left(Px\right)^2 \T_8 = -T_1 + T_2 + T_5 - T_6 + 2 T_7 + 2 T_8 \,.
\eea
\end{widetext}

In the following we present the calculation of the $O(x^2)$ corrections
to the light-cone expansion of the three-quark operator in \Gl{zerl} in a simplified 
situation, where positions of two of the three quarks coincide. This approximation is 
sufficient for the derivation of LCSRs to the tree-level accuracy.  
The strategy is based on the approach developed in \cite{BB89, BB91, Ball:1998ff}. 
We present a slightly more expanded derivation than it was
given in \cite{Braun:2001tj}, where only $\V_1^{M}$ was obtained. 
Note that considering  the vector and axial-vector Lorentz projections is sufficient
since the tensor ones can be  determined with the help of isospin relations~\cite{BFMS}. 

Consider first the case where positions of the two $u$-quarks coincide. 
This situation occurs when the $d$-quark interacts with the (electromagnetic) probe, hence we refer to it 
as the $d$-quark contribution:  
\begin{widetext}
\bea{dquark}  
x^\al \bra{0} \ep^{ijk} \left[u^i  
C\gamma_\alpha u^j\right](0) d_\ga^{k}(x)
          \ket{P} &=&   - x^\al \Bigg[  
\left({\cal V}_1  + \frac{x^2m_N^2}{4} \V_1^{M(d)}\right) P_\al \left(\ga_5 N\right)_\ga +  
\V_2 m_N P_\al  
\left(\!\not\!{x} \ga_5 N\right)_\ga  
\nn \\ &&{}\hspace*{-2cm} +  
\V_3 m_N  \left(\ga_\al \ga_5 N\right)_\ga  
+ \V_4 m_N^2   x_\al \left(\ga_5 N\right)_\ga  
+ \V_6 m_N^3  x_\al \left(\!\not{x} \ga_5 N\right)_\ga  
\Bigg],  
\nn
\\
x^\al \bra{0} \ep^{ijk} \left[u^i  
C\gamma_\alpha \gamma_5 u^j\right](0) d_\ga^{k}(x)
          \ket{P} &=&   - x^\al \Bigg[  
\left({\cal A}_1  + \frac{x^2m_N^2}{4} \A_1^{M(d)}\right) P_\al \left(\ga_5 N\right)_\ga +  
\A_2 m_N P_\al  
\left(\!\not\!{x} \ga_5 N\right)_\ga  
\nn \\ &&{}\hspace*{-2cm} +  
\A_3 m_N  \left(\ga_\al \ga_5 N\right)_\ga  
+ \A_4 m_N^2   x_\al \left(\ga_5 N\right)_\ga  
+ \A_6 m_N^3  x_\al \left(\!\not{x} \ga_5 N\right)_\ga  
\Bigg].  
\eea  
\end{widetext}

We remind that $\V_1, \A_1$ start at leading twist-3, and hence $\V_1^{M(d)}, \A_1^{M(d)}$
are of twist-5. 
Strictly speaking, since we are not taking into account twist-6
contributions induced by $O(x^2)$ corrections to $\V_2,\A_2$ and $\V_3,\A_3$, in order
to be consistent we have to discard the contribution of $\V_6$ and $\A_6$ altogether.
This contribution to the sum rules appears to be numerically negligible, 
however.

{}For definiteness, consider the vector projection.
The meaning of the separation between $\V_1$ and $\V_1^{M(d)}$ is most easily 
understood upon the short distance expansion $x_\mu \to 0$. In this 
way, the nonlocal ``light-ray''  operator in the l.h.s. of \Gl{dquark} is 
Taylor-expanded in a series of local operators with three quark fields 
and the increasing number of (covariant) derivatives acting on the d-quark.
The separation of the leading twist part of each local operator 
corresponds to the symmetrisation over all Lorentz indices and  
the subtraction of traces. Without loss of generality, we can consider the 
matrix element contracted with an additional factor $x_\alpha$, 
see \Gl{dquark}, so that the symmetrisation is achieved. To subtract
the traces, we formally write 
\begin{widetext}
\bea{dexpansion}  
\left. x_\al d(x) \right|_{\rm lt}
= \sum_{n= 0}^\infty \frac{1}{n!}  
\Bigg[x_\al x_{\mu_1} \ldots x_{\mu_n} - \frac{x^2}{4}  
\left(\frac{2}{n+1}\right) \sum_{\mu_i,\mu_j}  
\left(x_{\al} \ldots g_{\mu_i \mu_j} \ldots x_{\mu_n}\right)\Bigg]  
\partial^{\mu_1} \ldots \partial^{\mu_n} d(0) \,,  
\eea  
\end{widetext}
where `lt' stands for the leading-twist part.
Observing that $\frac{1}{n+1} = \int_0^1 \dd t \,  t^n $ 
the subtracted contributions $O(x^2)$   
can be reassembled in the form of a non-local string operator:
\bea{tracesubtract}  
\lefteqn{\bra{0} \ep^{ijk} \left[u^i C \!\not\!{x} u^j\right](0) d_\ga^{k}(x)  
          \ket{P}=}
\nn\\  
&=&  
\bra{0} \left[  
\ep^{ijk} \left[u^i  
C \!\not\!{x} u^j\right](0) d_\ga^{k}(x)  
\right]_{\rm l-t}  
\ket{P}  
\\ &+&  
\frac{x^2}{4}\! \int_0^1\!\! \dd t  
\frac{\partial^2}{\partial x_\al \partial x^\al}  
\bra{0} \ep^{ijk} \left[u^i  
C \!\not\!{x} u^j\right](0) d_\ga^{k}(t x)  
          \ket{P}.
\nn
\eea  
Alternatively, the same result can be obtained by observing  \cite{BB89} 
that the leading-twist light-ray operator has to satisfy   
the homogeneous Laplace equation
\beq{laplace}  
\frac{\partial^2}{\partial x_\la \partial x^\la}  
\bra{0} \left[  
\ep^{ijk} \left[u^i  
C \!\not\!{x} u^j\right](0) d_\ga^{k}(x)  
\right]_{\rm lt}  
\ket{P}  
 = 0 \, .  
\eeq  
Using QCD equations of motion the third line in \Gl{tracesubtract}  
can be simplified to   
\bea{eqom1}  
\lefteqn{\frac{\partial^2}{\partial x_\al \partial x^\al}  
\ep^{ijk} \left[u^i  
C \!\not\!{x} u^j\right](0) d_\ga^{k}(t x)  
=}
\nn\\&=&  
2 t \,  
\ep^{ijk} \left[u^i C \ga^\al u^j\right](0) D_\al d_\ga^{k}(t x)  
+ \mbox{\rm gluons}  
\nn \\  
&=&  2 t \, \partial_\al  
\ep^{ijk} \left[u^i C \ga^\al u^j\right](0) d_\ga^{k}(t x)  
+ \mbox{\rm gluons}\,, \phantom{--}{}
\eea  
where $\partial_\alpha$ is a derivative with respect to the overall
translation \cite{BB89};  
for the matrix element we can make the substitution  
$\partial_\al \to - i P_\al$.  
Inserting this result in \Gl{tracesubtract} we finally obtain  
\bea{tracesubtract2}  
\lefteqn{\bra{0} \ep^{ijk} \left[u^i  
C \!\not\!{x} u^j\right](0) d_\ga^{k}(x)  
          \ket{P}=}
\nonumber\\
&=&  \bra{0} \left[  
\ep^{ijk} \left[u^i  
C \!\not\!{x} u^j\right](0) d_\ga^{k}(x)  
\right]_{\rm lt}  
\ket{P}   + \frac{x^2}{4} (-i 2 P_\al) 
\nonumber\\&& 
 ×\int_0^1 \dd t \, t  \bra{0} \ep^{ijk} \left[u^i  
C \ga^\al u^j\right](0) d_\ga^{k}(t x)  
          \ket{P} 
\nonumber\\&&{}
+ \mathrm{gluons} \,.  
\eea  
Notice that the r.h.s. only involves (up to corrections with additional 
gluons) the already known distribution amplitudes.  
This equation therefore allows us to determine $\V^{M(d)}$  
--- which appears on the l.h.s. of \Gl{tracesubtract2} ---  
up to gluonic corrections.  

Consider the first term on the r.h.s. of  \Gl{tracesubtract2}.
We can write  
\bea{maelements}  
\lefteqn{\bra{0} \left[  \ep^{ijk} \left[u^i  
C \!\not\!{x} u^j\right](0) d_\ga^{k}(x)  
\right]_{\rm lt} \ket{P}=}
\nonumber\\  
 &=&{} -   
\int\! \D x \left[e^{-i P\cdot x x_3} (P x) \right]_{\rm lt}  
{\cal V}_1  
\left(\ga_5 N\right)_\ga 
\nn \\
&&{}-  
\int\! \D x \left[e^{-i P\cdot x x_3} \left(\!\not\!{x} \ga_5 N\right)_\ga  
\right]_{\rm lt}  
(Px) \V_2 m_N  
\nn \\ && -  
\int\! \D x \left[e^{-i P\cdot x x_3}  \left(\!\not\!{x}\ga_5 N\right)_\ga  
\right]_{\rm lt} \V_3 m_N   
+ \ldots  
\eea  
where  
$\left[e^{-i P\cdot x x_3} (P\cdot x) \right]_{\rm lt}$ and  
$\left[e^{-i P\cdot x x_3} \!\not\!{x}\right]_{\rm lt}$ are   
the leading-twist components for the free fields, defined as the 
 solutions of the corresponding homogeneous Laplace equation \cite{BB91}.
Note that the factor $Px$ in the second line in \Gl{maelements}
is not included under the $[\ldots]_{\rm lt}$ bracket since 
$(Px)\, \V_2 = 1/2 ( V_1 - V_2 - V_3)$ is a function of 
momentum fractions only and does not contain
any dependence on the position vector $x$. 
To the required ${\cal O}(x^2)$ accuracy, the leading-twist 
exponents $\left[e^{-i P\cdot x x_3} \ldots \right]_{\rm lt}$ can be obtained from the expression given in \cite{BB91}:
\bea{traceless0}  
\left[e^{-i P\cdot x x_3} \right]_{\rm lt}  
&=&  
e^{-i P\cdot x x_3} + \frac{x^2 m_N^2 x_3^2}{4}  
\int_0^1 \dd t \, e^{-i P\cdot x x_3 t} \,,  
\nn   
\eea  
by taking the derivative with respect to $x_3$ and with respect to
$P_\mu$. One gets
\bea{traceless}  
\lefteqn{\hspace*{-1.5cm}\left[e^{-i P\cdot x x_3} (P \cdot x) \right]_{\rm lt}  
 \,=\, (P x) \Big[ e^{-i P\cdot x x_3}} 
\nonumber\\&&{}+\frac{x^2 m_N^2 x_3^2}{4}  
\int_0^1 \dd t \, e^{-i P\cdot x x_3 t}\Big],
\\  
\lefteqn{\hspace*{-1.5cm}\left[e^{-i P\cdot x x_3} \!\not\!{x}\right]_{\rm lt}  
\,=\,  
\!\not\!{x}\;\Big[  
e^{-i P\cdot x x_3} }
\nonumber\\&&{}+ \frac{1}{4} x^2 m_N^2 x_3^2  
\int_0^1 \dd t \, t^2  e^{-i P\cdot x x_3 t}\Big]
\nonumber\\&&{}  
+  
\frac{i}{2} \!\not\!{P} {x_3 x^2} \int_0^1 \dd t \, t e^{-i P\cdot x x_3 t}  
\,.     
\eea  
Note that we have corrected a misprint in \cite{Braun:2001tj} 
in the last term of the second equation, where a factor 1/2 arises
instead of 1/4.

The corresponding contribution to $\V_1^{M(d)}$ is proportional to the 
nucleon mass squared and involves the leading twist distribution amplitude,
being an exact analogue of the Nachtmann power suppressed correction 
in deep inelastic scattering. The second contribution on the 
r.h.s. in \Gl{tracesubtract2} is special for the exclusive kinematics since
it involves a derivative over the total translation that vanishes for 
forward matrix elements. Its explicit form  
is easily found by contracting the three-quark matrix element  
in \Gl{dquark} with $P_\al$ instead of $x_\al$ and inserting the resulting 
expression  in  \Gl{tracesubtract2}. One gets 
\bea{translationterm}  
 \lefteqn{\hspace*{-0.2cm}\frac{x^2}{4} (-2i P_\al)\!\! \int_0^1\!\!\!\! \dd t \, t  
 \bra{0} \ep^{ijk} \! \left[u^i C 
\ga^\al u^j\right](0) d_\ga^{k}(t x) \ket{P}     
=}
\nn\\&=&
\frac{x^2 m_N^2 }{4} i\! \int\! {\cal D}x \! \int_0^1\! \dd t \, t\,  
e^{- i P\cdot x x_3 t} (V_1 + V_5) (\gamma_5 N)_\gamma
\nn\\[-2mm]&&{} + \ldots  \,,  
\eea 
where the ellipses stand for other Lorentz structures that do not contribute  
to $\V_1^{M(d)}$.  
Inserting everything into \Gl{tracesubtract} we arrive at  
\bea{endergebnis}  
&&(P x) \int \dd x_3 e^{-i x_3 P\cdot x}\V_1^{M(d)}(x_3) =
\nonumber\\[-2mm] &&{}\hspace*{0.3cm} = 
(Px) \int {\cal D} x \, x_3^2 \int_0^1 \dd t \,  
e^{- i P\cdot x x_3 t}\, V_1  
\nonumber\\[-2mm] &&{}\hspace*{0.3cm}  - i  
\int {\cal D} x\,  x_3  \int_0^1 \dd t \, t  
e^{- itx_3 P\cdot x} (V_1 -V_2)  
\nonumber\\[-2mm]  &&{}\hspace*{0.3cm}  
+ \frac{1}{Px} \int {\cal D} x\,  
e^{- i x_3 P\cdot x } (- 2 V_1 + V_3 + V_4 + 2 V_5)  
   \, .  
\nn\\[-2mm]
\eea  
In order to solve this equation we expand both sides
at short distances and obtain the moments of  $\V_1^{M(d)}$
with respect to $x_3$ expressed through 
moments of the  distribution amplitudes  
defined as $V_i^{(d)(n)} = \int {\cal D} x \, x_3^n V_i(x_i)$.  
One finds  
\bea{Vd-moments}  
\lefteqn{\int \dd  x_3 \,x_3^n \,\V_1^{M(d)}(x_3) \,=\, 
  \frac{1}{n+1} \bigg[V_1^{(d)(n+2)}}
\nn\\&& 
- \frac{1}{n+3}(V_1 -V_2)^{(d)(n+2)}- \frac{1}{n+3} (V_1 +V_5)^{(d)(n+1)}
 \nn \\ &&{}+
\frac{1}{n+2} (- 2 V_1 + V_3 + V_4 + 2 V_5)^{(d)(n+2)}\bigg],
\eea
up to contributions of multiparton distribution amplitudes with extra
gluons that have been neglected. 
The corresponding expression for $\V_1^{M(d)}(x_3)$ in the momentum fraction space is easily 
obtained by inserting the conformal expansions for   
$V_1, \ldots, V_6$ and inverting the moment equation, see below.

The analysis of the $u$-quark contribution is performed in a 
similar way. We consider the matrix element
\begin{widetext}
\bea{uquark}  
x^\al \bra{0} \ep^{ijk} \left[u^i(0)  
C\gamma_\alpha u^j(x)\right] d_\ga^{k}(0)  
          \ket{P} &=&  - x^\al \Bigg[  
\left({\cal V}_1  + \frac{x^2 m_N^2}{4} 
\V_1^{M(u)}\right) P_\al \left(\ga_5 N\right)_\ga
+\V_2 m_N P_\al  \left(\!\not\!{x} \ga_5 N\right)_\ga    
\nn \\ && \hspace*{-2cm}{}+  
\V_3 m_N  \left(\ga_\al \ga_5 N\right)_\ga  
+ \V_4 m_N^2   x_\al \left(\ga_5 N\right)_\ga  
+ \V_6 m_N^3  x_\al \left(\!\not{x} \ga_5 N\right)_\ga  
\Bigg]  
\eea  
and find repeating the same steps that lead to \Gl{tracesubtract}:  
\bea{tracesubtractu}  
\lefteqn{\bra{0} \ep^{ijk} \left[u^i(0)  
C \!\not\!{x} u^j(x)\right] d_\ga^{k}(0)  
          \ket{P}  \,=\,  
\bra{0} \left[  
\ep^{ijk} \left[u^i(0)  
C \!\not\!{x} u^j(x)\right] d_\ga^{k}(0)  
\right]_{\rm lt}  
\ket{P}}  
 \\ &&{}  
+ \frac{x^2}{4} \int_0^1 \dd t  
\frac{\partial^2}{\partial x_\al \partial x^\al}  
\bra{0} \ep^{ijk} \left[u^i(0)  
C \!\not\!{x} u^j(t x)\right] d_\ga^{k}(0)  
          \ket{P}  
 \,=\, 
\bra{0} \left[  
\ep^{ijk} \left[u^i(0)  
C \!\not\!{x} u^j(x)\right] d_\ga^{k}(0)  
\right]_{\rm lt}  
\ket{P}  
+ \mbox{\rm gluons},
\nn
\eea  
\end{widetext}
the only difference being that the term  
corresponding to a total translation does not arise in this case.   
{}For the  moments with respect to $x_2$ we get
\bea{Mu-moments}  
\lefteqn{\int \dd x_2  \,x_2^n \,\V_1^{M(u)}(x_2) 
\,=\, 
  \frac{1}{n+1} \bigg[V_1^{(u)(n+2)}}
\nn\\&&{} 
- \frac{1}{n+3}(V_1 -V_2)^{(u)(n+2)}
 \nn \\ &&{}+
\frac{1}{n+2} (- 2 V_1 + V_3 + V_4 + 2 V_5)^{(u)(n+2)}  
\bigg].
\nn \\ &&
\eea
Inserting the conformal expansions for   
$V_1, \ldots, V_6$ and inverting the moment equations we find
\bea{VM}
{\cal V}_1^{M(u)} (x_2) &=& \int \limits_0^{1-x_2} dx_1 V_1^{M}(x_1, x_2,1-x_1-x_2)
\nn\\ 
      & = & \frac{x_2^2}{24} \left( f_N C_{f}^u + \lambda_1 C_{\lambda}^u  \right)\,,
\\
 {\cal V}_1^{M(d)} (x_3) &=& \int \limits_0^{1-x_3} dx_1 V_1^{M}(x_1,1-x_1-x_3,x_3)
\nn\\ 
      & = & \frac{x_3^2}{24} \left( f_N C_{f}^d + \lambda_1 C_{\lambda}^d  \right)
\end{eqnarray}
with
\begin{eqnarray}
C_{f}^u       & = & (1 - x_2)^3 \big[113 + 495x_2 - 552x_2^2 
\nn\\&&{}- 10A_1^u(1 - 3x_2)
\nn\\&&{} +  2V_1^d(113 - 951x_2 + 828x_2^2) \big], 
\nonumber
\\
C_{\lambda}^u & = & - (1 \!-\! x_2)^3 
              \big[13 - 20f_1^d + 3x_2 + 10f_1^u(1 \!-\! 3x_2)\big], 
\nonumber
\\
C_{f}^d       & = & - (1 \!-\! x_3) \big[1441 + 505x_3 - 3371x_3^2 + 3405x_3^3 
\nn\\&&{}- 1104x_3^4  - 24V_1^d\big(207 \!-\! 3x_3 \!-\! 368x_3^2 \!+\! 412x_3^3 
\nn\\&&{}- 138x_3^4\big) \big] - 12(73 - 220V_1^d) \ln[x_3],
\nonumber
\\
C_{\lambda}^d & = &  - (1-x_3) \Big[11 + 131x_3 - 169x_3^2 + 63x_3^3 
\nn\\{}&&- 
    30f_1^d(3 + 11x_3 - 17x_3^2 + 7x_3^3) \big]  
\nonumber
\\
& & 
- 12(3 - 10f_1^d) \ln[x_3]\,.
\end{eqnarray}
This result agrees with \cite{Braun:2001tj}.

Similarly we get for the axial-vector functions:
\bea{Ad-moments}  
\lefteqn{\int \dd  x_3 \,x_3^n \,\A_1^{M(d)}(x_3)  
\,=\, 
\frac{1}{n+1} \Big[
 A_1^{(d)(n+2)}} 
\nn\\&&{}
- \frac{1}{n+3} (A_1 -A_2)^{(d)(n+2)} 
\nn
\\&&{}
+\frac{1}{n+2} (- 2 A_1 - A_3 - A_4 + 2 A_5)^{(d)(n+2)}
\nn\\&&{}  
- \frac{1}{n+3}(A_1 +A_5)^{(d)(n+1)}
\Big],
\nn 
\\
\lefteqn{\int \dd  x_2 \,x_2^n \,\A_1^{M(u)}(x_2)  
\,=\, 
\frac{1}{n+1} \Big[
 A_1^{(u)(n+2)} }
\nn\\&&{}- \frac{1}{n+3} (A_1 -A_2)^{(u)(n+2)} 
\nn
\\
&&{}
+\frac{1}{n+2} (- 2 A_1 - A_3 - A_4 + 2 A_5)^{(u)(n+2)}  
\Big].
\nn 
\eea
which is solved by 
\begin{eqnarray}
\A_1^{M(u)} (x_2) &=& \int_0^{1-x_2} dx_1 A_1^{M}(x_1, x_2,1-x_1-x_2)
\nn\\ 
      & = & \frac{ x_2^2}{24} (1 - x_2)^3 \left( f_N D_{f}^u  + \lambda_1 D_{\lambda}^u  \right)\,,
\nonumber \\
\A_1^{M(d)} (x_3) &=& \int_0^{1-x_3} dx_1 A_1^{M}(x_1,1-x_1-x_3,x_3)
\nn\\ 
       &=&  0\,,
\end{eqnarray}
with
\begin{eqnarray}
D_{f}^u       & = & 11 + 45 x_2 - 2 A_1^u (113 - 951 x_2 + 828 x_2^2 ) \nn\\&&{}
+ 10 V_1^d (1 - 30 x_2)\,,
\nn\\
D_{\lambda}^u & = &  29 - 45 x_2 - 10 f_1^u (7 - 9 x_2) - 20 f_1^d (5 - 6 x_2)\,.
\nn\\
\end{eqnarray}
Finally, using the isospin relation in Eq.~(2.20) of \cite{BFMS} one obtains 
the  $T_1^M$ functions in terms of $V_1^M$ and $A_1^M$:
\begin{eqnarray}
\T_1^{M(u)} (x) & = &  \frac{1}{2} 
\left[ V_1^{M(d)} (x) + V_1^{M(u)} (x) -  A_1^{M(u)} (x) \right],
\nn \\
\T_1^{M(d)} (x) & = &  V_1^{M(u)} (x) +  A_1^{M(u)} (x) \,.
\end{eqnarray}
Inserting the above expressions we get:
\begin{eqnarray}
\T_1^{M(u)} (x)  & = & \frac{ x^2}{48}  \left( f_N E_{f}^u  + \lambda_1 E_{\lambda}^u  \right),
\nonumber \\
\T_1^{M(d)} (x)  & = & \frac{ x^2 (1\!-\!x)^3}{6}  \left( f_N E_{f}^d  + \lambda_1 E_{\lambda}^d  \right)
\end{eqnarray}
with
\begin{eqnarray}
E_{f}^u & = & -\Big[
(1 - x)
(1339 + 259x - 2021x^2 + 1851x^3 
\nn\\&&- 552x^4 - 72A_1^u(1 - x)^3(3 - 23x) 
\nonumber \\ &&
 - 24V_1^d(216 - 99x - 134x^2 + 196x^3 - 69x^4))
\Big] 
\nn\\&&{}
- 12(73 - 220V_1^d) \ln[x]\,,
\nonumber \\
E_{\lambda}^u & = & -\Big[
(1 - x) 
(53 + 5x - 43x^2 + 21x^3 
\nn\\&&{}- 30 (7 - x - 5x^2 + 3x^3) f_1^d - 60(1 - x)^3 f_1^u)
\Big] 
\nonumber \\ &&
- 12 (3 - 10 f_1^d) \ln[x]\,,
\nonumber \\
E_{f}^d & = & 31 + 135 x - 138 x^2 
\nn\\&&{}
- (59 - 483 x + 414 x^2) (A_1^u - V_1^d)\,,
\nonumber \\
E_{\lambda}^d & = & 4 (1- 3 x) - 10 (2 - 3 x) (f_1^d+ f_1^u)\,. 
\end{eqnarray}
Note that the $x^2$-corrections do not depend on $\lambda_2$. Our results
agree with the ones obtained in \cite{Huang:2004vf}.

%
\section{Asymptotic Distribution Amplitudes}
%
\setcounter{equation}{0}
{}For completeness we present the set of DAs that is obtained by setting 
contributions of higher conformal spin operators to zero, cf. (\ref{asymptotic}).
The subtlety is that conformal symmetry is broken by nucleon mass corrections.
As a consequence, ``kinematic'' higher twist corrections of higher spin have 
to be retained in order to satisfy EOM. For the relevant parameters we get:  
for twist-3
\bea{tw3-asy}
&&\phi_3^0 = f_N \,, \quad  \phi_3^- = 0\,, \quad \phi_3^+ = 0\,;  
\eea
for twist-4
\bea{tw4-asy}
&&\phi_4^0 = \frac{1}{2} \left(f_N + \la_1 \right) ,\!\quad
\phi_4^+ = - \frac{1}{12} f_N \,,\!\quad
\phi_4^- =- \frac{5}{4}   f_N,
\nn \\
&&\psi_4^0 = \frac12\left(f_N - \la_1 \right)  ,\!\quad
\psi_4^+ =  - \frac{1}{12} f_N \, , \!\quad \psi_4^- =  \frac{5}{4}f_N , 
\nn \\ 
&&\xi_4^0 = \frac{1}{6} \la_2\,, \quad
\xi_4^+ = 0\,,\quad \xi_4^- = 0  \, ;
\eea
for twist-5
\bea{tw5-asy}
&&\phi_5^0 = \frac{1}{2} \left(f_N + \la_1 \right), \quad \psi_5^0 =  \frac12\left(f_N - \la_1 \right)\,
\nn\\  
&&\phi_5^+ = - \frac{1}{18} \left(65 f_N   + 3 \la_1 \right),
\quad
\phi_5^- = - \frac{1}{3} \left(5 f_N - \la_1 \right),
\nn \\
&& \psi_5^+ = -\frac{1}{18} \left(65 f_N  - 3 \lambda_1\right),
\quad 
\psi_5^- = \frac{1}{3} \left(5 f_N  + \la_1  \right),
\nn \\
&&\xi_5^0 = \frac{1}{6} \la_2\,,\quad
\xi_5^+ = - \frac{1}{18} \lambda_2\,,\quad
\xi_5^- = - \frac{1}{3} \la_2 \,,
\eea
and for twist-6 
\bea{tw6-asy}
&&\phi_6^0 = f_N ,\quad 
\phi_6^+ = - \frac{1}{30}\left(5 f_N  + 6 \la_1 \right),\quad
\nn\\&&
\phi_6^- = \frac{1}{10} \left(5 f_N  - 2 \la_1  \right).
\eea
The corresponding twist-3 DAs are:
\bea{Asy-twist-3} 
&&V_1(x_i) = 120 \, x_1 x_2 x_3 f_N\,,\quad
A_1(x_i) = 0 \,,
\nn\\&& 
T_1(x_i) =  120 \, x_1 x_2 x_3 f_N\,;
\eea
twist-4:
\bea{Asy-twist-4} 
&&V_2(x_i)  = 
2 \, x_1 x_2 \left[ 5 (1 + x_3) f_N + 6 \lambda_1  \right] \,,
\nn\\&&
A_2(x_i)  = 30 \, x_1 x_2 (x_1 - x_2) f_N \,,
\nn \\
&&T_2(x_i) = 4 \,x_1 x_2 \lambda_2 \,,
\nn \\
&&V_3(x_i) = 
x_3 \Big[
5 \Big(1 + 2 x_1 x_2 - 4 x_3 
\nn\\&&{}\hspace*{0.3cm}+ 3 (x_1^2 + x_2^2 + x_3^2) \Big) f_N
                   - 6 (1 - x_3) \lambda_1 \Big], 
\nn \\
&&A_3(x_i) = - 2 (x_1-x_2) x_3 
\left[5 (2 - 3 x_3) f_N - 3 \lambda_1
\right],
\nn \\
&&T_3(x_i)  = x_3 \Big[
5 \Big(1 + 2 x_1 x_2 + 2 x_3 
\nn\\&&{}\hspace*{0.3cm}- 3 \left(x_1^2 + x_2^2 + x_3^2 \right) \Big) f_N
                  + (1 - x_3) \lambda_2\Big], 
\nn \\
&&T_7(x_i)  = x_3 \Big[
5 \Big(1 + 2 x_1 x_2 + 2 x_3 
\nn\\&&{}\hspace*{0.3cm}- 3 \left(x_1^2 + x_2^2 + x_3^2 \right) \Big) f_N
                  - (1 - x_3) \lambda_2
\Big]; 
\eea
twist-5:
\bea{Asy-twist-5} 
&&V_4(x_i) = 
- \frac{1}{3} \big(28 - 65 (x_1^2 + x_2^2)- 30 x_1 x_2 - 13 x_3 
\nn\\&&{}\hspace*{0.2cm}- 15 x_3^2 \big) f_N 
- \Big( 1 + (x_1- x_2)^2 - x_3^2 \Big) \lambda_1
\,,
\nn\\
&&A_4(x_i) = \phantom{-}\frac{1}{3} (x_1 -x_2)
\left[ (37 - 80 x_3) f_N -6 \lambda_1 \right]\,,
\nn \\
&&T_4(x_i) = - \frac{1}{6} \Big[
2 \big(28 - 65 (x_1^2 + x_2^2) + 30 x_1 x_2 - 43 x_3 
\nn\\&&{}\hspace*{0.2cm}
+ 15 x_3^2\big) f_N 
- (1 \!+\! x_1^2 \!-\! 6 x_1 x_2 \!+\! x_2^2 \!+\! 2 x_3 \!-\! 3 x_3^2) \lambda_2
\Big],
\nn \\
&&T_8(x_i) = - \frac{1}{6} \Big[
2 (28 - 65 (x_1^2 + x_2^2) + 30 x_1 x_2 - 43 x_3 
\nn\\&&{}\hspace*{0.2cm}
+ 15 x_3^2) f_N 
+ (1 \!+\! x_1^2 \!-\! 6 x_1 x_2 \!+\! x_2^2 \!+\! 2 x_3 \!-\! 3 x_3^2) \lambda_2
\Big],
\nn \\
&&V_5(x_i) = x_3 
\left[ - \frac{2}{3} (28 - 65 x_3) f_N + 2 (1 + x_3) \lambda_1 \right], 
\nn\\
&&A_5(x_i) = 2 (x_1-x_2) x_3 \left(5 f_N - \lambda_1 \right), 
\nn\\
&&T_5(x_i) = \frac{2}{3} x_3 (1 + x_3) \lambda_2 \,.
\eea
and twist-6:
\bea{Asy-twist-6} 
&&V_6(x_i) = \frac{1}{3} (5 + 3 x_3) f_N - \frac{2}{5} (1 - 3 x_3) \lambda_1 \,,
\nn\\ && A_6(x_i) = - \frac{1}{5} (x_1 - x_2) (5 f_N - 2 \lambda_1) \,,
\nn \\ &&
T_6(x_i) = \frac{1}{3} (8 - 6 x_3) f_N\,.
\eea
The corresponding expressions for the 
$x^2$-corrections read:
\bea{Asy-x2} 
&&\V_1^{M(d)}(x) = \frac{x^2}{24} \Big[
(1 - x) \Big( (215 - 529 x + 427 x^2 
\nn\\&&{}\hspace*{0.2cm}
- 109 x^3) + 4 \ln[x] \Big) f_N 
+ 16 (1 - x)^3 \lambda_1 
\Big],
\nn \\
&&\V_1^{M(u)}(x) = \frac{x^2}{72} (1 - x)^3 \left[(565 - 417 x) f_N - 24 \lambda_1 \right],
\nn \\
&&\A_1^{M(u)}(x) = \frac{x^2}{72} (1 - x)^3 \left[(43 + 105 x) f_N  - 24 \lambda_1 \right],
\nn \\
&&\T_1^{M(d)}(x) = \frac{x^2}{9} (1 - x)^3 \left[(76 - 39x) f_N - 6 \lambda_1 \right],
\nn \\
&&\T_1^{M(u)}(x) = \frac{x^2}{48}
\Big[\Big((1 - x)(389 - 1051 x + 949 x^2 
\nn\\&&{}\hspace*{0.2cm}
- 283 x^3) + 4 \ln[x]\Big) f_N
+
16(1 - x)^3 \lambda_1
\Big].
\eea

%
\section{QCD sum rules}
%

In this Appendix we update the QCD sum rules for the shape parameters 
of the higher-twist DAs, which are defined as \cite{BFMS,Braun:2001tj}
\bea{local-operators-2} 
\lefteqn{\hspace*{-1.3cm}\bra{0} \left(u(0) C\ga_\mu u(0)\right) 
\!\not\!{z} \ga_5 \ga^\mu (i z \vecr{D} d)(0) \ket{P}
=}
\nn\\&&{}\phantom{-}\hspace*{2.3cm}=\la_1 f_1^d (pz) M \!\not\!{z} N(P)\,,  
\nn \\
\lefteqn{\hspace*{-1.3cm}\bra{0} \left(u^a(0) C\si_{\mu\nu} u(0)\right) 
 \!\not\!{z} \ga_5 \si^{\mu\nu}
(i z \vecr{D} d)(0) \ket{P}=}
\nn \\ &&{}\hspace*{2.3cm} \phantom{-} =\la_2 f_2^d (pz) M \!\not\!{z} N(P)\,,  
\nn \\
\lefteqn{\hspace*{-1.3cm}\bra{0}\left(u(0) C \ga_\mu \ga_5 i z\Dlr u(0)\right)
\!\not\!{z} \ga^\mu d(0) \ket{P}=} 
\nn \\ &&{} \hspace*{2.2cm}\phantom{-}= \la_1 f_1^u (pz) M \!\not\!{z} N(P)\,, 
\eea
where we have used the notation $i z\cdot\Dlr = i z\cdot (\vecr{D} - \vecl{D})$ and for brevity 
omitted color indices. 

The QCD sum rule estimates for $f_1^d, f_2^d, f_1^u$ are derived from the 
consideration of the two-point correlation functions
\begin{widetext}
\bea{cor1}
&&i \int \dd^4 x \, e^{i p x}
\bra{0} T \{ \ep^{ijk} \left(u^i(x) C \ga_\mu u^j(x)\right) \,
\ga_5 \ga^\mu (i z \vecr{D} d)^k(x) \;
\bar{\eta}_{1}(0) \} \ket{0} 
= \frac{f_1^d |\la_1|^2 M^2 p\cdot z ( \!\not\!{p} + M)}{M^2 - p^2}
+ \ldots
\,, \nn \\
&&i \int \dd^4 x \, e^{i p x} 
\bra{0} T \{ \ep^{ijk} 
\left(u(x) C \ga_\mu \ga_5 i z\Dlr u(x)\right)^{ij} \,
\ga^\mu d^k(x) \;
\bar{\eta}_{1}(0) \} \ket{0}
= \frac{f_1^u |\la_1|^2 M^2 p\cdot z (\!\not\!{p} + M)}{M^2 - p^2}
+ \ldots
\,, \nn \\
&& i \int \dd^4 x \, e^{i p x}
\bra{0} T \{ \ep^{ijk} \left(u^i(x) C\si_{\mu\nu} u^j(x)\right) 
\,\ga_5 \si^{\mu\nu} (i z \vecr{D} d)^k(x) \;
\bar{\eta}_{2}(0) \} \ket{0} 
= \frac{f_2^d |\la_2|^2 M^2 p\cdot z (\!\not\!{p}+M)}{M^2 - p^2}
+ \ldots
\,.
\nn \\[-2mm]
\eea
The dots refer to contributions of excited states 
and different Lorentz structures that we do not consider.
We have calculated the correlation functions in (\ref{cor1})
using a more consistent factorization approximation for the contribution of 
dimension-8 operators compared to \cite{BFMS}, which takes into
account the contribution of nonplanar diagrams. For example, we use
\bea{fac1}
\epsilon^{ijk}\epsilon^{i'j'k'}\langle 0| (u^iC\gamma_\mu \left[
         \vecr{D}_\alpha \vecr{D}_\beta +  \vecr{D}_\beta \vecr{D}_\alpha\right]^{kl}\!u^l)
       (\bar u^{i'}\gamma_\nu C \bar u^{k'})|0\rangle &=& 
       \frac{m_0^2  \langle \bar u u\rangle^2}{216}\left[
        19 g_{\mu\nu}g_{\alpha\beta} -2 (g_{\mu\alpha}g_{\nu\beta}-g_{\nu\alpha}g_{\mu\beta})\right],
\nn\\
\epsilon^{ijk}\epsilon^{i'j'k'}\langle 0| (u^iC\sigma_{\mu\nu} \left[
         \vecr{D}_\xi \vecr{D}_\eta +  \vecr{D}_\eta \vecr{D}_\xi\right]^{kl}\!u^l)
       (\bar u^{i'}\sigma_{\alpha\beta} C \bar u^{k'})|0\rangle &=&
      \frac{5m_0^2 \langle \bar u u\rangle^2}{72}g_{\xi\eta}\left(g_{\mu\alpha}g_{\nu\beta}-g_{\nu\alpha}g_{\mu\beta}\right),
\eea
where $\langle \bar u u\rangle$ is the $u$-quark condensate and 
$m_0^2 = \langle \bar u \sigma g G u\rangle/\langle \bar u u\rangle$.
{}Following the standard procedure and replacing $|\lambda_1|^2$ and $|\lambda_2|^2$ by the corresponding 
sum rules derived from the diagonal correlation functions of the $\eta_1$ and $\eta_2$ currents, respectively:
\bea{sumrule1}
2 (2\pi)^4 m_N^2 |\la_1|^2 &=& \exp(m_N^2/M^2) \left\{ 
M^6 E_3(s_0/M^2)
+ \frac{b}{4} 
M^2 E_1(s_0/M^2) 
+ \frac{a^2}{3} \left(4 - \frac43 \frac{m_0^2}{M^2}\right)\right\},
\nn\\
2 (2\pi)^4 m_N^2\frac{|\la_2|^2}{6} &=& \exp(m_N^2/M^2) \left\{ M^6
E_3(s_0/M^2)
+ \frac{b}{4} M^2 E_1(s_0/M^2) 
\right\},
\eea
we obtain 
\bea{oursumrules}
f_1^d&=&{}  
\frac{ \frac{3}{10} M^6 E_3(s_0/M^2) 
+ \frac{b}{24} M^2 E_1(s_0/M^2) 
+ \frac{a^2}{3} \left(4 - \frac{31}{9} \frac{m_0^2}{M^2}\right)
}{ M^6 E_3(s_0/M^2) + \frac{b}{4} M^2 E_1(s_0/M^2) 
+ \frac{a^2}{3} \left(4 - \frac{4}{3} \frac{m_0^2}{M^2}\right)} \,,
\nn \\
f_1^u &=&  
\frac{ \frac{1}{10} M^6 E_3(s_0/M^2) 
+ \frac{b}{8} M^2 E_1(s_0/M^2) 
- \frac{a^2}{3} \frac{m_0^2}{M^2}
}{ M^6 E_3(s_0/M^2) + \frac{b}{4} M^2 E_1(s_0/M^2) 
+ \frac{a^2}{3} \left(4 - \frac43 \frac{m_0^2}{M^2}\right)} 
\, ,
\nn \\
f_2^d &=&  
\frac{ \frac{8}{5} M^6 E_3(s_0/M^2)}{ 
6 M^6 E_3(s_0/M^2) + \frac{3 b}{2} M^2 E_1(s_0/M^2)} 
\, .
\eea
\end{widetext}
where
\beq{continuum} 
E_n(s_0/M^2) = 1 - e^{(-s_0/M^2)} \sum_{k=0}^{n-1} 
\frac{1}{k!} \left(\frac{s_0}{M^2}\right)^k \,.
\eeq
In all sum rules $M$ is the Borel parameter; we use the interval $1 {\rm GeV^2} \leq M^2 \leq 2 {\rm GeV^2}$, 
with the continuum threshold $\sqrt{s_0} \sim 1.5 \;{\rm GeV}$ and 
also values of the condensates at the scale $\mu^2 = 1~{\rm GeV^2}$ 
\bea{condensates}
a  = - (2\pi)^2 \langle \bar q q \rangle 
&\simeq& 0.55 \; {\rm GeV^3} \,,
\nn \\
b = (2\pi)^2 \langle \frac{\al_S}{\pi} G^2\rangle 
&\simeq& 0.47 \; {\rm GeV^4} \,,
\nn \\
m_0^2 = \frac{\langle \bar q g G q \rangle}{\langle \bar q
q\rangle} 
&\simeq& 0.65 \; {\rm GeV^2} \,.
\eea
With these inputs we find the numbers quoted in (\ref{numericsCZ2}).
These results have smaller errors and supersede the corresponding 
estimates in Ref.~\cite{BFMS}.


%
%
%
%


%

\begin{thebibliography}{99}  

\bibitem{Mcallister:1956ng}
  R.~W.~Mcallister and R.~Hofstadter,
  Phys.\ Rev.\  {\bf 102}, 851 (1956).

\bibitem{Brodsky:1973kr}
  S.~J.~Brodsky and G.~R.~Farrar,
  Phys.\ Rev.\ Lett.\  {\bf 31}, 1153 (1973).

\bibitem{Matveev:1973ra}
  V.~A.~Matveev, R.~M.~Muradian and A.~N.~Tavkhelidze,
  Lett.\ Nuovo Cim.\  {\bf 7}, 719 (1973).


\bibitem{Chernyak:1977as}
  V.~L.~Chernyak and A.~R.~Zhitnitsky,
  JETP Lett.\  {\bf 25}, 510 (1977)
  Sov.\ J.\ Nucl.\ Phys.\  {\bf 31}, 544 (1980); 
  V.~L.~Chernyak, A.~R.~Zhitnitsky and V.~G.~Serbo,
  JETP Lett.\  {\bf 26}, 594 (1977)
  Sov.\ J.\ Nucl.\ Phys.\  {\bf 31},  552 (1980).


\bibitem{Radyushkin:1977gp}
  A.~V.~Radyushkin, JINR report R2-10717 (1977), 
  arXiv:hep-ph/0410276 (English translation); \\ 
  A.~V.~Efremov and A.~V.~Radyushkin,
  Theor.\ Math.\ Phys.\  {\bf 42}, 97 (1980)
  Phys.\ Lett.\ B {\bf 94}, 245 (1980).


\bibitem{Lepage:1979zb}
  G.~P.~Lepage and S.~J.~Brodsky,
  Phys.\ Lett.\ B {\bf 87}, 359 (1979);
  Phys.\ Rev.\ D {\bf 22}, 2157 (1980).

\bibitem{Duncan:1979hi}
  A.~Duncan and A.~H.~Mueller,
  Phys.\ Rev.\ D {\bf 21} (1980) 1636.

\bibitem{Duncan:1979ny}
  A.~Duncan and A.~H.~Mueller,
  Phys.\ Lett.\ B {\bf 90} (1980) 159.

\bibitem{Milshtein:1981cy}
  A.~I.~Milshtein and V.~S.~Fadin,
  Yad.\ Fiz.\  {\bf 33} (1981) 1391.

\bibitem{Milshtein:1982js}
  A.~I.~Milshtein and V.~S.~Fadin,
  Yad.\ Fiz.\  {\bf 35} (1982) 1603.


\bibitem{Isgur:1984jm}
  N.~Isgur and C.~H.~Llewellyn Smith,
  Phys.\ Rev.\ Lett.\  {\bf 52}, 1080 (1984).
  
\bibitem{Isgur:1988iw}
  N.~Isgur and C.~H.~Llewellyn Smith,
  Nucl.\ Phys.\ B {\bf 317}, 526 (1989),
   Phys.\ Lett.\ B {\bf 217} (1989) 535. 

\bibitem{Kroll:1995pv}
  P.~Kroll, M.~Schurmann and P.~A.~M.~Guichon,
  Nucl.\ Phys.\ A {\bf 598}, 435 (1996).

\bibitem{Radyushkin:1998rt}
  A.~V.~Radyushkin,
  Phys.\ Rev.\ D {\bf 58}, 114008 (1998).
  
  
\bibitem{Diehl:1998kh}
  M.~Diehl, T.~Feldmann, R.~Jakob and P.~Kroll,
  Eur.\ Phys.\ J.\ C {\bf 8}, 409 (1999).

\bibitem{Goeke:2001tz}
  K.~Goeke, M.~V.~Polyakov and M.~Vanderhaeghen,
  Prog.\ Part.\ Nucl.\ Phys.\  {\bf 47}, 401 (2001).
  
\bibitem{Diehl:2003ny}
  M.~Diehl,
  Phys.\ Rept.\  {\bf 388}, 41 (2003).
  
\bibitem{Belitsky:2005qn}
  A.~V.~Belitsky and A.~V.~Radyushkin,
  Phys.\ Rept.\  {\bf 418}, 1 (2005). 

  
  
\bibitem{Belitsky:2003nz}
  A.~V.~Belitsky, X.~d.~Ji and F.~Yuan,
  Phys.\ Rev.\ D {\bf 69}, 074014 (2004).
  
\bibitem{Diehl:2004cx}
  M.~Diehl, T.~Feldmann, R.~Jakob and P.~Kroll,
  Eur.\ Phys.\ J.\ C {\bf 39}, 1 (2005).
 
\bibitem{Guidal:2004nd}
M.~Guidal, M.~V.~Polyakov, A.~V.~Radyushkin and M.~Vanderhaeghen,
Phys.\ Rev.\ D {\bf 72}, 054013 (2005).


\bibitem{Ioffe:1982qb}
  B.~L.~Ioffe and A.~V.~Smilga,
  Nucl.\ Phys.\ B {\bf 216}, 373 (1983).


\bibitem{Nesterenko:1982gc}
  V.~A.~Nesterenko and A.~V.~Radyushkin,
  Phys.\ Lett.\ B {\bf 115}, 410 (1982).

  
\bibitem{Shifman:1978bx}
  M.~A.~Shifman, A.~I.~Vainshtein and V.~I.~Zakharov,
  Nucl.\ Phys.\ B {\bf 147}, 385,  448 (1979).

\bibitem{Belyaev:1992xf}
  V.~M.~Belyaev and I.~I.~Kogan,
  Int.\ J.\ Mod.\ Phys.\ A {\bf 8}, 153 (1993).

\bibitem{Castillo:2003pt}
  H.~Castillo, C.~A.~Dominguez and M.~Loewe,
  JHEP {\bf 0503}, 012 (2005).


\bibitem{Bakulev:1991ps}
  A.~P.~Bakulev and A.~V.~Radyushkin,
  Phys.\ Lett.\ B {\bf 271}, 223 (1991).

\bibitem{Nesterenko:1983ef}
  V.~A.~Nesterenko and A.~V.~Radyushkin,
  Phys.\ Lett.\ B {\bf 128}, 439 (1983).


\bibitem{Meissner}
  P.~Mergell, U.~G.~Meissner and D.~Drechsel,
  Nucl.\ Phys.\ A {\bf 596}, 367 (1996);\\
  H.~W.~Hammer, U.~G.~Meissner and D.~Drechsel,
  Phys.\ Lett.\ B {\bf 385}, 343 (1996);\\
  H.~W.~Hammer and U.~G.~Meissner,
  Eur.\ Phys.\ J.\ A {\bf 20}, 469 (2004);\\
  M.~A.~Belushkin, H.~W.~Hammer and U.~G.~Meissner,
  Phys.\ Lett.\ B {\bf 633}, 507 (2006).


\bibitem{Baldini:2005vn}
  R.~Baldini, M.~Mirazita, S.~Pacetti, C.~Bini, P.~Gauzzi and M.~Negrini,
  Nucl.\ Phys.\ A {\bf 755}, 286 (2005).

\bibitem{Gockeler:2003ay}
  M.~G{\"o}ckeler, T.~R.~Hemmert, R.~Horsley, D.~Pleiter, P.~E.~L.~Rakow,
  A.~Sch{\"a}fer and G.~Schierholz
                  [QCDSF Collaboration],
  Phys.\ Rev.\ D {\bf 71} (2005) 034508.

\bibitem{Alexandrou:2004xn}
  C.~Alexandrou, P.~de Forcrand, H.~Neff, J.~W.~Negele, W.~Schroers and A.~Tsapalis,
  Phys.\ Rev.\ Lett.\  {\bf 94} (2005) 021601.




\bibitem{Gockeler:2004vx}
  M.~G{\"o}ckeler {\it et al.}  [QCDSF Collaboration],
  Nucl.\ Phys.\ Proc.\ Suppl.\  {\bf 140} (2005) 399.

\bibitem{Gockeler:2004mn}
  M.~G{\"o}ckeler {\it et al.}  [QCDSF Collaboration],
  Few Body Syst.\  {\bf 36} (2005) 111.

\bibitem{Braun:2001tj}
  V.~M.~Braun, A.~Lenz, N.~Mahnke and E.~Stein,
  Phys.\ Rev.\ D {\bf 65}, 074011 (2002).

\bibitem{Balitsky:1989ry}
  I.~I.~Balitsky, V.~M.~Braun and A.~V.~Koles­ni­chenko,
  Nucl.\ Phys.\ B {\bf 312}, 509 (1989).
  
\bibitem{Chernyak:1990ag}
  V.~L.~Chernyak and I.~R.~Zhitnitsky,
  Nucl.\ Phys.\ B {\bf 345}, 137 (1990).

\bibitem{Braun:1999uj}
  V.~M.~Braun, A.~Khodjamirian and M.~Maul,
  Phys.\ Rev.\ D {\bf 61}, 073004 (2000).
 
\bibitem{Bijnens:2002mg}
  J.~Bijnens and A.~Khodjamirian,
  Eur.\ Phys.\ J.\ C {\bf 26}, 67 (2002).
 
\bibitem{Braun:1997kw}
  V.~M.~Braun,
  arXiv:hep-ph/9801222.
 
\bibitem{Colangelo:2000dp}
  P.~Colangelo and A.~Khodjamirian,
  arXiv:hep-ph/0010175.
 
\bibitem{Wang:2006uv}
 Z.~G.~Wang, S.~L.~Wan and W.~M.~Yang,
 arXiv:hep-ph/0601025.

 \bibitem{Wang:2006su}
 Z.~G.~Wang, S.~L.~Wan and W.~M.~Yang,
 arXiv:hep-ph/0601060.

\bibitem{Lenz:2003tq}
  A.~Lenz, M.~Wittmann and E.~Stein,
  Phys.\ Lett.\ B {\bf 581}, 199 (2004).

\bibitem{Huang:2004vf}
  M.~q.~Huang and D.~W.~Wang,
  Phys.\ Rev.\ D {\bf 69}, 094003 (2004).

\bibitem{Braun:2005be}
 V.~M.~Braun, A.~Lenz, G.~Peters and \hfill \\ A.V.~Radyushkin,
 Phys.\ Rev.\ D {\bf 73} (2006) 034020.

%
\bibitem{PDG}
S.~Eidelman {\it et al.}  [Particle Data Group Collaboration],
Phys.\ Lett.\ B {\bf 592} (2004) 1.
%


\bibitem{Walker94}  
R.~C.~Walker {\it et al.},  
Phys.\ Rev.\ D {\bf 49} (1994) 5671.  

\bibitem{Andivahis94}  
L.~Andivahis {\it et al.},  
Phys.\ Rev.\ D {\bf 50} (1994) 5491.  

\bibitem{Litt70}  
J.~Litt {\it et al.},
Phys.\ Lett.\ B {\bf 31} (1970) 40.

\bibitem{Berger71}
C.~Berger, V.~Burkert, G.~Knop, B.~Langenbeck and K.~Rith,
Phys.\ Lett.\ B {\bf 35} (1971) 87.

\bibitem{Janssens66}  
T.~Janssens, R.~Hofstadter, E.~B.~Huges and M.~R.~Yearian,  
Phys.\ Rev.\ {\bf 142}, 922 (1966).

\bibitem{Arrington:2003df}
J.~Arrington,
Phys.\ Rev.\ C {\bf 68} (2003) 034325.

\bibitem{Christy:2004rc}
  M.~E.~Christy {\it et al.}  [E94110 Collaboration],
  Phys.\ Rev.\ C {\bf 70} (2004) 015206.

\bibitem{Qattan:2004ht}
  I.~A.~Qattan {\it et al.},
  Phys.\ Rev.\ Lett.\  {\bf 94} (2005) 142301.


\bibitem{Lung93}  
A.~Lung {\it et al.},
Phys.\ Rev.\ Lett.\  {\bf 70} (1993) 718.
\bibitem{Kubon02}  
G.~Kubon {\it et al.},
Phys.\ Lett.\ B {\bf 524} (2002) 26.
\bibitem{Anklin98}  
H.~Anklin {\it et al.},
Phys.\ Lett.\ B {\bf 428} (1998) 248.




\bibitem{Zhou01}  
H.~Zhou {\it et al.},   
Phys.\ Rev.\ Lett.\ {\bf 87} (2001), 081801-1.

\bibitem{Rohe99}  
D.~Rohe,   
Phys.\ Rev.\ Lett.\ {\bf 83} (1999), 4257.





\bibitem{Maximon:2000hm}
  L.~C.~Maximon and J.~A.~Tjon,
  Phys.\ Rev.\ C {\bf 62} (2000) 054320.

\bibitem{Blunden:2003sp}
  P.~G.~Blunden, W.~Melnitchouk and J.~A.~Tjon,
  Phys.\ Rev.\ Lett.\  {\bf 91} (2003) 142304.

\bibitem{Arrington:2003ck}
  J.~Arrington,
  Phys.\ Rev.\ C {\bf 69} (2004) 032201.

\bibitem{Chen:2004tw}
  Y.~C.~Chen, A.~Afanasev, S.~J.~Brodsky, C.~E.~Carlson and M.~Vanderhaeghen,
  Phys.\ Rev.\ Lett.\  {\bf 93} (2004) 122301.

\bibitem{Arrington:2004ae}
  J.~Arrington,
  Phys.\ Rev.\ C {\bf 71} (2005) 015202.

\bibitem{Tomasi-Gustafsson:2004ms}
  E.~Tomasi-Gustafsson and G.~I.~Gakh,
  Phys.\ Rev.\ C {\bf 72} (2005) 015209.

\bibitem{Afanasev:2005mp}
  A.~V.~Afanasev, S.~J.~Brodsky, C.~E.~Carlson, Y.~C.~Chen and M.~Vanderhaeghen,
  Phys.\ Rev.\ D {\bf 72} (2005) 013008.

\bibitem{Kondratyuk:2005kk}
  S.~Kondratyuk, P.~G.~Blunden, W.~Melnitchouk and J.~A.~Tjon,
  Phys.\ Rev.\ Lett.\  {\bf 95} (2005) 172503.

\bibitem{Blunden:2005ew}
  P.~G.~Blunden, W.~Melnitchouk and J.~A.~Tjon,
  Phys.\ Rev.\ C {\bf 72} (2005) 034612.



\bibitem{JLab1}
   M.~K.~Jones {\it et al.}  [Jefferson Lab Hall A Collaboration],
   Phys.\ Rev.\ Lett.\  {\bf 84} (2000) 1398;\\
   %
  V.~Punjabi {\it et al.},
  Phys.\ Rev.\ C {\bf 71}, 055202 (2005)
  [Erratum-ibid.\ C {\bf 71}, 069902 (2005)].


\bibitem{JLab2}
O.~Gayou {\it et al.},
Phys.\ Rev.\ C {\bf 64} (2001) 038202;
%
\bibitem{JLab3}
O.~Gayou {\it et al.}  [Jefferson Lab Hall A Collaboration],
Phys.\ Rev.\ Lett.\  {\bf 88} (2002) 092301.


\bibitem{GACC1}
  S.~J.~Barish {\it et al.},
  Phys.\ Rev.\ D {\bf 16} (1977) 3103.

\bibitem{GACC2}
  N.~J.~Baker {\it et al.},
  Phys.\ Rev.\ D {\bf 23} (1981) 2499.

\bibitem{GACC3}
  K.~L.~Miller {\it et al.},
  Phys.\ Rev.\ D {\bf 26} (1982) 537.

\bibitem{GACC4}
  T.~Kitagaki {\it et al.},
  Phys.\ Rev.\ D {\bf 28} (1983) 436.

\bibitem{Budd}
  H.~Budd, A.~Bodek and J.~Arrington,
  arXiv:hep-ex/0308005.

\bibitem{Mainz}
 A.~Liesenfeld {\it et al.}  [A1 Collaboration],
  Phys.\ Lett.\ B {\bf 468} (1999) 20.

\bibitem{Bernard}
  V.~Bernard, L.~Elouadrhiri and U.~G.~Meissner,
  J.\ Phys.\ G {\bf 28} (2002) R1.

\bibitem{GPdata}
S.~Choi {\it et al.},
  Phys.\ Rev.\ Lett.\  {\bf 71} (1993) 3927.

\bibitem{Ahrens}
  L.~A.~Ahrens {\it et al.},
  Phys.\ Lett.\ B {\bf 202} (1988) 284.

\bibitem{BD65} 
J.~D.~Bjorken and S.~D.~Drell,
{\em Relativistic Quantum Fields} (McGraw-Hill, New York, 1965).


\bibitem{BFMS}  
 V.~Braun, R.~J.~Fries, N.~Mahnke and E.~Stein,
 Nucl.\ Phys.\ B {\bf 589} (2000) 381
 [Erratum-ibid.\ B {\bf 607} (2001) 433].


\bibitem{Ioffe:1982ce}
  B.~L.~Ioffe,
  Z.\ Phys.\ C {\bf 18} (1983) 67.

\bibitem{Ioffe:1981kw}
  B.~L.~Ioffe,
  Nucl.\ Phys.\ B {\bf 188} (1981) 317
  [Erratum-ibid.\ B {\bf 191} (1981) 591].

\bibitem{Dosch:1988vv}
  H.~G.~Dosch, M.~Jamin and S.~Narison,
  Phys.\ Lett.\ B {\bf 220}, 251 (1989).

\bibitem{Chung:1981cc}
  Y.~Chung, H.~G.~Dosch, M.~Kremer and D.~Schall,
  Nucl.\ Phys.\ B {\bf 197} (1982) 55.


%
\bibitem{Che84}  
V.L.\ Chernyak and I.R.\ Zhitnitsky, Nucl. Phys. {\bf B246} (1984) 52.

\bibitem{Zhitnitsky:1985dd}
  A.~R.~Zhitnitsky, I.~R.~Zhitnitsky and V.~L.~Chernyak,
  Sov.\ J.\ Nucl.\ Phys.\  {\bf 41} (1985) 284
  [Yad.\ Fiz.\  {\bf 41} (1985) 445].

\bibitem{Huang:1998gp}
  T.~Huang and Z.~H.~Li,
  Phys.\ Rev.\ D {\bf 57} (1998) 1993.


\bibitem{KS87}  
I.D.\ King and C.T.\ Sachrajda, Nucl. Phys. {\bf B279} (1987) 785.

\bibitem{COZ88}  
V.L.\ Chernyak, A.A.\ Ogloblin and I.R.\ Zhitnitsky,  
Sov.\ J.\ Nucl.\ Phys. {\bf 48} (1988) 536;  
 Z.\ Phys.\ C {\bf {42}} (1989) 583.  

\bibitem{Bolz:1996sw}
  J.~Bolz and P.~Kroll,
  Z.\ Phys.\ A {\bf 356} (1996) 327.

%
\bibitem{BB89}  
 I.~I.~Balitsky and V.~M.~Braun,  
 Nucl.\ Phys.\ B {\bf 311}, 541 (1989).  

\bibitem{BB91}
 I.~I.~Balitsky and V.~M.~Braun,
 Nucl.\ Phys.\ B {\bf 361} (1991) 93.

\bibitem{Ball:1998ff}
 P.~Ball and V.~M.~Braun,
 Nucl.\ Phys.\ B {\bf 543} (1999) 201.


\end{thebibliography}
\end{document}